\documentclass[twocolumn,aps,prl,showpacs,superscriptaddress]{revtex4}
\usepackage[dvips]{graphicx}
\usepackage{bm}
\begin{document}
\title{Spin Anisotropy Effects in Dimer Single Molecule Magnets}
\author{Dmitri V. Efremov}
\email{efremov@theory.phy.tu-dresden.de} \affiliation{Institut
f{\"u}r Theoretische Physik, Technische Universit{\"a}t Dresden,
01062 Dresden, Germany}\author{Richard A. Klemm}
\email{klemm@phys.ksu.edu} \affiliation{Department of Physics,
Kansas State University, Manhattan, KS 66506 USA}
\date{\today}
\begin{abstract}
We  present a model of  equal spin $s_1$ dimer single molecule
magnets. The spins within each dimer interact
 via the Heisenberg and the most general set of four quadratic  anisotropic spin
 interactions with respective strengths $J$ and $\{J_j\}$,  and with the magnetic induction ${\bf B}$.
We solve the model exactly for $s_1=1/2$, 1, and 5/2, and for
antiferromagnetic Heisenberg couplings ($J<0$), present ${\bm
M}({\bm B})$ curves at low $T$ for these cases. Low-$T$ $C_V({\bm
B})$ curves for $s_1=1/2$ and electron paramagnetic susceptibility
$\chi({\bm B},\omega)$ for $s_1=1$ are also provided. For weakly
anisotropic dimers, the Hartree approximation yields rather simple
analytic formulas for ${\bm M}({\bm B})$ and $C_V({\bm B})$ at
arbitrary $s_1$ that accurately fit the exact solutions at
sufficiently low $T$ or large $B$. Low-$T$, large-$B$ formulas for
the inelastic neutron scattering cross-section $S({\bm B},{\bm
q},\omega)$ and $\chi({\bm B},\omega)$ with arbitrary $s_1$ and
${\bm B}$ in an extended Hartree approximation are also given. We
show that the various anisotropy interactions give rise to new,
detectable electron paramagnetic resonances and to new peaks in
$S({\bm q},\omega)$, which can provide  precise determinations of
the particular set of microscopic anisotropy interaction
strengths. For antiferromagnetic Heisenberg couplings ($J<0$) and
weak anisotropy interactions ($|J_j/J|\ll 1$), we provide analytic
formulas for the $2s_1$ level-crossing magnetic inductions
$B_{s,s_1}^{\rm
  lc}(\theta,\phi)$, at which
 the  low-$T$ magnetization ${\bm M}({\bm B})$ exhibits
 steps
  and the low-$T$ specific
  heat  $C_V({\bm B})$ exhibits zeroes, surrounded by double peaks
  of uniform height. Strong anisotropy
interactions drastically alter these behaviors, however.  Our
results are discussed with regard to existing experiments on
$s_1=5/2$ Fe$_2$ dimers, suggesting the presence of single-ion
anisotropy in one one them, but apparently without any sizeable
global spin anisotropy.  Further experiments on single crystals of
these and some $s_1=9/2$ [Mn$_4$]$_2$ dimers are therefore
warranted, and we particularly urge electron paramagnetic
resonance and inelastic neutron scattering experiments to be
performed.
\end{abstract}
 \vskip0pt\vskip0pt
\maketitle

\section{I. Introduction}

Single molecule magnets (SMM's) have been under intense study
recently, due to their potential uses in magnetic storage and
quantum computing.\cite{background,sarachik,loss}  The materials
consist of insulating crystalline arrays of identical SMM's  1-3
nm in size, each containing two or more magnetic ions. Since the
magnetic ions in each SMM are surrounded by non-magnetic ligands,
the intermolecular magnetic interactions are usually negligible.
Although the most commonly studied SMM's are the high-spin
Mn$_{12}$ and Fe$_8$,\cite{background,sarachik,loss,Fe8,WS}  such
SMM's contain a variety of ferromagnetic (FM) and
antiferromagnetic (AFM) intramolecular interactions, rendering
unique fits to a variety of experiments difficult.\cite{Fe8spin9}

In addition, there have been many studies of AFM Fe$_n$ ring
compounds, where $n=6, 8, 10, 12,$
etc.\cite{Fering1,Fe6ring,Fe8ring,Fering2}  In these studies,
analyses of inelastic neutron diffraction data and
 the magnetic induction ${\bm B}$ dependence of the
low-temperature $T$ specific heat $C_V$ and magnetization ${\bm
M}$ steps were made, using the isotropic Heisenberg near-neighbor
exchange interaction, the Zeeman interaction, and various
near-neighbor spin anisotropy
interactions.\cite{Fering1,Fe6ring,Fe8ring,Fering2} However, the
rings were so complicated that  analyses of the data using those
simple models were inaccessible to present day
computers.\cite{Fe6ring,Fe8ring} Thus, those authors used either
simulations or phenomenological fits to a first-order perturbation
expansion with different spin anisotropy values for each global
ring spin value.\cite{Fering2,Fe6ring,Fe8ring}

Here we focus on the simpler cases of equal spin $s_1=s_2$
magnetic dimers, for which the full spin anisotropy effects can be
evaluated analytically,  investigated in detail numerically, and
compared with experiment.
 AFM dimers with  $s_1=1/2, 3/2$,\cite{V2neutron,Gudel,V2P2O9,ek} and various
forms of Fe$_2$ with $s_1=5/2$
 were
 studied recently.\cite{Fe2,Fe2mag,Fe2Cl,taft,Fe2Cl3,Fe2Clnew} Several
  Fe$_2$ dimers and effective
$s_1=9/2$ dimers  of the type
[Mn$_4$]$_2$,\cite{Mn4dimer,Mn4dimerDalal} have magnetic
interactions weak enough that their effects can be probed at
$T\approx$ 1K with presently available  ${\bm B}$. A comparison of
our  results with ${\bm M}({\bm B})$ step data on a Fe$_2$ dimer
 strongly suggests  a
substantial presence  of local spin anisotropy.\cite{Fe2Cl}

The paper is organized as follows.  In Section II, we present the
model in the crystal representation and give exact formulas for
the matrix elements.  The general thermodynamics are presented in
Section III, along with the universal behavior of the $M(B)$ and
$C_V(B)$ behavior associated with the energy level crossing.  In
Section IV, we solve the model exactly for $s_1=1/2$, giving
analytic expressions for ${\bm M}$ and the specific heat $C_V$. In
Section V, we discuss the exact solution for $s_1=1$, present the
equations from which the eigenvalues are readily obtained, and
give  examples of the low-$T$ ${\bm M}({\bm B})$ curves.  In
Section VI, we present  examples of the low-$T$ $s_1=5/2$ ${\bm
M}({\bm B})$ curves.  In Section VII, we present our induction
representation results for the eigenstates to first order in the
anisotropy energies.  These are used to obtain the asymptotic
Hartree expressions for the thermodynamic quantities ${\bm M}$ and
$C_V$, and for the inelastic neutron scattering cross-section
$S({\bm B},{\bm q},\omega)$ and the electron paramagnetic
resonance susceptibility $\chi({\bm B},\omega)$  for arbitrary
$s_1$, that are accurate at sufficiently low $T$ and/or large $B$.
In addition, an analytic formula accurate to second order in the
four independent anisotropy exchange energies is provided for
$B_{s,s_1}^{\rm lc}(\theta,\phi)$, the ${\bm B}$ values at which
the $s$th energy level crossing occurs.  In Section VIII, we
consider FM anisotropic interactions sufficiently strong as to
remove the level crossing effects.  Finally, in Section IX, we
summarize our results and discuss them  with regard to
 experiments on Fe$_2$ dimers.

\section{II. The Model in the Crystal Representation}

 We represent the $s_1=s_2$ dimer
quantum states, $|\psi_s^m\rangle$ in terms of the global (total)
spin and magnetic quantum numbers $s$ and $m$, where ${\bm S}={\bm
S}_1+{\bm S}_2$ and $S_z={\bm S}\cdot\hat{\bm z}$ satisfy ${\bm
S}^2|\psi_s^m\rangle=s(s+1)|\psi_s^m\rangle$ and
$S_z|\psi_s^m\rangle=m|\psi_s^m\rangle$, where
$s=0,1,\ldots,2s_1$, $m=-s,\ldots, s$, and we  set $\hbar=1$.
 We also have $S_{\pm}|\psi_s^m\rangle=A_s^{\pm m}|\psi_s^{m\pm1}\rangle$, where
$S_{\pm}=S_x\pm iS_y$ and
\begin{eqnarray}
A_s^m&=&\sqrt{(s-m)(s+m+1)}.\label{Asm}
\end{eqnarray} For an arbitrary  ${\bm B}$,  we assume the Hamiltonian has the
form ${\cal H}={\cal H}_0+{\cal H}_a+{\cal H}_b+{\cal H}_d+{\cal
H}_e$, where
\begin{eqnarray}{\cal H}_0&=&-J{\bm
S}^2/2-\gamma{\bm S}\cdot{\bm B} \end{eqnarray}
 contains the
Heisenberg exchange and Zeeman interactions, the gyromagnetic
ratio $\gamma=g\mu_B$,  $g\approx2$ and $\mu_B$ is the Bohr
magneton. The global axial and azimuthal anisotropy terms
\begin{eqnarray}{\cal H}_b&=&-J_bS_z^2\label{Hb} \end{eqnarray}
and
\begin{eqnarray}{\cal H}_d&=&-J_d(S_x^2-S_y^2),\label{Hd}
\end{eqnarray} respectively, only involve components of ${\bm S}$, but have been the main
anisotropy terms discussed in the SMM
literature,\cite{WS,Mn4dimer} so we have included them for
comparison. Such terms have been commonly studied as an effective
Hamiltonian for a singlet orbital ground state, in which the
tensor global spin interaction  with a fixed spin quantum number
$s$ has the form ${\bm S}\cdot\tensor{\bm\Lambda}\cdot{\bm S}$,
resulting in the principal axes $x,y$ and $z$.\cite{WaldmannNi}
For a dimer, we take $\tensor{\bm \Lambda}$ to be diagonal in the
orientation pictured in Fig. 1. Then
$J=-(\Lambda_{xx}+\Lambda_{yy})/2$,
$J_b=-\Lambda_{zz}+(\Lambda_{xx}+\Lambda_{yy})/2$, and
$J_d=(\Lambda_{yy}-\Lambda_{xx})/2$.  Taking $|J_d/J|\ll1$ and
$|J_b/J|\ll1$ still leaves $J_d/J_b$ unrestricted.  The single-ion
axial and azimuthal anisotropy terms,
\begin{eqnarray}
{\cal H}_a&=&-J_a\sum_{i=1}^2S_{iz}^2\label{lz}\\
 \noalign{\rm and} {\cal
H}_e&=&-J_e\sum_{i=1}^2\Bigl(S_{ix}^2-S_{iy}^2\Bigr),\label{le}
\end{eqnarray}
respectively, arise from spin-orbit interactions of the local
crystal field with the individual spins. These terms have usually
been  neglected in the SMM literature, but have been studied with
regard to complexes containing a single magnetic ion, such as
Ni$^{+2}$,\cite{WaldmannNi} and with regard to clusters of larger
numbers of identical magnetic ions.\cite{WaldmannNi,Almenar}

The local axially and azimuthally anisotropic exchange
interactions
\begin{eqnarray} {\cal
H}_f&=&-J_fS_{1z}S_{2z}\label{Hf},\\
{\cal H}_c&=&-J_c(S_{1x}S_{2x}-S_{1y}S_{2y}),\label{Hc}
\end{eqnarray}
 satisfy
 \begin{eqnarray}
 2{\cal H}_f/J_f&=&{\cal
H}_b/J_b-{\cal H}_a/J_a,\label{Hfequiv}\\
2{\cal H}_c/J_c&=&{\cal H}_d/J_d-{\cal
 H}_e/J_e,\label{Hcequiv}
 \end{eqnarray}
so we need only include either either ${\cal H}_a$ or ${\cal H}_f$
and ${\cal H}_e$ or ${\cal H}_c$, respectively.\cite{WaldmannNi}
That is, if we stick to the Hamiltonian ${\cal H}$, we may
incorporate the effects of ${\cal H}_f$ and ${\cal H}_c$ by
letting $J_b\rightarrow J_b+J_f/2$, $J_a\rightarrow J_a-J_f/2$,
and $J_d\rightarrow J_d+J_c/2$, $J_e\rightarrow J_e-J_c/2$,
respectively. Since  ${\cal H}_a$ and ${\cal H}_e$ describe the
axial and azimuthal anisotropy each single ion attains from its
surrounding environment, they are the physically relevant local
anisotropy interactions.   We note that ${\cal H}_b$ and ${\cal
H}_a$ are symmetric under $\hat{\bm x}\leftrightarrow\hat{\bm y}$,
whereas ${\cal H}_d$ and ${\cal H}_e$ are antisymmetric under
$\hat{\bm x}\leftrightarrow\hat{\bm y}$, independent of $s_1$.

For the case of Fe$_2$,\cite{Fe2} a constituent of the high-spin
SMM Fe$_8$ and the AFM Fe$_n$ rings,\cite{Fe8,Fe6ring,Fe8ring} the
exchange between the Fe$^{+3}$ $s_1=5/2$ spins occurs via two
oxygen ions, and these four ions essentially lie in the same
($xz$) plane.\cite{Fe2,taft} We set the $z$ axis parallel to the
dimer axis, as pictured in Fig. 1.  Since the quantization axis is
along the dimer axis, which is fixed in a crystal, we  denote this
representation as the crystal representation.

We generally expect each of the $J_j$ for $j=a,b,d,e$ to satisfy
$|J_i/J|\ll1$, but there are not generally any other restrictions
upon the various magnitudes of the $J_j$.  Since all dimers known
to date have predominantly AFM couplings ($J<0$), and also because
their magnetizations and specific heats are particularly
interesting, we shall only consider AFM dimers.  In addition,
since no studies on unequal-spin dimers have been reported to our
knowledge, we shall only treat the equal-spin $s_1=s_2$ case.  We
note that for equal spin dimers, the group symmetry of the dimer
environment is $C_{2v}$, so that Dzaloshinski{\u\i}-Moriya
interactions do not arise.\cite{WaldmannNi}  Hence, our
Hamiltonian ${\cal H}$ is the most general quadratic anisotropic
spin Hamiltonian of an equal-spin dimer.

\begin{figure}
\includegraphics[width=0.15\textwidth]{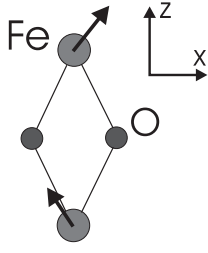}\vskip5pt
\caption{ Sketch of an Fe$_2$ dimer, with two bridging O$^{-2}$
ions (O). Ligands (not pictured) are
 attached to the  Fe$^{+3}$ ions (Fe).  The arrows signify
 spins.}\label{fig1}\end{figure}

For ${\bm B}=B(\sin\theta\cos\phi,\sin\theta\sin\phi,\cos\theta)$,
we have
\begin{eqnarray}
{\cal H}_0|\psi_s^m\rangle&=&E^m_{s}|\psi_s^m\rangle+\delta
E\sum_{\sigma=\pm1}e^{-i\sigma\phi}A_s^{\sigma
m}|\psi_s^{m+\sigma}\rangle,\nonumber\\
& &\\ {\cal
H}_b|\psi_s^m\rangle&=&-J_bm^2|\psi_s^m\rangle,\label{Hbpsi}\\
\noalign{\rm and}\nonumber\\ {\cal
H}_d|\psi_s^m\rangle&=&\frac{-J_d}{2}\sum_{\sigma=\pm1}F_s^{\sigma
m}|\psi_s^{m+2\sigma}\rangle,\label{Hdpsi} \\ \noalign{\rm
where}\nonumber\\
E^m_s&=&-Js(s+1)/2-mb\cos\theta,\\
\delta E&=&-\frac{1}{2}b\sin\theta, \\
b&=&\gamma B.\label{b}\\
\nonumber{\rm
and}\nonumber\\
F_{s}^{\sigma m}&=&A_s^{\sigma m}A_s^{1+\sigma m}.\label{Fsx}
\end{eqnarray}
 ${\cal H}_a$ and ${\cal H}_c$
 contain the individual spin operators $S_{iz}$ and
$S_{i\pm}$ for $i=1,2$. After some Clebsch-Gordon algebra
involving the Wigner-Eckart theorem, for arbitrary  $(s_1,s,m)$,
\begin{eqnarray}
S_{i\pm}|\psi_s^m\rangle&=&\frac{1}{2}A_s^{\pm
m}|\psi_s^{m\pm1}\rangle\mp\frac{1}{2}(-1)^{i}\Bigl(C_{s,s_1}^{\pm
m}|\psi_{s-1}^{m\pm1}\rangle\nonumber\\
& &\qquad-C_{s+1,s_1}^{-1\mp m}|\psi_{s+1}^{m\pm1}\rangle\Bigr),\label{sipm}\\
S_{iz}|\psi_s^m\rangle&=&\frac{m}{2}|\psi_s^m\rangle-\frac{1}{2}(-1)^{i}\Bigl(D_{s,s_1}^m|\psi_{s-1}^m\rangle\nonumber\\
& &\qquad +D_{s+1,s_1}^m|\psi_{s+1}^m\rangle\Bigr),\label{siz}
\end{eqnarray}
 where
 \begin{eqnarray}
 C_{s,s_1}^m&=&\eta_{s,s_1}\sqrt{(s-m)(s-m-1)},\label{Css1m}\\
  D_{s,
s_1}^m&=&\eta_{s,s_1}\sqrt{(s^2-m^2)},\label{Dss1m} \end{eqnarray}
and
\begin{eqnarray}
\eta_{s,s_1}&=&\sqrt{[(2s_1+1)^2-s^2]/(4s^2-1)}.\label{eta}
\end{eqnarray}
The generalizations of these results to dimers with general
$s_1,s_2$ are given in Appendix A.

  For $s=0$, we require $m=0$, for which
$C_{0,s_1}^0=D_{0,s_1}^0=0$.  We then find
\begin{eqnarray}
{\cal
H}_a|\psi_s^m\rangle&=&\frac{-J_a}{2}\Bigl(G_{s,s_1}^m|\psi_s^m\rangle\nonumber\\
& &\qquad
+\sum_{\sigma'=\pm1}H_{s,s_1}^{m,\sigma'}|\psi_{s+2\sigma'}^m\rangle\Bigr),\\
{\cal
H}_e|\psi_s^m\rangle&=&-\frac{J_e}{4}\sum_{\sigma=\pm1}\Bigl(L_{s,s_1}^{\sigma
m}|\psi_s^{m+2\sigma}\rangle\nonumber\\
& &\qquad+\sum_{\sigma'=\pm1}K_{s,s_1}^{\sigma
m,\sigma'}|\psi_{s+2\sigma'}^{m+2\sigma}\rangle\Bigr),\label{Hepsi}\\
G_{s,s_1}^m&=&m^2+\bigl(D_{s,s_1}^m\bigr)^2+\bigl(D_{s+1,s_1}^m\bigr)^2\nonumber\\
&=&
s(s+1)-1\nonumber\\
& &+[2m^2+1-2s(s+1)]\alpha_{s,s_1},\label{Gss1m}\\
H_{s,s_1}^{m,\sigma'}&=&D_{s+(\sigma'+1)/2,s_1}^mD_{s+(3\sigma'+1)/2,s_1}^m,\label{Hss1msigma}\\
K_{s,s_1}^{x,\sigma'}&=&C_{s+(\sigma'+1)/2,s_1}^{-\sigma'x-(1+\sigma')/2}C_{s+(3\sigma'+1)/2,s_1}^{-\sigma'x-(3\sigma'+1)/2},\label{Kss1xsigma}\\
L_{s,s_1}^x&=&2F_s^x\alpha_{s,s_1},\label{Lss1}\\
\noalign{and}
\alpha_{s,s_1}&=&\frac{3s(s+1)-4s_1(s_1+1)-3}{(2s-1)(2s+3)}.\label{alphass1}
\end{eqnarray}
We note that $2\alpha_{s,s_1}=1-\eta_{s,s_1}^2-\eta_{s+1,s_1}^2$
and that Eq. (\ref{alphass1}) holds for $s\ge0$. Equations
(\ref{sipm}) and (\ref{siz}) allow for an exact solution to the
most general Hamiltonian of arbitrary order in the individual spin
operators. In all previous treatments of more complicated spin
systems with similar anisotropic interactions, it was only
possible to obtain numerical solutions, and therefore the full
anisotropy of the magnetization and specific heat was not
calculated.\cite{WaldmannNi,Almenar}
 The operations of ${\cal H}_0$, ${\cal H}_b$ and
${\cal H}_d$  satisfy the selection rules $\Delta s=0$, $\Delta
m=0,\pm1,\pm2$.  The local anisotropy interactions ${\cal H}_a$
and ${\cal H}_e$
 allow transitions satisfying $\Delta s=0,\pm2$,
$\Delta m=0$, and  $\Delta s=0,\pm2$, $\Delta m=\pm2$,
respectively, so in the presence of either of these interactions,
$s$ is no longer a good quantum number, unless $s_1=1/2$.

\section{III. General Thermodynamics}

In order to obtain the thermodynamic properties, we first
calculate the canonical partition function, $Z={\rm
Tr}\exp(-\beta{\cal H})$. Since ${\cal H}$ is not diagonal in the
$(s,m)$ representation, we must construct the wave function from
all possible spin states. We then write
\begin{eqnarray}
Z&=&{\rm Tr}\langle \Psi_{s_1}|e^{-\beta{\cal
H}}|\Psi_{s_1}\rangle,
\end{eqnarray}
where $|\Psi_{s_1}\rangle$ is constructed from the
$\{|\psi_s^m\rangle\}$ basis as
\begin{eqnarray}
\langle\Psi_{s_1}|&=&\Bigl(\langle\psi_{2s_1}^{2s_1}|,
\langle\psi_{2s_1}^{2s_1-1}|, \ldots, \langle\psi_1^0|,
\langle\psi_1^{-1}|, \langle\psi_0^0|\Big),\nonumber\\
\end{eqnarray}
where $\beta=1/(k_BT)$ and  $k_B$ is Boltzmann's constant. To
evaluate the trace, it is  useful to diagonalize the $\langle
\Psi_{s_1}|{\cal H}|\Psi_{s_1}\rangle$ matrix. To do so, we let
$|\Psi_{s_1}\rangle={\bm U}|\Phi_{s_1}\rangle$, where
\begin{eqnarray}
\langle\Phi_{s_1}|&=&\Bigl(\langle\phi_{n_{s_{1}}}|,\langle\phi_{n_{s_{1}}-1}|,\ldots,\langle\phi_1|\Bigr),
\end{eqnarray}
is constructed from the new orthonormal basis
$\{|\phi_n\rangle\}$, and ${\bm U}$ is a unitary matrix of rank
$n_{s_{1}}=(2s_1+1)^2$. Choosing ${\bm U}$ to diagonalize ${\cal
H}$, $U{\cal H}U^{\dag}=\tilde{\cal H}$, we generally obtain
$\tilde{\cal H}|\phi_n\rangle=\epsilon_n|\phi_n\rangle$ and the
partition function for a SMM dimer,
\begin{eqnarray}
Z&=&\sum_{n=1}^{n_{s_{1}}}\exp(-\beta\epsilon_n).
\end{eqnarray}
The specific heat $C_V=k_B\beta^2\partial^2\ln Z/\partial\beta^2$
is then easily found at all $T, {\bm B}$, \begin{eqnarray}
C_V&=&\frac{k_B\beta^2}{Z^2}\biggl[Z\sum_{n=1}^{n_{s_1}}\epsilon_n^2e^{-\beta\epsilon_n}
-\Bigl(\sum_{n=1}^{n_{s_1}}\epsilon_ne^{-\beta\epsilon_n}\Bigr)^2\biggr].\label{sh}
\end{eqnarray}
  The magnetization
\begin{eqnarray}
{\bm M}&=&-\frac{1}{Z}\sum_{n=1}^{n_{s_{1}}}{\bm\nabla}_{\bm
B}(\epsilon_n)\exp(-\beta\epsilon_n),\label{mag}
\end{eqnarray}
 requires ${\bm\nabla}_{\bm
B}(\epsilon_n)$ for each ${\bm B}$.  As $T\rightarrow0$, at most
two eigenstates are relevant.  For most $B$ values, only one
$\epsilon_n$ is important.  But near the reduced $s$th
level-crossing induction  $b_s^{*}=\gamma B_{s,s_1}^{\rm
lc}(\theta,\phi)$ at which $\epsilon_s=\epsilon_{s-1}$, two
eigenstates are relevant. We then  obtain for these two energies
\begin{eqnarray}
C_V(b)/k_B&\approx &\frac{(\beta\Delta\epsilon_s/2)^2}{\cosh^2(\beta\Delta\epsilon_s/2)},\\
M(b)&\approx &\frac{1}{2}(M_s+M_{s-1})\nonumber\\
& &+\frac{1}{2}(M_s-M_{s-1})\tanh(\beta\Delta\epsilon_s/2),\\
\noalign{\rm where}
\Delta\epsilon_s&=&\epsilon_s(b)-\epsilon_{s-1}(b),
\end{eqnarray}
and $M_s=\gamma\nabla_b\epsilon_s(b)$ is the magnetization of the
$s$th eigenstate only.
 We then expand in powers of $b-b_s^{*}$,
\begin{eqnarray}
\epsilon_s(b)&=&\epsilon_s(b_s^{*})+(b-b_s^{*})a_{1,s}+\frac{1}{2}(b-b_s^{*})^2a_{2,s}+\ldots,\label{epsilonsb}\nonumber\\
\end{eqnarray}
and we find
\begin{eqnarray} C_{V}(b_s^{*})&{\rightarrow\atop{T\rightarrow0}}&0,\label{CVzeroes}\\
C_V\Bigl[b_s^{*}\pm
\frac{2c}{\beta(a_{1,s}-a_{1,s-1})}\Bigr]&{\rightarrow\atop{k_BT\ll|J|}}&C_V^{\rm peak}\label{peak}\\
C_V^{\rm peak}/k_B&= &\Bigl(\frac{c}{\cosh
c}\Bigr)^2\nonumber\\
&\approx &0.439229,\label{peakheight}
 \end{eqnarray} and
\begin{eqnarray}
M(b_s^{*})/\gamma&{\rightarrow\atop{T\rightarrow0}}&
\frac{1}{2}(a_{1,s}+a_{1,s-1})\nonumber\\
&=&s-1/2+{\cal O}(J_j/J)^2,\label{Msteps}\\
\frac{dM}{\gamma^2db}\Bigr|_{b_s^{*}}&{\rightarrow\atop{T\rightarrow0}}&\frac{\beta}{4}(a_{1,s}-a_{1,s-1})^2\nonumber\\
&=&\frac{\beta}{4}\Bigl(1+{\cal O}(J_j/J)^2\Bigr),\label{slope}
\end{eqnarray}  where $s=1,\ldots,2s_1$ and $c\approx 1.19967864$ is the solution to $\tanh
c=1/c$. The coefficients $a_{1,s}=s+{\cal O}(J_j/J)^2$ and
$a_{2,s}={\cal O}(J_j/J)^2$.   The easiest way to see this is to
first rotate the crystal so the quantization axis is along ${\bm
B}$, as discussed in Appendix B. An expression for $a_{1,s}$ to
second order in the $J_j/J$ is given in Appendix D. We note that
$b_s^{*}=\gamma B_{s,s_1}^{\rm lc}(\theta,\phi)$ depends upon $s,
s_1$, and the direction of ${\bm B}$ when anisotropic interactions
are present.

The $C_V(B)$ double peaks are double Schottky anomalies at the
reduced induction values $b_s^{*}\pm \frac{2c}{\beta}+{\cal
O}(J_j/J)^2$.
 Hence, to ${\cal }(J_j/J)$, the heights
and midpoint slopes of the $2s_1$ low-$T$ $M(B)$ steps are uniform
and the same as for the isotropic case, but the step positions and
hence their plateaus are not. Correspondingly, the heights and
positions of the $2s_1$  $C_V(B)$ double peaks are uniform and the
same as for the isotropic case, but positions of the zeroes are
not.  Hence, the non-universal level-crossing inductions
$B_{s,s_1}^{\rm lc}(\theta,\phi)$ fully determine the low-$T$
thermodynamics of weakly anisotropic AFM dimers.

In the next three sections, we consider the special cases of
$s_1=1/2,1$ and 5/2.  Then, in Section VII and Appendix D, we
present out general expression for $B_{s,s_1}^{\rm
lc}(\theta,\phi)$ accurate to second order in each of the $J_j$.
We remark that a double peak in the low-$T$ $C_V(B)$ curve has
been seen experimentally in a much more complicated Fe$_6$ ring
compound, and was attributed to level crossing.\cite{Fe6ring}

\section{IV. Analytic results for spin 1/2}

Plots of $C_V/k_B$ and  $M/\gamma$ versus $\gamma B/|J|$ for the
isotropic spin 1/2 dimer were given previously.\cite{ek}  For
$s_1=1/2$ with an arbitrary ${\bm B}$ and $J_j$ for $j=a,b,d,e$,
the rank 4 Hamiltonian matrix is block diagonal, since $s=0,1$ is
a good quantum number.
 The eigenvalues  are given by
\begin{eqnarray}
\epsilon_1&=&-\frac{J_a}{2},\label{epsilon1}\\
\epsilon_n&=&-\frac{J_a}{2}-J+\lambda_n,\qquad
n=2,3,4,\label{epsilon234}
\end{eqnarray}
where
\begin{eqnarray}
0&=&\lambda^3_n+2\lambda^2_nJ_b-\lambda_n[J_d^2-J_b^2+b^2]\nonumber\\
& &-b^2\sin^2\theta[J_b-J_d\cos(2\phi)].\label{halfeigen}
\end{eqnarray}
 For the special cases
  ${\bm B}||\hat{\bm
i}$ for $i=x,y,z$, the $\lambda_n^i$ satisfy
\begin{eqnarray}
\lambda_n^z&=&0,-J_b\pm F_z,\label{lambdaz}\\
\lambda_n^{x,y}&=&-2J_{y,x}, \>\>-J_{x,y}\pm
F_{x,y},\label{lambdaxy}
\end{eqnarray}
where \begin{eqnarray}
 F_i&=&\sqrt{b^2+J_i^2},\label{Fz}\\
 J_{x,y,z}&=&(J_b\pm J_d)/2,\>\> J_d,\label{Jxy}
\end{eqnarray}
respectively, and where $J_x$ ($J_y$) corresponds to the upper
(lower) sign.

 The magnetization for ${\bm B}||\hat{\bm i}$ is given
by
\begin{eqnarray}
M_i&=&\frac{\gamma b\sinh(\beta F_i)}{F_i{\cal D}_i},\label{magi} \\
{\cal D}_i&=&\cosh(\beta F_i)+\Delta_i/2,\label{Di}\\
\Delta_x&=&\exp(-\beta J_{x})[\exp(-\beta J)+\exp(2\beta J_{y})],\\
\Delta_y&=&\exp(-\beta J_{y})[\exp(-\beta J)+\exp(2\beta J_{x})],\\
\Delta_z&=&\exp(-\beta J_b)[\exp(-\beta J)+1],
\end{eqnarray}
 respectively.
  When the
interactions are written in terms of the less physical $J_f, J_b,
J_d$, and $J_c$, then $J_b\rightarrow J_b+J_f$ and $J_d\rightarrow
J_d+J_c/2$, so that for $s_1=1/2$, $J_f$ and $J_c$ merely
renormalize $J_b$ and $J_d$.  Neither of the single-ion spin
anisotropy terms ${\cal H}_a$ and ${\cal H}_e$ affect the
thermodynamics for $s_1=1/2$. We note that $M_y(J_d)=M_x(-J_d)$
for each $B$, as expected from ${\cal H}_d=-J_d(S_x^2-S_y^2)$.

 From Eqs.
(\ref{epsilon1}), (\ref{epsilon234}), (\ref{lambdaz}), and
(\ref{lambdaxy}),  the single level crossing induction ($s=1$) for
am $s_1=1/2$ dimer  with ${\bm B}||\hat{\bm i}$ occurs at
\begin{eqnarray}
 \gamma  B^{\rm
lc}_{1,1/2}&=&\left\{\begin{array}{cc}\sqrt{J^2+J(J_b\pm
J_d)},&{\bm
B}||\hat{\bm x},\hat{\bm y}\\
\sqrt{(J+J_b)^2-J_d^2},&{\bm B}||\hat{\bm z}\end{array}\right.
,\label{Bstephalf}
\end{eqnarray}
 provided that  $J+J_{x,y}<0$ and $J+J_b<0$,
respectively.

To distinguish the different effects of the  global anisotropy
interactions $J_b$ and $J_d$ that affect the magnetization of
$s_1=1/2$ dimers, in Figs. 2 and 3,  we have respectively plotted
the low-$T$ $ M/\gamma$ versus $\gamma B/|J|$ with $J_b=0.1J,
J_d=0$ and $J_d=0.1J, J_b=0$ for ${\bm B}|||\hat{\bm z}$ (solid)
and ${\bm B}||\hat{\bm x}$ (dashed), along with the isotropic case
$J_b=J_d=0$ (dotted).   From Fig. 2, $J_b<0$ and $J_d=0$ causes a
greater shift to higher $B$ values at the magnetization step with
${\bm B}||\hat{\bm z}$ than for ${\bm B}||\hat{\bm x}$, consistent
with Eq. (\ref{Bstephalf}).  In addition, $J_d$ finite with
$J_b=0$ has a very different effect upon the anisotropy of the
magnetization step, as shown in Fig. 3. Although for ${\bm
B}||\hat{\bm z}$, $B$ at the step is slightly reduced from its
isotropic interaction value, for ${\bm B}||\hat{\bm x}$, the
magnetization step occurs at a larger $B$. These results are
consistent with Eq. (\ref{Bstephalf}).  The midpoint slopes are
universal, in accordance with Eq. (\ref{slope}).

\begin{figure}
\includegraphics[width=0.45\textwidth]{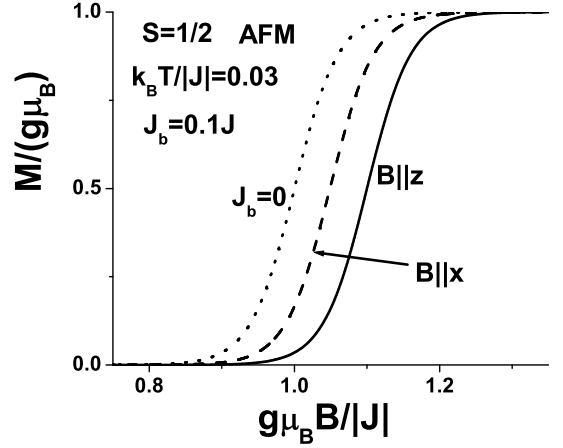}
\caption{Plots of $M/\gamma$ versus $\gamma B/|J|$ at
$k_BT/|J|=0.03$ for the AFM spin 1/2 dimer with $J_b=0.1J$,
$J_d=0$, with ${\bm B}||\hat{\bm z}$ (solid), ${\bm B}||\hat{\bm
x}$ (dashed), along with the isotropic case $J_b=J_d=0$ (dotted).}
\label{fig2}
\end{figure}

\begin{figure}
\includegraphics[width=0.45\textwidth]{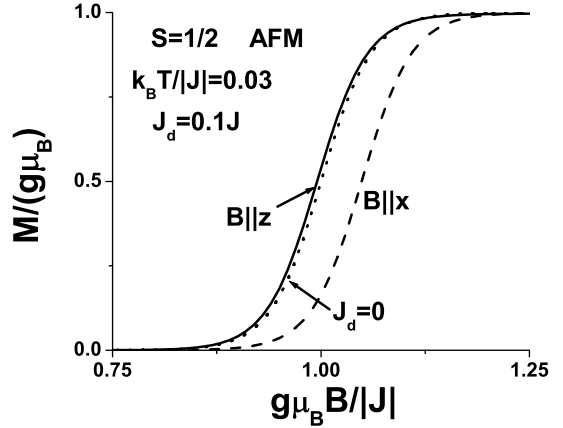}
\caption{Plots of $M/\gamma$ versus $\gamma B/|J|$ at
$k_BT/|J|=0.03$ for the AFM spin 1/2 dimer with $J_d=0.1J$,
$J_b=0$, with ${\bm B}||\hat{\bm z}$ (solid), ${\bm B}||\hat{\bm
x}$ (dashed), along with the isotropic case $J_b=J_d=0$ (dotted).}
\label{fig3}
\end{figure}

The specific heat of an $s_1=1/2$ dimer with ${\bm B}||\hat{\bm
i}$ is
\begin{eqnarray}
C_{Vi}&=&\frac{k_B\beta^2{\cal N}_i}{{\cal D}_i^2},\end{eqnarray}
 where the ${\cal D}_i$ are given by Eq. (\ref{Di}), and
 the ${\cal N}_i$ are given in Appendix A.
Plots at
 low $T$ of $C_V/k_B$ versus $\gamma B/|J|$ for $s_1=1/2$ dimers
with the corresponding global anisotropies $J_b=0.1J$ and
$J_d=0.1J$ are shown in Figs. 4 and 5, respectively.  We note the
universal curve shapes, but non-universal level-crossing
positions, in quantitative agreement with Eqs.
(\ref{CVzeroes})-(\ref{peakheight}). In Fig. 4, the positions of
the maxima and the central minimum in $C_V$ track that of the
magnetization step in Fig. 2 with the same parameters. With
$J_d=0.1J$, the behaviors in $C_V$ and $M$ for ${\bm B}||\hat{\bm
x}$ are also very similar. However, there is a slight difference
in the behaviors for ${\bm B}||\hat{\bm z}$. Note that $M$ (Fig.
3) shows a slight reduction for ${\bm B}||\hat{\bm z}$ in the
induction required for the step, whereas $C_V$ (Fig. 5) shows a
slight increase in the positions of the peaks.  This detail  only
appears when $J_d\ne0$, for which the effective temperature is
slightly higher than for $J_d=0$.

\begin{figure}
\includegraphics[width=0.45\textwidth]{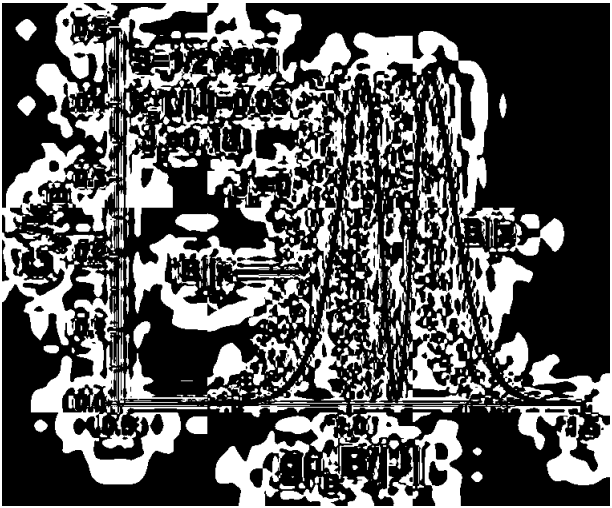}
\caption{Plot of $C_V/k_B$ versus $\gamma B/|J|$ for the AFM spin
1/2 dimer with $J_b=0.1J, J_d=0$ at $k_BT/|J|=0.03$ with the same
curve notation as in Fig. 2.}\label{fig4}
\end{figure}

\begin{figure}
\includegraphics[width=0.45\textwidth]{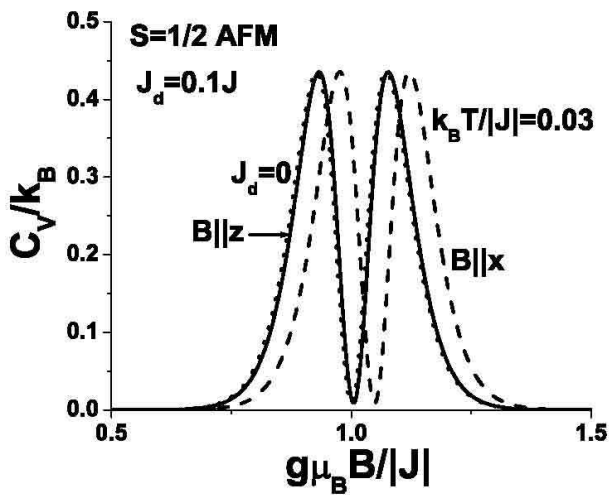}
\caption{Plot of $C_V/k_B$ versus $\gamma B/|J|$ for the AFM spin
1/2 dimer with $J_d=0.1J$, $J_b=0$  at $k_BT/|J|=0.03$ with the
same curve notation as in Fig. 3.}\label{fig5}
\end{figure}

\section{V. Analytic and numerical results for spin 1}

For dimers with $s_1=1$, the allowed $s$ values are $s=0,1,2$. The
three $s=1$ states are decoupled from the six remaining $s=0,2$
states. They satisfy a cubic equation given in Appendix A. For
${\bm B}||\hat{\bm i}$ with $i=x,y,z$, this cubic equation
simplifies to a linear and a quadratic equation, as for the $s=1$
eigenstates of $s_1=1/2$ dimers.

The remaining six eigenstates corresponding nominally to $s=0,2$
are in general all mixed.  The  matrix  leading to the hexatic
equation from which the six eigenvalues can be obtained is given
in Appendix A. That is sufficient to evaluate the eigenvalues for
$s_1=1$ using symbolic manipulation software. When combined with
the three $s=1$ eigenvalues, one can then use Eqs. (\ref{sh}) and
(\ref{mag}) to obtain the resulting exact magnetization and
specific heat at an arbitrary ${\bm B}$. The combined nine
eigenvalues depend upon all four anisotropy parameters $J_j$ for
$j=a,b,d,e$.

 To the extent that the eigenvalues can be obtained from the
solutions to either linear or quadratic equations, the expressions
for the $B_{s,1}^{\rm lc}$ are simple, and are given in Appendix
A. In the global anisotropy case $J_a=J_e=0$, the first level
crossing induction $B_{1,1}^{\rm lc}$ with ${\bm B}||\hat{\bm i}$
is identical to $B_{1,1/2}^{\rm lc}$, the level crossing with
$s_1=1/2$ given by Eq. (\ref{Bstephalf}). For comparison, we
expand the first and second level crossing inductions to first
order in each of the $J_j$ for $j=a,b,d,e$ for ${\bm B}||\hat{\bm
i}$ where $i=x,y,z$,
\begin{eqnarray}
\gamma B_{1,1,z}^{{\rm
lc}(1)}&=&-J+\frac{J_a}{3}-J_b,\label{b11z}\\
\gamma B_{1,1,x,y}^{{\rm lc}
(1)}&=&-J-\frac{J_a}{6}-\frac{J_b}{2}\mp\frac{1}{2}(J_d-J_e),\label{b11xy}\\
\gamma B_{2,1,z}^{{\rm lc} (1)}&=&-2J-J_a-3J_b,\label{b21z}\\
\gamma B_{2,1,x,y}^{{\rm lc}
(1)}&=&-2J+\frac{J_a}{2}-\frac{J_b}{2}\mp\frac{5J_d}{2}\mp\frac{3J_e}{2},\label{b21xy}
\end{eqnarray}
where the upper (lower) sign is for ${\bm B}||\hat{\bm x}$ (${\bm
B}||\hat{\bm y}$), respectively.

From these simple first-order results, it is possible to
understand the qualitatively different  behavior obtained with
local, single-ion, anisotropy from that obtained with global
anisotropy. In the isotropic case $J_j=0\>\>\forall j$, the first
and second level crossings occur at $-J$ and $-2J$, respectively.
For each induction direction,  the signs of the $J_b$ and $J_d$
contributions to $B_{2,1}^{{\rm lc}(1)}+2J$ and $B_{1,1}^{{\rm
lc}(1)}+J$ are the same, whereas the signs of the $J_a$ and $J_e$
contributions to  $B_{2,1}^{{\rm lc}(1)}+2J$ and $B_{1,1}^{{\rm
lc}(1)}+J$ are the {\it opposite}.

In Figs. 6-10, we plot $ M/\gamma$  versus $\gamma B/|J|$ for
  five low-$T$ cases of AFM $s_1=1$ dimers, taking
 $k_BT/|J|=0.03$.  These curves all exhibit the universal features predicted by Eqs. (\ref{Msteps}) and (\ref{slope}).
 The corresponding $C_V/k_B$ versus $\gamma B/|J|$ curves, which also obey Eqs. (\ref{CVzeroes})-(\ref{peakheight}),
 are presented elsewhere.\cite{ekcondmat}
 In these figures, only one of the five
 anisotropy interactions $J_j$ is non-vanishing, and we take
 $J_j/J=0.1$, for $j= b, d, a, e,$ and $c$, respectively.  Unlike the case of $s_1=1/2$
 dimers,  for which $J_c$ merely renormalizes $J_d$, for $s_1=1$
 each of these interactions leads to distinct anisotropy
 effects in the low-$T$ thermodynamics arising from the $B_{s,1}^{\rm lc}(\theta,\phi)$.

 \begin{figure}
\includegraphics[width=0.45\textwidth]{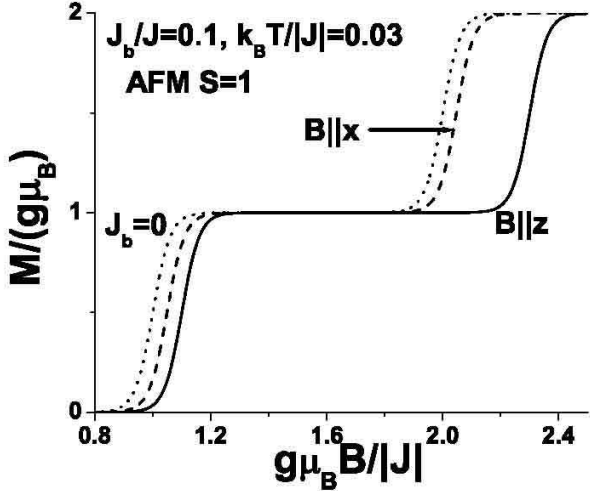}
\caption{Plot at $k_BT/|J|=0.03$ and $J_b/J=0.1$ of $M/\gamma$
versus $\gamma B/|J|$ for the AFM spin 1 dimer, with the same
curve notation as in Fig. 2.}\label{fig6}
\end{figure}

 We first show the results for the global anisotropy interactions.
  In Fig. 6, we plot $M(B)$ for $J_b=0.1J$ (and all other $J_j=0$) at
  $k_BT/|J|=0.03$ for the AFM $s_1=1$ dimer.  $B_{s,1}^{\rm lc}(\theta,0)$ is largest and increases monotonically
  with level-crossing number $s$ for $\theta=0$, but is independent of $s$ for $\theta=\pi/2$,  nearly quantitatively
  consistent with Eqs. (\ref{b11z})-(\ref{b21xy}).
    In Fig. 7, the
  corresponding results for $J_d=0.1J$ (and the other $J_j=0$) are
  shown.  Again, $B_{s,1}^{\rm lc}(\theta,0)$ is largest and
  increases monotonically for $\theta=\pi/2$, and is nearly independent
  of $s$ for $\theta=0$, as for $s=1/2$,
   nearly quantitatively consistent with Eqs.
   (\ref{b11z})-(\ref{b21xy}). Note that $B_{s,1}^{\rm lc}(0,0)$
   is nearly indistinguishable from the isotropic case.

\begin{figure}
\includegraphics[width=0.45\textwidth]{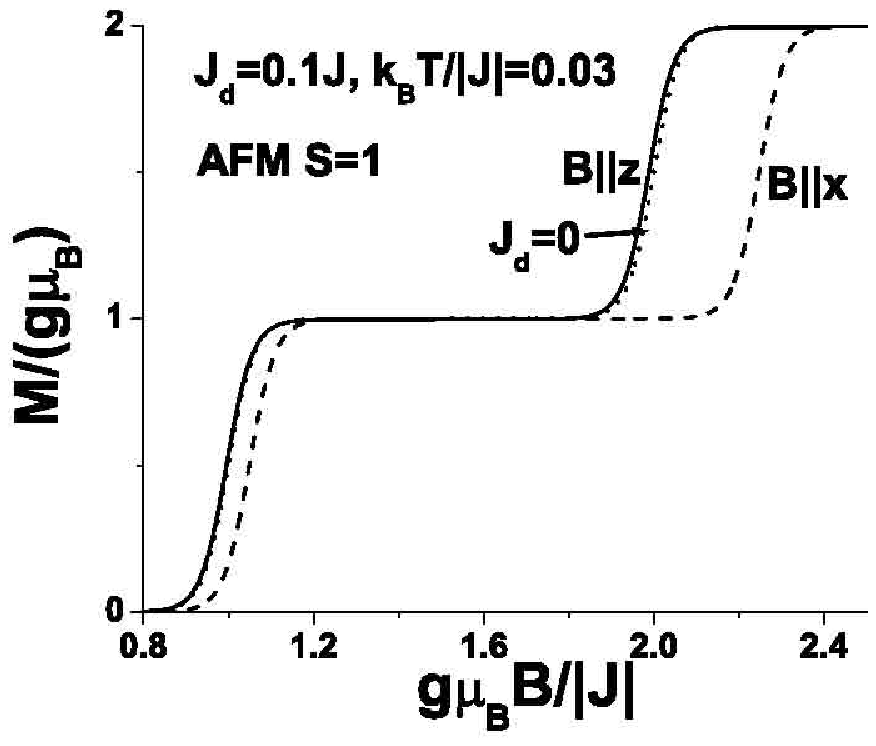}
\caption{Plot at $k_BT/|J|=0.03$ and $J_d/J=0.1$ of ${\bm
M}/\gamma$ versus $\gamma B/|J|$ for the AFM spin 1 dimer, with
the same curve notation as in Fig. 3.}\label{fig7}
\end{figure}

  Next, in Figs. 8 and 9, we show the corresponding curves for
  $M(B)$  at low $T$ for the local anisotropies
  $J_a=0.1J$ and $J_e=0.1J$, respectively, with the other $J_j=0$.  In contrast
  to the global anisotropies,  $B_{s,1}^{\rm lc}(\theta,0)+sJ$
  changes sign for both $\theta=0,\pi/2$, nearly quantitatively
  consistent with Eqs. (\ref{b11z})-(\ref{b21xy}).  We note that
  $B_{s,1}^{\rm lc}(0,0)$ is nearly indistinguishable from the
  isotropic case for $J_e\ne0$ shown in Fig. 9.

  \begin{figure}
\includegraphics[width=0.45\textwidth]{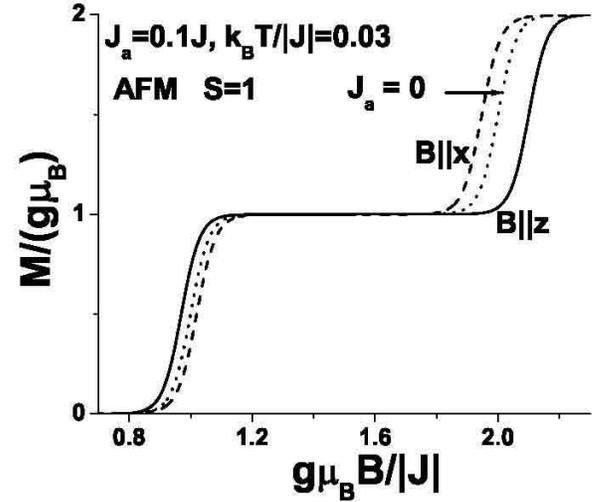}
\caption{Plot at $k_BT/|J|=0.03$ and $J_a/J=0.1$ of $M/\gamma$
versus $\gamma B/|J|$ for the AFM spin 1 dimer.  The curve
notation is the same as in Fig. 2, except that the isotropic case
(dotted) curve is for $J_a=0$.}\label{fig8}
\end{figure}

\begin{figure}
\includegraphics[width=0.45\textwidth]{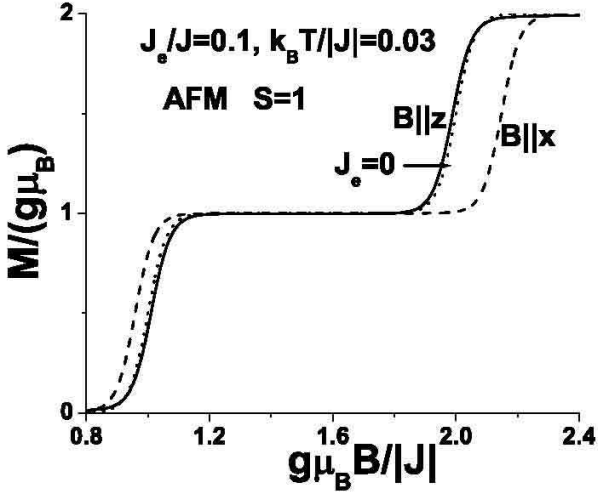}
\caption{Plot at $k_BT/|J|=0.03$ and $J_e/J=0.1$ of $M/\gamma$
versus $\gamma B/|J|$ for the AFM spin 1 dimer.  The curve
notation is the same as in Fig. 2, except that the isotropic case
(dotted) curve is for $J_e=0$.}\label{fig9}
\end{figure}

  Finally, in Fig. 10 , we show the corresponding $M(B)$  curves for
  $J_c=0.1J$.
    With this combination,
  $B_{s,1}^{\rm lc}(\theta,0)$ is independent of $s$ for both
  $\theta=0,\pi/2$, but is larger for $\theta=\pi/2$, nearly
  quantitatively consistent with Eqs. (\ref{b11z})-(\ref{b21xy})
  with $J_d, J_e\rightarrow \pm J_c/2$, respectively.

  In short, the case $s_1=1$ is sufficient to exhibit the
  very different behaviors obtained from the single-ion, local
  spin anisotropy interactions from those obtained from the global
  spin anisotropy interactions.  As $s_1$ increases
  beyond 1, the situation becomes not only more complicated, but also
   more interesting, as shown
  in the following.

\begin{figure}
\includegraphics[width=0.45\textwidth]{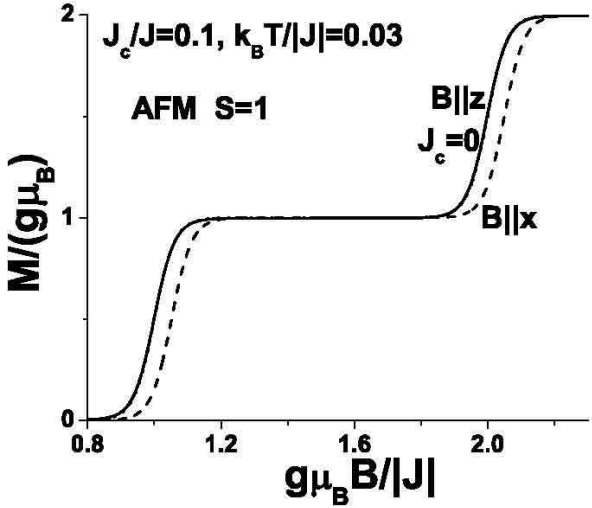}
\caption{Plot at $k_BT/|J|=0.03$ and $J_c/J=0.1$ of $M/\gamma$
versus $\gamma B/|J|$ for the AFM spin 1 dimer.  The curve
notation is the same as in Fig. 2, except that the isotropic case
(dotted) curve is for $J_c=0$.}\label{fig10}
\end{figure}

\section{VI. Exact Numerical Results for spin 5/2}

For $s_1=5/2$, one of the cases of greatest experimental interest,
when ${\cal H}_a$ and ${\cal H}_e$ are present, none of the
allowed $s, m$ values is a true quantum number.  That is, ${\cal
H}_a$ and ${\cal H}_e$ cause all of the states with nominally odd
or even $s$  to mix with one another. For ${\bm B}||\hat{\bm i}$
for $i=x, y, z$, this simplifies in the crystal representation, as
for $s_1=1$, since only states with odd or even  $m$ values in the
appropriately chosen representation can mix. By using symbolic
manipulation software, it is possible to solve for the exact
eigenvalues of the $s_1=5/2$ dimer. However, because the analytic
expressions for the eigenvalues are much more complicated than
those for $s_1=1$ presented in Appendix A, we shall not attempt to
present them, but will instead focus upon their numerical
evaluation for specific cases.

\begin{figure}
\includegraphics[width=0.45\textwidth]{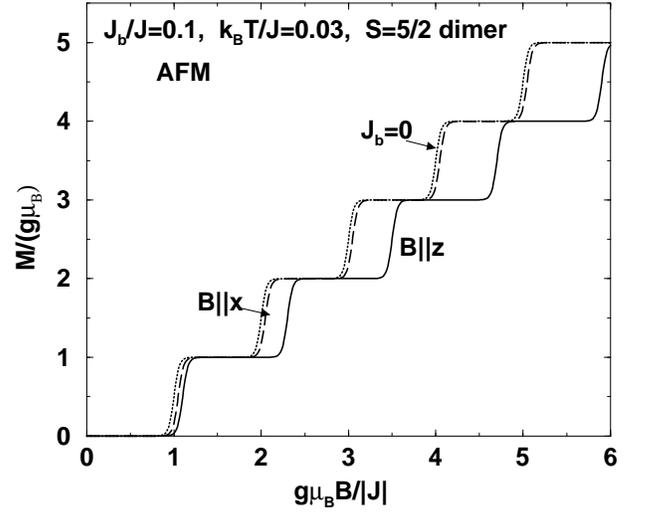}
\caption{Plot  at $k_BT/|J|=0.03$ of $M/\gamma$ versus $\gamma
B/|J|$ for the AFM spin 5/2 dimer with $J_b/J=0.1$.  The curve
notation is the same as in Fig. 2.}\label{fig11}
\end{figure}

\begin{figure}
\includegraphics[width=0.45\textwidth]{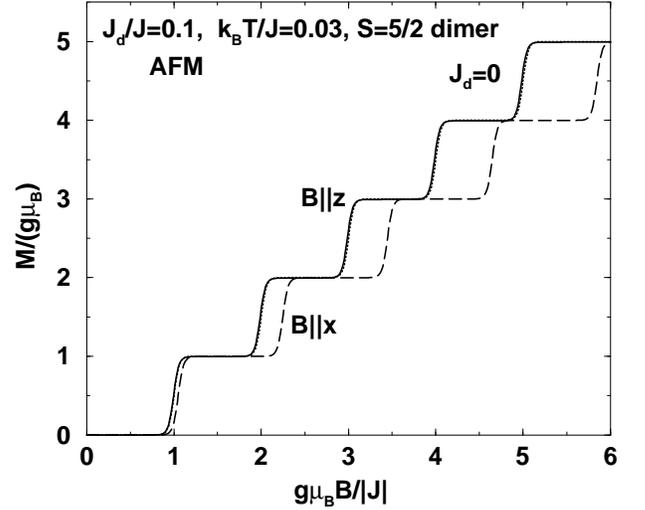}
\caption{Plot  of $M/\gamma$ versus $\gamma B/|J|$ for the AFM
spin 5/2 dimer at $k_BT/|J|=0.03$ with $J_d/J=0.1$. The curve
notation is the same as in Fig. 3.}\label{fig12}
\end{figure}

To first order in the $J_j$, the first three level crossings for
${\bm B}||\hat{\bm i}$ with $i=x, y, z$ are
\begin{eqnarray}
\gamma B_{1,5/2,z}^{{\rm lc}(1)}&=&-J+\frac{32 J_a}{15}-J_b,\label{B52step1}\\
\gamma B_{1,5/2,x,y}^{{\rm
lc}(1)}&=&-J-\frac{16J_a}{15}-\frac{J_b}{2}\mp\frac{J_d}{2}
\pm\frac{16J_e}{5},\\
\gamma B_{2,5/2,z}^{{\rm lc}(1)}&=&-2J-\frac{8J_a}{35}-3J_b,\\
\gamma B_{2,5/2,x,y}^{{\rm
lc}(1)}&=&-2J+\frac{4J_a}{35}-\frac{J_b}{2}\mp\frac{5J_d}{2}\mp\frac{12J_e}{35},\\
\gamma B_{3,5/2,z}^{{\rm lc}(1)}&=&-3J-\frac{106J_a}{63}-5J_b,\\
\gamma B_{3,5/2,x,y}^{{\rm
lc}(1)}&=&-3J+\frac{53J_a}{63}-\frac{J_b}{2}\mp\frac{9J_d}{2}\mp\frac{53J_e}{21}.\label{B52step3}
\end{eqnarray}

In Figs. 11-14, we plot $M/\gamma$ and $C_V(B)$ versus $\gamma
B/|J|$ for four low-$T$ cases of AFM $s_1=5/2$ dimers, $J_b=0.1J$,
$J_d=0.1J$, $J_a=0.1J$, and $J_c=0.1J$, respectively, and the
other $J_j=0$, taking $k_BT/|J|=0.03$.  Each of these curves
exhibit the universal step behavior predicted in Eqs.
(\ref{Msteps}) and (\ref{slope}).  Corresponding $C_V/k_B$ versus
$\gamma B/|J|$ curves also exhibit the universal double peak
behaviors predicted in Eqs. (\ref{CVzeroes})-(\ref{peakheight}),
and are shown elsewhere.\cite{ekcondmat}  In Fig. 15, examples of
$M(\theta)$ at fixed $B$ and $\phi=0$ are shown. Figures 13-15 are
sufficient to distinguish the more interesting local spin
anisotropy effects in AFM dimers with higher $s_1$ values from the
non-existent or less interesting ones present  with $s_1=1/2, 1$,
respectively. In Figs. 11-14, the solid and dashed curves
represent the ${\bm B}||\hat{\bm z}$ and ${\bm B}||\hat{\bm x}$
cases, and the dotted curve represents the isotropic case,
$J_j=0\forall j$, as in Figs. 2-10.

 We first examine the global spin anisotropy
effects of ${\cal H}_b$ and ${\cal H}_d$ in Figs. 11 and 12. These
figures exhibit the same behavior shown for $s_1=1/2,1$ in the
corresponding Figs. 2, 6 and 3, 7.  Note that  $B_{s,5/2}^{\rm
lc}(\theta,0)$ is largest for $\theta=\pi/2$ with $J_b=0.1J$ and
for $\theta=0$ with $J_d=0.1J$, and increases monotonically with
$s$, as for $s_1=1$ in both cases, nearly quantitatively
consistent with Eqs. (\ref{B52step1})-(\ref{B52step3}). By
contrast, $B_{s,52}^{\rm lc}(\theta,0)$ for $\theta=\pi/2$ with
$J_b=0.1J$ and for $\theta=0$ for $J_d=0.1J$ are nearly
indistinguishable from the isotropic case, also nearly
quantitatively consistent with Eqs.
(\ref{B52step1})-(\ref{B52step3}).

\begin{figure}
\includegraphics[width=0.45\textwidth]{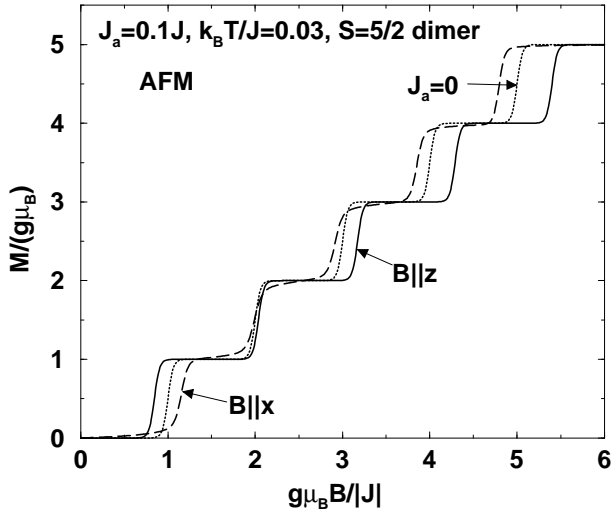}
\caption{Plot at $k_BT/|J|=0.03$ and $J_a/J=0.1$ of $M/\gamma$
versus $\gamma B/|J|$ for the AFM spin 5/2 dimer.  The curve
notation is the same as in Fig. 8.}\label{fig13}
\end{figure}

\begin{figure}
\includegraphics[width=0.45\textwidth]{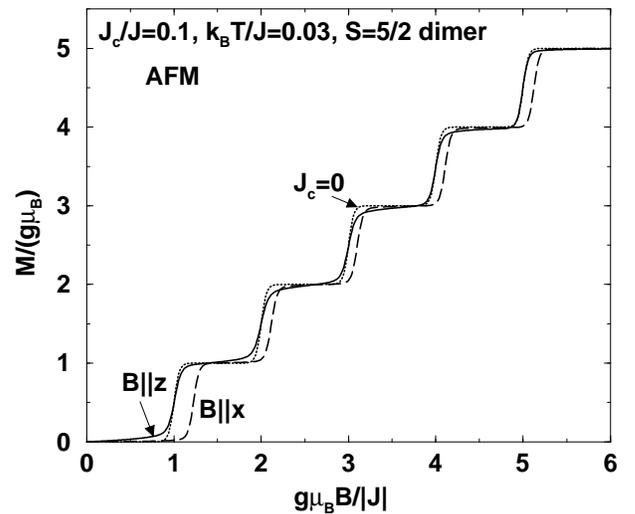}
\caption{Plot at $k_BT/|J|=0.03$ and $J_c/J=0.1$ of $M/\gamma$
versus $\gamma B/|J|$ for the AFM spin 5/2 dimer. The curve
notation is the same as in Fig. 10.}\label{fig14}
\end{figure}

In contrast, the local field anisotropy interactions show very
different and much more interesting behaviors.  In Fig. 13, we
present our results for the effects of the single-ion axial
anisotropy interaction ${\cal H}_a$, Eq. (\ref{lz}). As in Fig. 8
for $s_1=1$, $B_{s,5/2}^{\rm lc}(\theta,0)+sJ$ changes sign with
increasing $s$.  $B_{1,5/2}^{\rm lc}(0,0)+J<0$, whereas for
$s\ge2$, $B_{s,5/2}^{\rm lc}(0,0)+sJ>0$ and increases
monotonically with  $s$.   For $\theta=\pi/2$, nearly the opposite
situation occurs. $B_{s,5/2}^{\rm lc}(\pi/2,0)+sJ$ is positive for
$s=1$ and decreases monotonically with increasing $s$.  In both
cases,$|B_{s,5/2}^{\rm lc}(\theta,0)+sJ|$ is a minimum for $s=2$.
 These local axial anisotropy effects,
consistent with Eqs. (\ref{B52step1})-(\ref{B52step3}), are very
different than the global anisotropy ones pictured in Figs. 11 and
12.  They are also much richer and interesting than the
corresponding case for $s_1=1$ pictured in Fig. 8.

In Fig. 14, we  present our $M/\gamma$ versus $\gamma B/|J|$
results for the case of the local azimuthally anisotropic exchange
interaction, ${\cal H}_c$, Eq. (\ref{Hc}), evaluated for AFM
dimers at $k_BT/|J|=0.03$ with $J_c=0.1J$ and the remaining
$J_j=0$.  $B_{s,5/2}^{{\rm lc}(1)}$ for this case is obtained from
Eqs. (\ref{B52step1})-(\ref{B52step3}) by setting $J_d, J_e=\pm
J_c/2$, respectively, and $J_a=J_b=0$.  $B_{s,5/2}^{\rm lc}(0,0)$
is nearly indistinguishable from the isotropic case, as in Fig. 10
for $s_1=1$, except for some minor curve shape effects far from
the step midpoints.  $B_{s,5/2}^{\rm lc}(\pi/2,0)+sJ$ is always
positive, as for $s_1=1$ shown in Fig. 10, but has a minimum at
$s=3$.   This is also in stark contrast to the monotonic global
anisotropy behavior for $s_1=5/2$ seen in Figs. 11 and 12.  Both
of these behaviors are nearly quantitatively consistent with Eqs.
(\ref{B52step1})-(\ref{B52step3}) as modified to include $J_c$.

Although not pictured explicitly, the behavior for $J_e=0.1J$ with
the other $J_j=0$ is rather  like that of the $J_a=0.1J$ curves
pictured in Fig. 13 with $\hat{\bm z}\leftrightarrow\hat{\bm x}$,
differing in ways similar to those differences between the
$J_a=0.1J$ and $J_e=0.1J$ curves pictured for $s_1=1$ in Figs. 8
and 9.  As indicated in Eqs. (\ref{B52step1})-(\ref{B52step3}),
the  $B_{s,5/2}^{\rm lc}(0,0)$ are nearly independent of $J_e$.
However, the $B_{s,5/2}^{\rm lc}(\pi/2,0)+sJ$ with $J_e=0.1J$  are
nearly three times as large as for $J_a=0.1J$ case, so that the
minimum  $|B_{s,5/2}^{\rm lc}(\pi/2,0)+sJ|$ is also a minimum for
$s=2$.  More details are given in Section VII and Appendix D.

\begin{figure}
\includegraphics[width=0.45\textwidth]{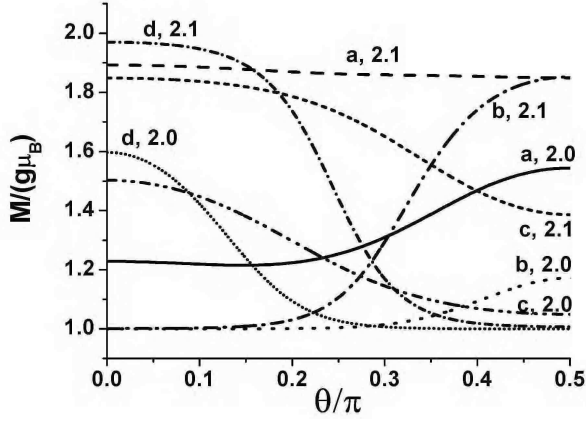}
\caption{ Plot  of $M/\gamma$ at $\phi=0$ versus $\theta/\pi$ near
the second step at $\gamma B/|J|=2.0, 2.1$ and $k_BT/|J|=0.03$ for
each $J_j/J=0.1$ with $i=a,b,c,d$. Curves are labelled with $j,
\gamma B/|J|$ values.}\label{fig15}
\end{figure}

In addition, the angular dependencies of $ M$ are different for
each of the four $J_j$ we presented for $s_1=5/2$. In Fig. 15, we
present the results for $M(B,\theta,\phi=0)/\gamma$ versus
$\theta/\pi$ near the second level crossing at $\gamma B/|J|=2.0,
2.1$ and $k_BT/|J|=0.03$ for each of the four $s_1=5/2$ AFM
magnetization cases pictured in Figs. 11-14. Note that
$M(B,\pi-\theta,0)=M(B,\theta,0)$. The $J_a$ and $J_b$ curves,
while rather similar at $\gamma B/|J|=2.0$, are very different at
$\gamma B/|J|=2.1$. Hence, ${\bm M}({\bm B})$ depends strongly
upon the particular type of spin anisotropy.

\section{VII. Analytic results for weakly anisotropic dimers of arbitrary spin}

\subsection{A. Induction representation eigenstates first order in the anisotropies}

Since the diagonalization of the Hamiltonian matrix is difficult
for an arbitrary magnetic field direction combined with an
arbitrary combination of spin anisotropy interactions, and must be
done separately for each value of $s_1$, it is useful to consider
a perturbation  in the relative strengths $J_j/J$ of the
anisotropy interactions. We nominally assume $|J_j/J|\ll1$ for
$j=a,b,d,e$. However, to compare with low-$T$ magnetization and
specific heat experiments at various applied field directions and
magnitudes, one cannot  take the magnitude of the magnetic
induction to be small. In order to incorporate the magnetic
induction of an arbitrary strength and direction accurately, we
therefore rotate the crystal axes $(\hat{\bm x}, \hat{\bm y},
\hat{\bm z})$ to $(\hat{\bm x}', \hat{\bm y}', \hat{\bm z}')$, so
that ${\bm B}=B\hat{\bm z}'$.
 The rotation matrix and a brief discussion of its ramifications are given in Appendix B.

In these rotated coordinates, the Zeeman interaction $-bS_{z'}$,
where $b=\gamma B$, is diagonal. We therefore denote this
representation as the induction representation.   The Hamiltonian
${\cal H}'$ in this representation is given in Appendix B.    In
the induction representation, we choose the quantum states to be
$|\varphi_s^m\rangle$.  In the absence of the four anisotropy
interactions $J_j$, ${\cal H}'={\cal H}_0'$ is diagonal,
\begin{eqnarray} {\cal
H}_0'|\varphi_s^m\rangle&=&E^{m,(0)}_{s}|\varphi_s^m\rangle,\\
\noalign{\rm where}\nonumber\\
 E^{m,(0)}_{s}&=&-Js(s+1)/2-mb.\label{E0}\end{eqnarray}

The operations of the remaining terms in ${\cal H}'$ on the
eigenstates $|\varphi_s^m\rangle$ are given in Appendix C.
  The first order correction to the energy in the
induction representation,
$E_{s,s_1}^{m,(1)}=\langle\varphi_s^m|{\cal
H}'|\varphi_s^m\rangle$, is found to be
\begin{eqnarray}
E_{s,s_1}^{m,(1)}&=&-\frac{J_b}{2}[2s(s+1)-1]-\frac{J_a}{2}[s(s+1)-1]\nonumber\\
& &+\frac{\tilde{J}_{b,a}^{s,s_1}}{2}[m^2+s(s+1)-1]\nonumber\\
&
&+\frac{1}{2}[s(s+1)-3m^2]\nonumber\\
& &\times\Bigl(\tilde{J}_{b,a}^{s,s_1}\cos^2\theta
+\tilde{J}_{d,e}^{s,s_1}\sin^2\theta\cos(2\phi)\Bigr),
\label{Esm1}
\end{eqnarray}
where
\begin{eqnarray}
\tilde{J}_{b,a}^{s,s_1}&=&J_b+\alpha_{s,s_1}J_a,\label{tildeJba}\\
\tilde{J}_{d,e}^{s,s_1}&=&J_d+\alpha_{s,s_1}J_e,
\label{tildeJde}\end{eqnarray} and $\alpha_{s,s_1}$ is given by
Eq. (\ref{alphass1}).

 Since the $\theta,\phi$ dependence of $E_s^{m,(1)}$ arises from
the term proportional to $\tilde{J}_{b,a}^{s,s_1}\cos^2\theta
 +\tilde{J}_{d,e}^{s,s_1}\sin^2\theta\cos(2\phi)$, it is tempting to
think that the thermodynamics with $\tilde{J}_{d,e}^{s,s_1}=0$ and
${\bm B}||\hat{\bm z}$ are equivalent to those with
$\tilde{J}_{b,a}^{s,s_1}=0$ and ${\bm B}||\hat{\bm x}$.  However,
as shown explicitly in the following, the
$\theta,\phi$-independent parts of Eq. (\ref{Esm1}) strongly break
this apparent equivalence, causing the $B_{s,s_1}^{\rm
lc}(\theta,\phi)$  for those two cases to differ. This implies
that $J_b$ and $J_d$ are inequivalent, as are $J_a$ and $J_e$,
even to first order in the anisotropy strengths.

\subsection{B. First order thermodynamics}

In the Hartree approximation,  the anisotropy interactions are
included to first order only.  In this approximation, $s$ and $m$
are still good quantum numbers, so the partition function
\begin{eqnarray}
Z^{(1)}&=&\sum_{s=0}^{2s_1}\sum_{m=-s}^se^{-\beta
E^m_{s,s_1}},\label{Zapprox}
\end{eqnarray}
where $E^m_{s,s_1}=E^{m,(0)}_{s}+E^{m,(1)}_{s,s_1}$. Although it
is difficult to perform the summation over the $m$ values
analytically,  it is nevertheless elementary to evaluate $Z$
numerically for an arbitrary $B, \theta,\phi$, and $T$ from the
eigenstate energies.     The magnetization in the Hartree
approximation is
\begin{eqnarray}
M^{(1)}(B,\theta,\phi)&=&\frac{\gamma}{Z^{(1)}}\sum_{s=0}^{2s_1}\sum_{m=-s}^sme^{-\beta
E^m_{s,s_1}}.
\end{eqnarray}

Similarly, the specific heat in the Hartree approximation is
\begin{eqnarray}
C_{V}^{(1)}(B,\theta,\phi)&\approx&\frac{k_B\beta^2}{\left(Z^{(1)}\right)^2}\Bigl[Z\sum_{s=0}^{2s_1}\sum_{m=-s}^s(E^m_{s,s_1})^2e^{-\beta
E^m_{s,s_1}}\nonumber\\
& &-\Bigl(\sum_{s=0}^{2s_1}\sum_{m=-s}^sE^m_{s,s_1}e^{-\beta
E^m_{s,s_1}}\Bigr)^2\Bigr].\label{CVapprox}
\end{eqnarray}

\begin{figure}
\includegraphics[width=0.45\textwidth]{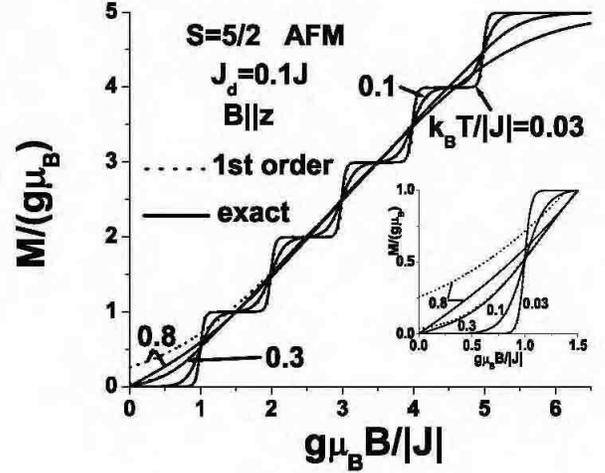}
\caption{Comparison of $M/\gamma$ versus $\gamma B/|J|$ obtained
using the Hartree asymptotic form (dotted) with the exact
calculation (solid), for the $s_1=5/2$ AFM dimer with $J_d=0.1J$,
$J_a=J_b=J_e=0$, at $k_BT/|J|=0.03,0.1,0.3,0.8$, as indicated.
Inset:  expanded view of the region $0\le \gamma B/|J|\le1.5$.}
\label{magtest}
\end{figure}

\begin{figure}
\includegraphics[width=0.45\textwidth]{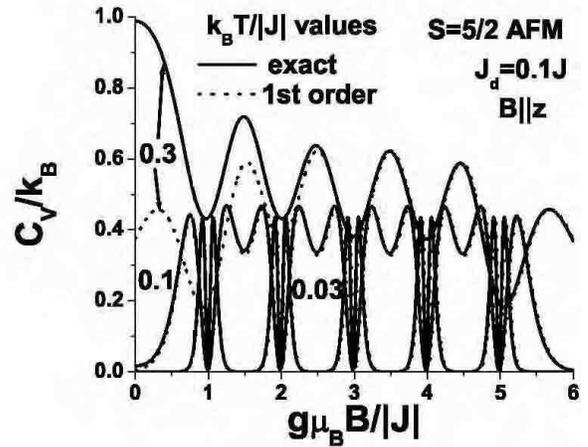}
\caption{Comparison of $C_V/k_B$ versus $\gamma B/|J|$ obtained
using the Hartree asymptotic form (dotted) with the exact
calculation (solid), for the $s_1=5/2$ AFM dimer with $J_d=0.1J$,
$J_a=J_b=J_e=0$, at $k_BT/|J|=0.03,0.1,0.3$, as indicated.}
\label{shtest}
\end{figure}

As a test of the accuracy of this Hartree calculation, we have
compared the  Hartree and exact $M(B)$ obtained for the $s_1=5/2$
dimer with $J_d=0.1J$ and ${\bm B}||\hat{\bm z}$ at various $T$
values in Fig. \ref{magtest}.  The corresponding comparison
between the Hartree and exact $C_V(B)$ is shown in Fig.
\ref{shtest}. We see that  the  curves evaluated using the Hartree
and the exact expressions for $M$ and $C_V$ with $s_1=5/2$ are
indistinguishable at $k_BT/|J|=0.03$. The $C_V$ curves are
noticeably different at $k_BT/|J|=0.1$ for $\gamma B/|J|<0.4$, and
at $k_BT/|J|=0.3$ they are noticeably different for $\gamma
B/|J|<2.6$. Corresponding noticeable differences in the $M$ curves
at the same $B$ values appear at $T$ values roughly three times as
high as in the $C_V$ curves.

At very low $T$, $k_BT/|J|\ll1$, the most important states in this
perturbative scheme are the minima for each $s$ value,
$E_{s,s_1}^s$, which determine the  level crossings in the Hartree
approximation. As $T\rightarrow0$, we can ignore all of the  $m\ne
s$ states in Eqs. (\ref{Zapprox})-(\ref{CVapprox}).  This
two-level approximation is the basis for the universal behavior
given by Eqs. (\ref{CVzeroes})-(\ref{slope}), which fits all of
the exact curves we presented reasonably well.

\subsection{C. First order response functions}
\subsubsection{1. Inelastic neutron scattering cross-section}

Two response functions relevant for the study of SMM's are the
inelastic neutron scattering cross-section $S({\bm B},{\bm
q},\omega)$ and the electron paramagnetic resonance (EPR)
susceptibility $\chi({\bm B},\omega)$ in strong magnetic
inductions ${\bm B}$.
 The low-$T$ inelastic neutron cross-section $S({\bm B},{\bm q},\omega)$
can be evaluated from
\begin{eqnarray}
S({\bm B},{\bm
q},\omega)&=&\sum_{\alpha,\beta=1}^3(\delta_{\alpha,\beta}-\hat{q}_{\alpha}\hat{q}_{\beta})\int\frac{dt}{2\pi}e^{i\omega
t}\nonumber\\
& &\times{\rm Tr}\Bigl(e^{-\beta{\cal H}}S_{\alpha}({\bm
q},t)S_{\beta}({\bm q},0)\Bigr),\label{Sqomega}\\
S_{\alpha}({\bm q},0)&=&S_{1,\alpha}e^{i{\bm q}\cdot{\bm
d}}+S_{2,\alpha}e^{-i{\bm q}\cdot{\bm d}},
\end{eqnarray}
where $2{\bm d}$ is the vector between the spins of a dimer and
$\hat{q}_{\alpha}$ is the $\alpha$th component of ${\bm q}/q$. It
is customary to write the operators in the crystal representation,
for which ${\bm d}=d\hat{\bm z}$. \cite{ek,white}  With inelastic
neutron scattering, one can probe the dimer with a strong magnetic
field at various directions with respect to both the dimer axis
and the scattering plane.  To the extent that the eigenstates
$\{|\phi_n\rangle\}$ and their energies $\epsilon_n$ can be
evaluated exactly,
\begin{eqnarray}
S({\bm B},{\bm
q},\omega)&=&\sum_{\alpha,\beta=1}^3(\delta_{\alpha,\beta}-\hat{q}_{\alpha}\hat{q}_{\beta})\sum_{n,n'=1}^{n_{s_1}}e^{-\beta\epsilon_n}\nonumber\\
& &\times\delta(\omega+\epsilon_n-\epsilon_{n'})\nonumber\\
& &\times\langle\phi_n|\tilde{S}_{\alpha}({\bm
q},0)|\phi_{n'}\rangle\langle\phi_{n'}|\tilde{S}_{\beta}^{\dag}({\bm
q},0)|\phi_n\rangle, \nonumber\\
& &\\ \tilde{S}_{\alpha}({\bm q},0)&=&US_{\alpha}({\bm
q},0)U^{\dag},
\end{eqnarray}
where $U$ is the unitary operator that diagonalized ${\cal H}$.
When exact expressions for $S({\bm B},{\bm q},\omega)$ are tedious
to obtain, it is straightforward to obtain it in the Hartree
approximation, $S^{(1)}({\bm B},{\bm q},\omega)$. But to do so, it
is easiest to define the axes in the induction representation, for
which $\{|\phi_n\rangle\}=\{|\tilde{\varphi}_s^m\rangle\}$ and the
$\hat{\bm e}_{\alpha}=\hat{\bm x}',\hat{\bm y}',\hat{\bm z}'$ for
$\alpha=1,2,3,$ respectively. In the Hartree approximation
obtained by setting the
$|\tilde{\varphi}_s^m\rangle=|\varphi_s^m\rangle$, the bare wave
functions, $S^{(1)}({\bm B},{\bm q},\omega)$ for an arbitrary
$B,\theta,\phi$ is then
\begin{eqnarray}
S_1^{(1)}&=&\sum_{s=0}^{2s_1}\sum_{m=-s}^se^{-\beta
E_{s,s_1}^m}\nonumber\\
& &\times\Bigl(\cos^2({\bm q}\cdot{\bm
d})\sin^2\theta_{b,q}{\cal F}_{1,s_1}^{m,s (0)}(\omega,\theta,\phi)\nonumber\\
& &+\sin^2({\bm q}\cdot{\bm
d})\sin^2\theta_{b,q}{\cal F}_{2,s_1}^{m,s (0)}(\omega,\theta,\phi)\nonumber\\
& &+\cos^2({\bm q}\cdot{\bm
d})\frac{2-\sin^2\theta_{b,q}}{4}{\cal F}_{3,s_1}^{m,s (0)}(\omega,\theta,\phi)\nonumber\\
& &+\sin^2({\bm q}\cdot{\bm
d})\frac{2-\sin^2\theta_{b,q}}{4}{\cal F}_{4,s_1}^{m,s (0)}(\omega,\theta,\phi)\Bigr),\label{SqomegaHartree}\nonumber\\
\end{eqnarray}
 where ${\bm q}\cdot{\bm d}=qd\cos\theta_q$ is invariant under
the rotation, $2{\bm d}$ is the vector separating the dimer spins,
$\theta_{b,q}$ is the angle between  ${\bm q}$ and ${\bm B}$, and
\begin{eqnarray}
{\cal F}_{1,s_1}^{m,s (0)}&=&m^2\delta(\omega),\label{F10}\\
{\cal F}_{2,s_1}^{m,s (0)}&=&\sum_{\sigma'=\pm1}\delta\bigl(\omega+E^m_{s,s_1}-E_{s+\sigma',s_1}^m\bigr)\nonumber\\
& &\times \Bigl(D_{s+(\sigma'+1)/2,s_1}^m\Bigr)^2,\\
{\cal F}_{3,s_1}^{m,s
(0)}&=&\sum_{\sigma=\pm1}\delta\bigl(\omega+E_{s,s_1}^m-E_{s,s_1}^{m+\sigma}\bigr)\bigl(A_s^{\sigma
m}\bigr)^2,\\
{\cal F}_{4,s_1}^{m,s
(0)}&=&\sum_{\sigma,\sigma'=\pm1}\delta\bigl(\omega+E_{s,s_1}^m-E_{s+\sigma',s_1}^{m+\sigma}\bigr)\nonumber\\
& &\qquad\times
\Bigl(C_{s+(\sigma'+1)/2,s_1}^{-(\sigma'+1)/2-\sigma\sigma'm}\Bigr)^2,\label{F40}
\end{eqnarray}
where the coefficients are given respectively by Eqs.
(\ref{Dss1m}), (\ref{Asm}), and (\ref{Css1m}), and  we have
suppressed the $\omega,\theta,\phi$ arguments of the ${\cal
F}_{n,s_1}^{m,s (i)}$ functions for simplicity of presentation. In
Eq. (\ref{SqomegaHartree}), the only dependencies upon the
anisotropy energies is in the first-order eigenstate energies
$E_{s,s_1}^m=E_s^{m,(0)}+E_{s,s_1}^{m,(1)}$ given by Eqs.
(\ref{E0}) and (\ref{Esm1}), respectively.

As described in detail in Appendix C,  corrections to the wave
functions first order in the anisotropy interactions lead to the
``extended Hartree'' approximation, $S^{(1{\rm e})}({\bm B},{\bm
q},\omega)$, which exhibits additional transitions with strengths
first and second order in the anisotropy energies. $S^{ (1{\rm
e})}({\bm B},{\bm q},\omega)$ contains terms arising from the
traceless diagonal and off-diagonal parts of the $\alpha,\beta$
component matrix, and from the contributions of the diagonal part
of the component matrix containing corrections to its trace. These
latter terms are proportional to the same four functions of ${\bf
q}$ as in Eq. (\ref{SqomegaHartree}), and can be separated
experimentally from the other terms by choosing the scattering
plane such that ${\bm q}\perp{\bm B}$ (or $\sin\theta_{b,q}=1$)
for all $(\theta,\phi)$ and by averaging over all azimuthal angles
$\phi_{b,q}$.  Then, the transitions from $(s,m)$ to $(s',m')$,
where $s'=s\pm2,3$ and $m'=m,m\pm1,m\pm2,m\pm3$ with leading
amplitudes second order in the anisotropy interactions upon depend
upon the the local anisotropy interactions $J_a$ and $J_e$ through
the three functions $f_n(\theta,\phi)$ with $n=3,6,7$ given in
Appendix D. Hence, if these transitions can be identified, a study
of the $(\theta,\phi)$  dependencies of their strengths can
provide direct measurements of those parameters.  Then, studying
the weak transitions $(s,m)\rightarrow(s',m')$ with $s'=s$,
$m\pm1,\pm2,\pm3$ and $s'=s\pm1,m'=m\pm2$ can provide further
measurements of the global anisotropy interactions $J_d$ and
$J_d$.

\subsubsection{2. Electron paramagnetic resonance susceptibility}

In an EPR experiment, one applies a strong magnetic field, plus a
weak oscillatory transverse field, leading to the overall
induction ${\bm B}=B\hat{\bm z}'+B_{\perp}[\hat{\bm x}'\cos(\omega
t)-\sigma\hat{\bm y}'\sin(\omega t)]$, where we assume the
oscillatory induction precesses clockwise (counterclockwise) for
$\sigma=\pm1$.\cite{white} For a weak transverse induction
$B_{\perp}$, one measures the resulting linear
 response $\chi_{-\sigma,\sigma}({\bm B},\omega)$ given by
 \begin{eqnarray}
 \chi_{-\sigma,\sigma}&=&\frac{i\gamma^2}{Z}\int_0^{\infty}\frac{dt}{2\pi}e^{i(\omega+i\eta) t}{\rm Tr}\Bigl(e^{-\beta{\cal
 H}}[S_{-\sigma}(t),S_{\sigma}(0)]\Bigr),\nonumber\\
 & &\\
 S_{\sigma}(t)&=&e^{i{\cal
 H}t}S_{\sigma}(0)e^{-i{\cal H}t},
 \end{eqnarray}
 where $\eta=0+$, the global spin raising and lowering operators $S_{\pm}$
 are defined in the induction representation, as described in
 Appendix B and $Z$ is the partition function, and $[a,b]=ab-ba$ is the commutator.  In principle, an exact expression for $\chi_{-\sigma,\sigma}({\bm B},\omega)$ can be obtained by
 starting in the induction representation with the basis $\{|\varphi_s^m\rangle\}$,
 diagonalizing ${\cal H}$ with the unitary operator $V$, $V{\cal H}V^{\dag}=\tilde{H}'$,
 and the new basis $\{|\tilde{\phi}_n\rangle\}$ is obtained from $|\tilde{\phi}_n\rangle=V|\varphi_s^m\rangle$, leading to
 $\tilde{\cal H}'|\tilde{\phi}_n\rangle=\tilde{\epsilon}_n|\tilde{\phi}_n\rangle$.  One then has
\begin{eqnarray}
\chi_{-\sigma,\sigma}({\bm B},\omega)&=&\frac{\gamma^2}{\pi Z}\sum_{n,n'=1}^{n_{s_1}}e^{-\beta\tilde{\epsilon}_n}\nonumber\\
&
&\times\langle\tilde{\phi}_n|\tilde{S}_{-\sigma}(0)|\tilde{\phi}_{n'}\rangle\langle\tilde{\phi}_{n'}|\tilde{S}_{\sigma}(0)|\tilde{\phi}_n\rangle\nonumber\\
&
&\times\Bigl(\frac{1}{\omega+\tilde{\epsilon}_n-\tilde{\epsilon}_{n'}+i\eta}\nonumber\\
&
&\qquad-\frac{1}{\omega-\tilde{\epsilon}_{n}+\tilde{\epsilon}_{n'}+i\eta}\Bigr),\label{chiexact}
\end{eqnarray}
where $\tilde{S}_{\sigma}(0)=VS_{\sigma}(0)V^{\dag}$.
  It is then elementary to obtain $\chi^{''}_{-\sigma,\sigma}(\omega)$ in the Hartree
 approximation valid to first order in the anisotropy parameters, which is well-behaved for all induction values.
 The imaginary part of $\chi_{-\sigma,\sigma}$ is then
\begin{eqnarray}
\chi^{(1)''}_{-\sigma,\sigma}({\bm
B},\omega)&=&\frac{\gamma^2}{Z^{(1)}}\sum_{s=0}^{2s_1}\sum_{m=-s}^s\exp[-\beta
E_{s,s_1}^m]\nonumber\\
& &[s(s+1)-m^2-\sigma
m]\nonumber\\
&
&\times\Bigl(\delta(E_{s,s_1}^m-E_{s,s_1}^{m+\sigma}+\omega)\nonumber\\
& &-\delta(E_{s,s_1}^{m+\sigma}-E_{s,s_1}^{m}+\omega)\Bigr),
\end{eqnarray}
where $E_{s,s_1}^m=E_s^{m,(0)}+E_{s,s_1}^{m,(1)}$ is given by Eqs.
(\ref{E0}) and (\ref{Esm1}) and $Z^{(1)}$ is given by Eq.
(\ref{Zapprox}).  The arguments of the $\delta$-function give rise
to the resonant frequencies first order in the anisotropy
energies, or  to the first order resonant magnetic inductions,
\begin{eqnarray} \gamma B_{\rm
res}^{(1)}&=&\pm\omega+\frac{(2m+\sigma)}{2}g_{s,s_1}(\theta,\phi),\\
g_{s,s_1}(\theta,\phi)&=&\tilde{J}_{b,a}^{s,s_1}(1-3\cos^2\theta)\nonumber\\
& &\qquad-3\tilde{J}_{d,e}^{s,s_1}\sin^2\theta\cos(2\phi),
\end{eqnarray}
where $\tilde{J}_{b,a}^{s,s_1}$ and $\tilde{J}_{d,e}^{s,s_1}$ are
given by Eqs. (\ref{tildeJba}) and (\ref{tildeJde}), respectively,
and we have ignored the processes that give rise to finite
transition widths.\cite{park}  By varying the direction and
magnitude of ${\bm B}$, it is possible to obtain sufficient
information to  fit all of the parameters in the model. At low $T$
the $m=s$ states dominate, so that only $\sigma=-1$ is allowed. By
increasing ${\bm B}$ past the $s$th level crossing, one can probe
the effective $s$th state of the AFM dimer using EPR.

In Appendix C, we extend these results to include the corrections
to the wave functions first order in the anisotropy interactions.
With these corrections, additional EPR transitions are present,
which   can provide additional information useful in experimental
identification of the anisotropy interactions.  It is shown that
there are  10 additional resonant magnetic induction strengths of
the forms
\begin{eqnarray}
\gamma B^{(1)}_{\rm
res}&=&\pm a_n\omega+b_ng_{s,s_1}(\theta,\phi)\nonumber\\
& &+c_nh_{s,s_1}(\theta,\phi),\label{Bres}\\
h_{s,s_1}(\theta,\phi)&=&-(J+2J_b+J_a)\nonumber\\
& &+\tilde{J}_{b,a}^{s,s_1}(1+\cos^2\theta)\nonumber\\
& &\qquad+\tilde{J}_{d,e}^{s,s_1}\sin^2\theta\cos(2\phi),
\end{eqnarray}
where the  $a_n, b_n,$ and $c_n$ are listed in Appendix C.  We
note that these additional resonant inductions are first order in
the anisotropy interactions, but have amplitudes that are second
order in the anisotropy interactions, so that in most cases, they
may be difficult to detect, but their detection in certain
materials would provide a clear signal of the presence of
significantly strong  anisotropy interactions.  In addition,
resonant inductions with $c_n\ne0$ are generally very large.

To illustrate one example of an effect of local spin anisotropy
upon the EPR transitions, we consider the simple case of an
$s_1=1$ AFM dimer with the only non-vanishing anisotropy energy
$J_a/J=0.1$ and $\theta=\pi/4$.  The energy levels for this system
are pictured in Fig. 18.  In Fig. 19, the EPR transition energies
versus $\gamma B/|J|$ for the $s_1=1$ AFM dimer with $J_a/J=0.1$
and $\theta=\pi/4$ are shown.  The widths of the lines are
proportional to the strengths of the EPR matrix elements.  Note
that the ground state at $B=0$ is approximately
$|\psi_0^0\rangle$, but its energy is slightly negative due to an
admixture of this state with the nominal $|\psi_2^m\rangle$ states
with even $m$, as shown in Appendix A. The EPR matrix elements are
very small at these $B$ values, as the transitions only exist due
to the admixture of the wave functions.  At higher $B$ strengths,
the ground state goes through two level crossings, with the first
level crossing being to the lowest energy $s=1$ state, which is
nominally $|\psi_1^1\rangle$, with some mixture of the other
$|\psi_1^m\rangle$ states.  The leading transition is to the
nominal $|\psi_1^0\rangle$ state.  Then, the second level crossing
causes the ground state to be the nominal $|\psi_2^2\rangle$
state, which is of course modified by the mixing with the nominal
$|\psi_0^0\rangle$ and the other nominal $|\psi_2^m\rangle$ for
$m=0,-2$. Note that near to $\gamma B/|J|=3$, there is a level
repulsion. The leading EPR transition is to the nominal
$|\psi_2^1\rangle$ state, and because of the strong matrix
elements, the splitting of the energies due to the level repulsion
should be observable in EPR experiments.

\begin{figure}
\includegraphics[width=0.45\textwidth]{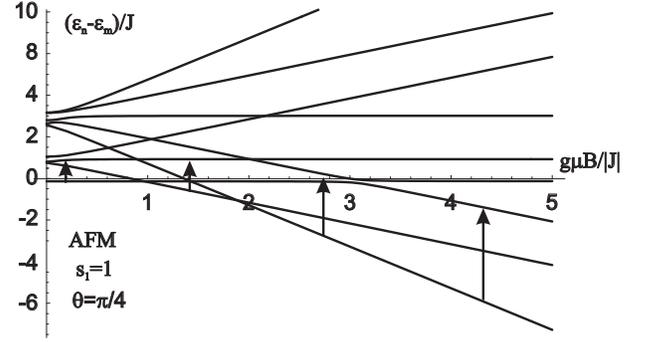}
\caption{Plot of the exact eigenvalues $\epsilon_n/|J|$ versus
$\gamma B/|J|$ for $s_1=1$ AFM dimers with $J_a/J=0.1$ and
$\theta=\pi/4$. The arrows represent the strongest EPR transitions
at their corresponding field strengths.}\label{fig18}
\end{figure}

\begin{figure}
\includegraphics[width=0.45\textwidth]{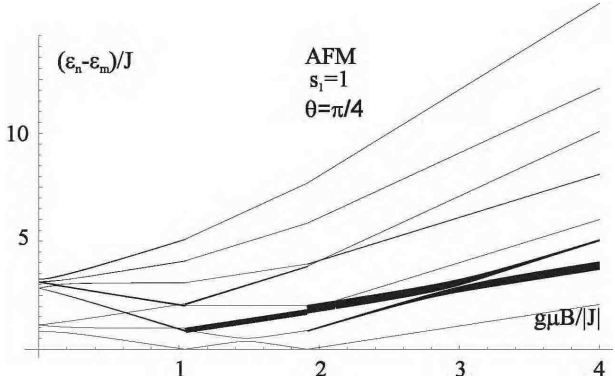}
\caption{Plot of the EPR transition energies
$(\epsilon_n-\epsilon_1)/|J|$  versus  $\gamma B/|J|$ from the
ground state for the $s_1=1$ AFM dimer with $J_a/J=0.1$ and
$\theta=\pi/4$. The widths of the lines are proportional to the
strengths of the matrix elements for the
transitions.}\label{fig19}
\end{figure}

\subsection{D. Level crossings first order in the anisotropy energies}

We can find an expression for the $s^{\rm th}$ AFM level crossing
at the induction $B^{{\rm lc}(1)}_{s,s_1}$ to first order in the
anisotropy interactions for a general $s_1$ spin dimer by equating
$E^{s,(0)}_{s}+E^{s,(1)}_{s,s_1}$ to
$E^{s-1,(0)}_{s-1}+E^{s-1,(1)}_{s-1,s_1}$, yielding
\begin{eqnarray}\gamma B^{{\rm lc}(1)}_{s,s_1}&=&-Js-J_b/2-c_{s,s_1}J_a\nonumber\\
& &-\frac{(4s-3)}{2}[J_b\cos^2\theta+J_d\sin^2\theta\cos(2\phi)]\nonumber\\
& &+3c_{s,s_1}[J_a\cos^2\theta+J_e\sin^2\theta\cos(2\phi)],\label{Bstep}\\
c_{s,s_1}&=&\frac{[3+3s-5s^2-4s^3+4s_1(s_1+1)]}{2(2s+1)(2s+3)}.
\end{eqnarray}
This expression is consistent with those obtained for
$s_1=1/2,1,5/2$ given by Eqs. (\ref{Bstephalf}),
(\ref{b11z})-(\ref{b21xy}), and (\ref{B52step1})-(\ref{B52step3}).
in Figs. 11-14. We note that $\gamma B_{s,s_1}^{{\rm lc}(1)}$
contains the $\theta,\phi$-independent terms,
$-Js-J_b/2-c_{s,s_1}J_a$, which distinguish $J_b$ from $J_d$ and
$J_a$ from $J_e$.

 In particular, we note
that the single-ion anisotropy interactions behave very
differently with increasing step number than do the global
anisotropy interactions, especially for large $s_1$.  For the
three cases we studied in detail, for $s_1=1/2$, $c_{1,1/2}=0$, so
that the local anisotropy terms are irrelevant, for $s_1=1$,
$c_{1,1}=\frac{1}{6}$ and  $c_{2,1}=-\frac{1}{2}$ have different
signs, and for $s_1=5/2$ as in Fe$_2$ dimers, the first three
$c_{s,5/2}$ coefficients are $\frac{16}{15}$, $-\frac{4}{35}$, and
$-\frac{53}{63}$, respectively, the second being an order of
magnitude smaller than the other two, and opposite in sign from
the first. For $s_1=9/2$ dimers such as [Mn$_4$]$_2$, the first
four $c_{s,9/2}$ are $\frac{16}{5}$, $\frac{4}{5}$,
$-\frac{1}{3}$, and $-\frac{37}{33}$, which changes sign between
$s=2$ and $s=3$, where its magnitude is a minimum.

This is in sharp contrast to the global anisotropy interactions,
for which the analogous coefficient $(4s-3)/2$ increases
monotonically with $s$, independent of $s_1$. These differences
should be possible to verify experimentally in careful low-$T$
experiments at high magnetic fields applied at various directions
on single crystals of those $s_1=5/2$ Fe$_2$ and $s_1=9/2$
[Mn$_4$]$_2$ dimers for which $|J|$ is sufficiently small.

\subsection{E. Level crossings to second order in the anisotropy energies}

 To aid in the
analysis of experimental data, we have extended this perturbative
calculation to second order in each of the four principle $J_j$.
Since we expect the $|J_j/J|\le0.1$ in most circumstances, this
extension ought to be sufficient to accurately analyze most
experimentally important samples. Since the $c_{s,s_1}$ magnitudes
can become larger than unity for $s_1=5/2$ and especially for
$s_1=9/2$, we expect second order corrections to be significant in
those cases.  To second order in the anisotropy interactions, the
eigenstate energies $E_{s,s_1}^{m,(2)}$ are given in Appendix C.
We note that the $E_{s,s_1}^{m(2)}$ contain divergences at $\gamma
B/|J|=0, 2s-1, 2s+3, s-1/2,$ and $s+3/2$, so that near to those
values, one would need to modify the perturbation expansion to
take proper account of the degeneracies. Hence, the expressions
for $E_{s,s_1}^{m(2)}$ cannot be used in the asymptotic
expressions for the thermodynamics, Eqs.
(\ref{Zapprox})-(\ref{CVapprox}). However, as the $s$th AFM level
crossing occurs approximately at $\gamma B/|J|=s$, which is far
from any divergences, we can safely use this second order
expansion to obtain an expression for the level crossings second
order in the anisotropy interaction energies.  We find
\begin{eqnarray}
\gamma B^{{\rm lc}(2)}_{s,s_1}&=&\gamma B^{{\rm
lc}(1)}_{s,s_1}+\Bigl(E^{s,(2)}_{s,s_1}-E_{s-1,s_1}^{s-1,(2)}\Bigr)\Bigr|_{B=-Js/\gamma},\label{Bstep2}\nonumber\\
\end{eqnarray}
where $\gamma B^{{\rm lc}(1)}_{s,s_1}$ is given by Eq.
(\ref{Bstep}).

The full expression for the second term in Eq. (\ref{Bstep2}) is
given in Appendix D. From this expression, it is easy to see that
for $s_1=1/2$, $E_{1,1/2}^{{\rm
lc}(2)}=\frac{}{2}f_1^{(2)}(\theta,\phi)+\frac{J}{8}f_4^{(2)}(\theta,\phi)$,
where $f_1^{(2)}$ and $f_4^{(2)}$ are given in Appendix C.  For
${\bm B}||\hat{\bm z}$, $E_{1,1/2}^{{\rm
lc}(2)}=-\frac{1}{2}J_d^2/|J|$, and with ${\bm B}||\hat{\bm
x},\hat{\bm y}$, $E_{1,1/2}^{{\rm lc}(2)}=-\frac{1}{8}(J_b\pm
J_d)^2/|J|$, in agreement with the expansion to second order in
the $J_j$ of our exact formulas in Eq. (\ref{Bstephalf}). However,
the second order functions have more complicated $\theta,\phi$
dependencies than do the first order $B^{{\rm lc}(1)}_{s,s_1}$ in
Eq. (\ref{Bstep}). We have also explicitly checked each formula in
Appendix D for $s_1=1$ and $s=1,2$.  Thus, Eq. (\ref{Bstep2}) is a
highly accurate expression for the full $\theta,\phi$ dependencies
of all $s=1,\ldots,2s_1+1$ level crossings of a single crystal
dimer of single ion spin $s_1$.

By superposing this non-universal level-crossing  formula, Eq.
(\ref{Bstep2}), combined with the universal behavior presented in
Eqs. (\ref{CVzeroes})-(\ref{slope}), it is easy to use our results
in accurate fits to experimental data at low temperatures and high
magnetic induction strengths.

\section{VIII. AFM dimers with strong anisotropy interactions}

Finally, we consider some cases of strong anisotropy interactions,
in which one or more of the $J_j$ is comparable to $J$ in
magnitude.  The cases of interest are those for which the
anisotropy interactions can remove the level crossing effects in
AFM dimers that give rise to the $2s_1+1$ $M(B)$ steps and
$C_V(B)$ double peaks. These cases generally occur for strong FM
anisotropy interactions, but there are some examples in which they
can occur with strong AFM anisotropy interactions. We first
consider the case of $s_1=1/2$ dimers, for which the situation can
be analyzed analytically, at least for ${\bm B}||\hat{\bm i}$.
Then, we consider the $s_1=5/2$ case numerically.

\subsection{A. Analytic and numerical results for $s_1=1/2$ dimers}

For the $s_1=1/2$ dimer, the energies for ${\bf B}||\hat{\bf i}$
are given by Eqs. (\ref{epsilon1})-(\ref{Jxy}). We are interested
in examining the cases in which the anisotropy is strong enough to
cause the level crossing to disappear.   Strong anisotropy with
$J_b$ and $J_d$ having the same sign as $J$ do not differ
significantly from the weak coupling cases, as they do not remove
the level crossing, and do not affect significantly the heights of
the magnetization step and the number and shapes of the $C_V(B)$
double peaks, but just shift their positions, as for weak
anisotropy. There are then three AFM ($J<0$) cases of interest.
These are: (1) $J_b=J_d=|J|/2$, (2) $J_b=|J|$, and (3) $J_d=|J|$.
The $M(B)$ and $C_V(B)$ curves for these cases with ${\bm
B}||\hat{\bm z}$ at the very low-$T$ value $k_BT/|J|=0.01$ are
shown as the solid, dotted, and dashed curves in Fig. 20 and 21,
respectively.

For case (1), $J_b=J_d=|J|/2$,   the same four eigenvalues are
obtained for ${\bm B}||\hat{\bm z}$ and ${\bm B}||\hat{\bm x}$.
Assuming $J<0$, these are ranked from highest to lowest as
\begin{eqnarray}
\epsilon_4&=&-J/2+\sqrt{b^2+J^2/4},\\
\epsilon_3&=&-J,\\
\epsilon_2&=&0,\\
\epsilon_1&=&-J/2-\sqrt{b^2+J^2/4}.
\end{eqnarray}
  Note that
$\epsilon_1=\epsilon_2$ only at $b=0$ (where
$\epsilon_3=\epsilon_4$), $\epsilon_1$ decreases with increasing
$b$, and $\epsilon_4$ increases with increasing $b$, so the levels
get further apart with increasing $b$.  In this case, $M(B)$ shown
as the dotted curve in Fig. 20 is broad, even at the very low $T$
value plotted.  In addition, $C_V(B)$, shown as the dotted curve
in Fig. 19, exhibits a single Schottky-like anomaly, but with a
 different shape than that of the standard Schottky case, which
arises from a  splitting of the two lowest energies linear in $b$
at $b=0$.  In this case, the peak value of $C_V/k_B$ is 0.439229,
the same as the uniform double peak value, but the peak position
is as $b=\sqrt{2c|J|/\beta}$, where $c=1.19967864$.

\begin{figure}
\includegraphics[width=0.45\textwidth]{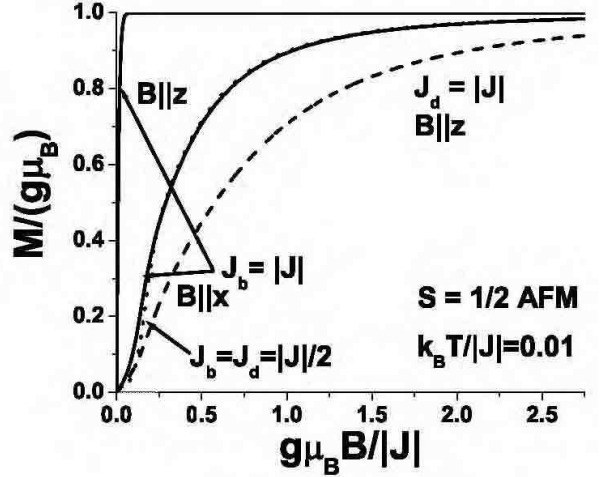}
\caption{Plots of $M/\gamma$ versus $\gamma B/|J|$ at
$k_BT/|J|=0.01$ with $J_b=|J|$ (solid) for both ${\bm B}||\hat{\bm
x},\hat{\bm z}$, $J_b=J_d=|J|/2$ (dotted), and $J_d=|J|$ with
${\bm B}||\hat{\bm z}$.(dashed) for the $s_1=1/2$ AFM
dimer.}\label{fig20}
\end{figure}

\begin{figure}
\includegraphics[width=0.45\textwidth]{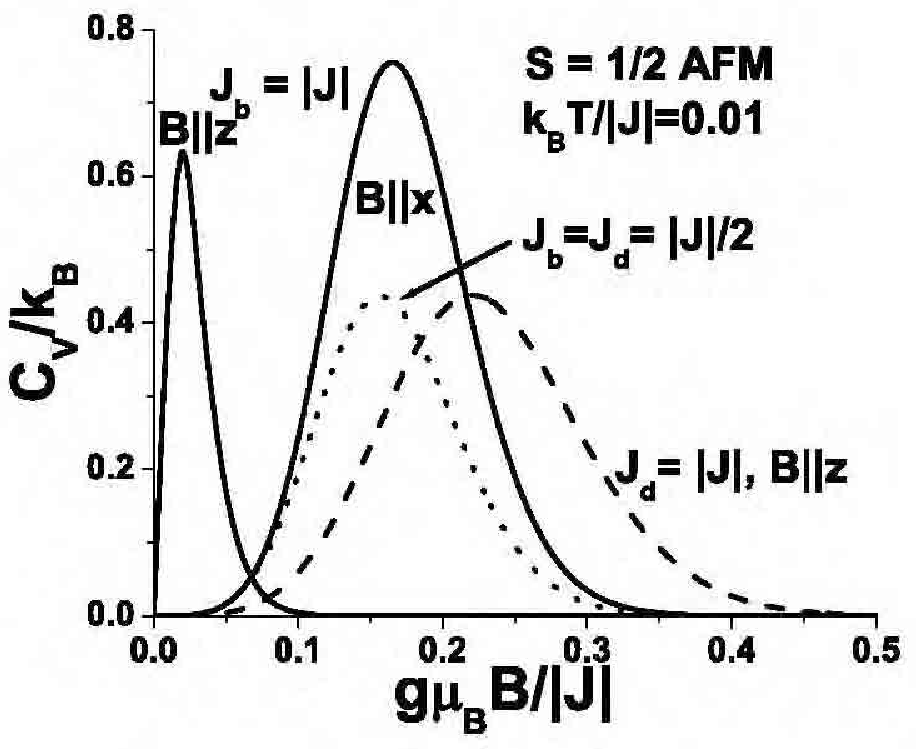}
\caption{Plots of $C_V/k_B$ versus $\gamma B/|J|$ at
$k_BT/|J|=0.01$ with  $J_b=|J|$ (solid) for both ${\bm
B}||\hat{\bm x},\hat{\bm z}$, $J_b=J_d=|J|/2$ (dotted), and
$J_d=|J|$ with ${\bm B}||\hat{\bm z}$ (dashed) for the $s_1=1/2$
AFM dimer.}\label{fig21}
\end{figure}

Next, we examine case (2), $J_b=|J|$, $J_d=0$.  In this case the
energies are different for the two field directions.  For ${\bf
B}||\hat{\bf z}$ that
\begin{eqnarray}
\epsilon_4&=&-J,\\
\epsilon_3&=&b,\label{case2z3}\\
\epsilon_2&=&0,\\
\epsilon_1&=&-b.\label{case2z1}
\end{eqnarray}
At $b=0$, these energies comprise a triply degenerate (or spin 1)
ground state at 0 and an excited $+|J|$ state.  With increasing
$b$, $\epsilon_1$ decreases linearly, and $\epsilon_4$ increases
linearly, crossing $\epsilon_4$ at $b=|J|$, which does not affect
the low-$T$ thermodynamics. In this case, $M(B)$ is a sharp step
at $B=0$, shown as a solid curve in Fig. 20, and $C_V(B)$ is a
conventional spin 1 Schottky anomaly at low $b$, as shown by the
left solid curve in Fig. 21.  From Eqs.
(\ref{case2z3})-(\ref{case2z1}), $C_V/k_b$ has a maximum value at
0.637203 at $b=1.8806775/\beta$.
 Note that the larger maximum value than in the other $C_V(B)$
cases arises from the higher degeneracy as $b\rightarrow0$.

 For ${\bf b}||\hat{\bf x}$, the energies for case (2) are
\begin{eqnarray}
\epsilon_4&=&-J/2+\sqrt{b^2+J^2/4},\\
\epsilon_3&=&0,\label{case2x3}\\
\epsilon_2&=&0,\\
\epsilon_1&=&-J/2-\sqrt{b^2+J^2/4}.\label{case2x1}
\end{eqnarray}
At $b=0$, the ground state is again triply degenerate with energy
0, and the excited state has energy $|J|$.  With increasing $b$,
$\epsilon_1$ separates from $\epsilon_2$ and $\epsilon_3$, curving
below them, and $\epsilon_4$ increases quadratically.  For this
induction direction there is no level crossing, but there is some
${\bm b}$ anisotropy at finite $T$, because of the quadratic
versus linear $b$ dependencies.  $C_V/k_B$ has the maximum value
0.761802 at $b=\sqrt{2c_1|J|/\beta}$, where $c_1=2.654658$, which
is noticeably different from the case with ${\bm B}||\hat{\bm z}$.
The $M(B)$ and $C_V(B)$ curves are the right solid curves in Figs.
20 and 21, respectively.

Now we consider case (3), $J_d=-J$, $J_b=0$.  Again, there are
slight differences for ${\bf B}||\hat{\bf z}$ and ${\bf
B}||\hat{\bf x}$. For ${\bf B}||\hat{\bf z}$, the eigenstate
energies are
\begin{eqnarray}
\epsilon_4&=&-J+\sqrt{b^2+J^2},\\
\epsilon_3&=&-J,\\
\epsilon_2&=&0,\\
\epsilon_1&=&-J-\sqrt{b^2+J^2}.
\end{eqnarray}
At $b=0$, the ground state is doubly degenerate at energy 0.  The
other two states have energies $|J|$ and $2|J|$, respectively. As
$b$ increases, $\epsilon_1$ decreases below $\epsilon_2$, and
$\epsilon_4$ increases monotonically.  Hence, there is no level
crossing, but at low $T$, two levels are relevant, in a fashion
slightly different from a standard Schottky specific heat anomaly.
The low-$T$ $M(B)$ and $C_V(B)$ curves for this induction
direction are shown as the dashed curves in Figs. 20 and 21,
respectively.  We note that they are rather similar from those of
case (1), but have the broadest $M(B)$ increase and $C_V(B)$ peak
of the three cases considered.  $C_V/k_B$ has the peak value
0.439229 which is the same as the uniform double-peak value, but
its position is at $b=2\sqrt{c|J|/\beta}$, where $c=0.19967864$.

For ${\bf B}||\hat{\bf x}$, we have
\begin{eqnarray}
\epsilon_4&=&-2J,\\
\epsilon_3&=&-J/2+\sqrt{b^2+J^2/4},\\
\epsilon_2&=&0,\\
\epsilon_1&=&-J/2-\sqrt{b^2+J^2/4}.
\end{eqnarray}
At $b=0$, the eigenstates are the same as for the field in the $z$
direction, with a doubly degenerate ground state at 0, and excited
states at $|J|$ and $2|J|$.  However, the field dependence is a
bit different than for the $z$ direction.  Again the ground state
decreases with increasing $b$, but the energy scale of the
curvature is $|J|/2$ instead of $|J|$.  In addition, $\epsilon_3$
can cross $\epsilon_4$ at $b=\sqrt{2}|J|$.  This level crossing is
irrelevant to the low-$T$ behavior, however, but modifies the
$M(B)$ curve somewhat at finite $T$.

\subsection{B. AFM $s_1=5/2$ dimers with strong anisotropy interactions}

\begin{figure}
\includegraphics[width=0.45\textwidth]{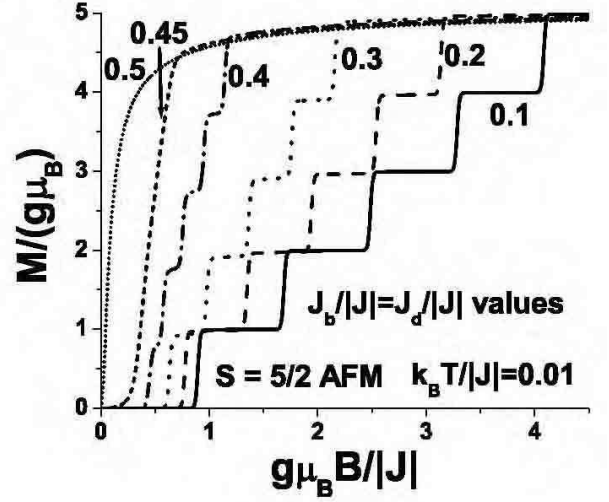}
\caption{Plots of $M/\gamma$ versus $\gamma B/|J|$ at
$k_BT/|J|=0.01$ with $J_b=J_d=c|J|$ for the AFM $s_1=5/2$ dimer.
The cases $c=0.1$ (solid), $c=0.2$(dashed), $c=0.3$ (dotted),
$c=0.4$ (dot-dashed), $c=0.45$ (short dashed), and $c=0.5$ (short
dotted) are shown.}\label{fig22}
\end{figure}

We now consider some cases of strong  anisotropy interactions in
the $s_1=5/2$ dimer.  In Fig. 22, we present low-$T$ plots of
$M(B)$ with $J_b=J_d=c|J|$, where $c=0.1, 0.2, 0.3, 0.4, 0.45$ and
0.5, as indicated.  For the weak coupling case $c=0.1$, the steps
have uniform height.  With increasing $c$, the curves shift
monotonically to lower $b$ values.  Slight deviations in the
uniformity of the step height are detectable in the first two
steps for $c=0.2$, but the nonuniformity in the step height is
progressively more pronounced for $c=0.3$ and 0.4, respectively.
However, for $c=0.45$, the individual steps have completely
disappeared, and are replaced by what appears to be a single,
broad step centered at $b/J\approx0.5$.  At $c=0.5$, $M(B)$ rises
to its maximum value with a very steep slope at $b=0$, not showing
any evidence for any level crossings.

\begin{figure}
\includegraphics[width=0.45\textwidth]{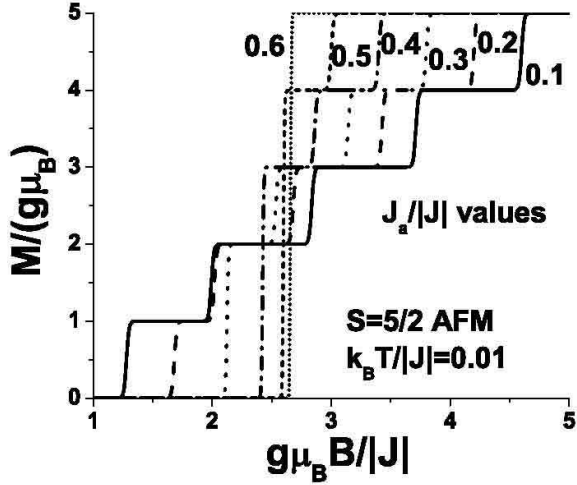}
\caption{Plots of $M/\gamma$ versus $\gamma B/|J|$ at
$k_BT/|J|=0.01$ with $J_a=c|J|$ for the AFM $s_1=5/2$ dimer. The
cases $c=0.1$ (solid), $c=0.2$ (dashed), $c=0.3$ (dotted), $c=0.4$
(dot-dashed),   $c=0.5$ (short dashed), and $c=0.6$  (short
dotted) are shown.}\label{fig23}
\end{figure}

\begin{figure}
\includegraphics[width=0.45\textwidth]{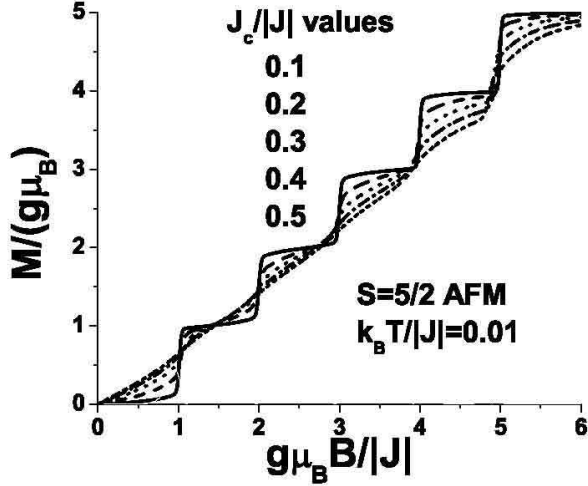}
\caption{Plots of $M/\gamma$ versus $\gamma B/|J|$ at
$k_BT/|J|=0.01$ with $J_c=c|J|$ for the AFM $s_1=5/2$ dimer. The
cases $c=0.1$ (solid), $c=0.2$ (dashed), $c=0.3$ (dotted), $c=0.4$
(dot-dashed),  and $c=0.5$ (short dotted) are shown.}\label{fig24}
\end{figure}

Next, we investigated the effects of large $J_a$.  In Fig. 23  we
plotted $M/\gamma$ versus $\gamma B/|J|$ for $J_a=c|J|$, with
$c=0.1$ (solid), 0.2 (dashed), 0.3 (dotted), 0.4 (dot-dashed), 0.5
(short dashed), and 0.6 (short dotted).  We see the the effects of
strong $J_a$ are very different from those of strong $J_b$ and
$J_d$ shown in Fig. 22.  Instead of the step positions decreasing
monotonically with increasing $J_b=J_d$, as $J_a$ increases, the
highest three step positions decrease with increasing $J_a$, but
the lowest two step positions increase with increasing $J_a$.

We now consider the case of large $J_c$.  In Fig. 24, we plotted
$M/\gamma$ versus $b/|J|$ at $k_BT/|J|=0.01$ for the $s_1=5/2$ AFM
dimer with $J_c=c|J|$, where $c=0.1, 0.2, 0.3, 0.4,$ and 0.5.
These results are shown respectively as the solid, dashed, dotted,
dot-dashed, and short dotted curves.  Unlike the strong global
anisotropy case pictured in Fig. 22 and the strong local axial
anisotropy case pictured in Fig. 23, these curves do not shift
significantly to lower $b$ values with increasing $J_c/|J|$, but
the step shapes are greatly altered. With $c=0.4$, the first two
steps are hard to discern, but rounded remnants of the three last
steps are evident. At $c=0.5$, the third step is hard to detect,
and the remnants of the last two steps are reduced in magnitude.
We remark that the curve with $c=-0.3$ is indistinguishable from
the dotted curve for $c=0.3$, so that in this case, strong $J_c/J$
of either sign can grossly alter the $M({\bm B})$ step behavior.

\begin{figure}
\includegraphics[width=0.45\textwidth]{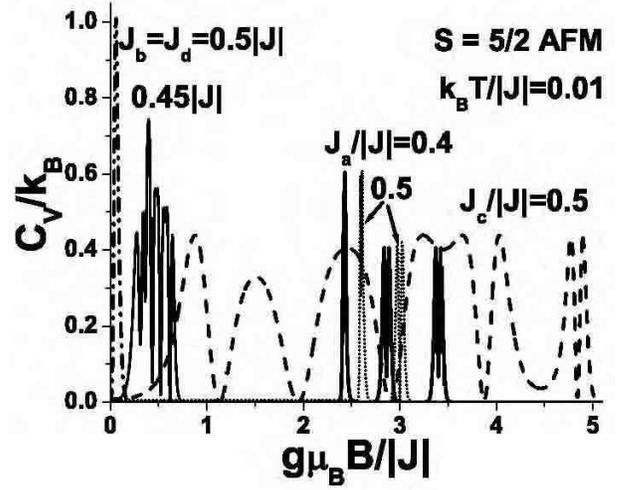}
\caption{Plots of $C_V/k_B$ versus $\gamma B/|J|$ at
$k_BT/|J|=0.01$  for the AFM $s_1=5/2$ dimer. The cases
$J_b=J_d=c|J|$ for $c=0.45$ (solid) and $c=0.5$ (dot-dashed),
$J_a=c|J|$ for $c=0.4$ (solid) and 0.5 (short-dashed) and
$J_c=0.5|J|$ (thick dashed) are shown.}\label{fig25}
\end{figure}

In Fig. 25, we show  $C_V/k_B$ versus $b/|J|$ low-$T$ curves for
some of the AFM $s_1=5/2$ cases pictured in Fig. 22-24.  The solid
and dotted curves on the left-hand side of the figure correspond
to $J_b=J_d=c|J|$ with $c=0.45$ and 0.5, respectively, the solid
and dot-dashed curves in the central portion of the figure
correspond to $J_a/|J|=0.4$ and 0.5, respectively, and the dashed
curve running throughout the domain pictured corresponds to
$J_c=0.5|J|$. We note that for the global anisotropy case shown,
strong anisotropy pushes the $C_V(B)$ peaks to lower $b$, and
squeezes them together, so that they tend to overlap.  At the
limiting case $J_b=J_d=0.5|J|$, the individual level crossings
have been eliminated, and all ten of the peaks are combined into a
single Schottky anomaly. For the local axial anisotropy
interaction, strong FM anisotropic interaction squeeze the
$C_V(B)$ together in the middle of the domain, with the limiting
case of a single peak at $J_a/|J|=0.6$.  For the strong azimuthal
anisotropic exchange interaction case, however, strong FM
anisotropy shifts the double peak positions only slightly, and
broadens the double peaks, so that only the highest set is clearly
discernable, even at the very low $T$ value plotted. In each case,
the deviation of the peak height from the two-level prediction ,
Eq. (\ref{peak}), is substantial, indicating more than two levels
are important at low $T$, as for the three-level system with
$s_1=1/2$ and $J_b=|J|$.  Thus, the different types of strong spin
anisotropy lead to clearly different low-$T$ $C_V(B)$ behaviors
for $s_1>1/2$ dimers.

\section{IX. Summary and conclusions}

In summary, we solved for the low-temperature  magnetization and
specific heat of equal spin $s_1$ antiferromagnetic dimer single
molecule magnets, including the most general set of anisotropic
spin exchange interactions quadratic in the spin operators. The
magnetization and specific heat  exhibit steps and zeroes,
respectively, at the non-universal level-crossing induction values
$B_{s,s_1}^{\rm lc}(\theta,\phi)$, but the magnetization steps and
their midpoint slopes, plus the two peaks surrounding the specific
heat zeroes all exhibit universal behavior at sufficiently low
temperatures to first order in the anisotropy interaction
strengths. Strong anisotropy interactions generally lead to highly
non-universal behavior in antiferromagnetic dimers. Local (or
single-ion) anisotropy interactions lead to low-temperature
magnetization step plateaus that have a much richer variation with
the magnetic induction ${\bm B}$ than do those obtained from
global anisotropy interactions, provided that
 $s_1>1/2$.  For the most general quadratic anisotropic spin
interactions at an arbitrary ${\bm B}$, we derived simple,
 asymptotic analytic expressions for the low-temperature
magnetization, specific heat,  inelastic neutron scattering
cross-section, and electron paramagnetic response susceptibility,
which are accurate for weak anisotropy.  We also derived  an
accurate expression for the level-crossing induction
$B_{s,s_1}^{\rm lc}(\theta,\phi)$, enabling fast and accurate fits
to experimental data.

There were two low-$T$ $M(B)$ studies of Fe$_2$
dimers.\cite{Fe2mag,Fe2Cl}  For
$\mu$-oxalatotetrakis(acetylacetonato)Fe$_2$, all five peaks in
$dM/dH$ were measured in pulsed magnetic fields $H$.  These evenly
spaced peaks
  indicated little, if any, spin
anisotropy effects.\cite{Fe2mag}  On the other hand, studies of
the first 2-3 $dM/dH$ peaks in powdered samples of
[Fe(salen)Cl]$_2$, where salen is
$N,N'$-ethylenebis(salicylideneiminato), were much more
interesting.\cite{Fe2Cl} These data showed a broad first peak at
$B=17-20$T that was only partially resolvable into two separate
peaks, followed by a sharp second peak at $B=36$ T, consistent
with  local axial anisotropy of strength $|J_a/J|\approx0.1$, as
 obtained from the derivatives of the curves shown in Fig. 13. From Eqs.
 (\ref{B52step1})-(\ref{B52step3}), one could also have $|(J_a\pm
3J_e)/J|\approx0.1$.
 These amounts might perhaps be combined with a smaller $|J_c/J|$, obtained from the derivatives of the curves
 pictured in Fig. 14.

 Without a detailed single
 crystal study with magnetic fields along different crystal
 directions, it is impossible to determine the relative amounts of
 $J_a$ and $J_e$ that would best fit the data.
However, the existing data on [Fe(salen)Cl]$_2$ appear to be
inconsistent with a predominant global anisotropy interaction of
either type, as obtained from the derivatives of the curves shown
in Figs. 11 and 12. Only single crystal studies could determine if
a small amount of such global anisotropy interactions were present
in addition to one or more presumably larger local spin anisotropy
interactions. In comparing the two materials cited above, it
appears that the interaction of a Cl$^{-}$ ion neighboring each
Fe$^{3+}$ ion leads to strong local (or single-ion) anisotropy
effects.  In order to verify this hypothesis and to elucidate the
details of the interactions, further experiments using single
crystals in different field orientations on this and related
Fe$_2$ dimers with 1-3 similarly bonded Cl$^{-}$ ions are
urged.\cite{Fe2Cl3,Fe2Clnew}   We also urge single crystal data on
some of the $s_1=9/2$ [Mn$_4$]$_2$ dimers,
\cite{Mn4dimer,Mn4dimerDalal,ek}, as well as on $s_1=1/2$ dimers
lacking in predicted local spin anisotropy effects.
\cite{V2neutron,Gudel,V2P2O9,ek}To aid in the fits, we derived
simple, useful formulas for the magnetization and specific heat at
low temperature and sufficiently large magnetic induction.  More
important, we derived accurate analytic formulas for the electron
paramagnetic resonance susceptibility $\chi({|bm B},\omega)$ and
the inelastic neutron scattering $S({\bm B},{\bm q},\omega)$,
which allow for a precise determination of the various microscopic
anisotropy energies.

With local anisotropy interactions, the total spin $s$ is not a
good quantum number, potentially modifying our understanding of
quantum tunneling processes in single molecule magnets. It might
also be possible to fit a variety of experimental results using a
smaller, consistent set of model parameters.\cite{Fe8spin9} We
emphasize that the study of dimer single molecule magnets, for
which the most general anisotropic quadratic exchange interactions
can be solved exactly, may be our best hope for attaining a more
fundamental understanding of the underlying physics of more
general single molecule magnets.

\section{Acknowledgments}

We thank the Max-Planck-Institut f{\"u}r Physik komplexer Systeme,
Dresden, Germany, the University of North Dakota, Grand Forks, ND,
USA, and Talat S. Rahman  for their kind hospitality and support.
This work was supported by the Netherlands Foundation for the
Fundamental Research of Matter and by the NSF under contract
NER-0304665.

\section{Appendix A}
\subsection{Single-ion operators for arbitrary $s_1,s_2$}
For completeness, we present the results of the single-ion
operators for general dimers consisting of magnetic ions with
spins $s_1$ and $s_2$. We find
\begin{eqnarray}
S_{i,z}|\psi_s^m\rangle&=&\frac{m}{2}\Bigl(1-(-1)^i\xi_{s,s_1,s_2}\Bigr)|\psi_s^m\rangle\nonumber\\
& &-\frac{1}{2}(-1)^i\Bigl(D_{s,s_1,s_2}^m|\psi_{s-1}^m\rangle\nonumber\\
& &\qquad\qquad+D_{s+1,s_1,s_2}^m|\psi_{s+1}^m\rangle\Bigr),\\
S_{i,\sigma}|\psi_s^m\rangle&=&\frac{A_s^{\sigma
m}}{2}\Bigl(1-(-1)^i\xi_{s,s_1,s_2}\Bigr)|\psi_s^{m+\sigma}\rangle\nonumber\\
& &-\frac{\sigma}{2}(-1)^i\Bigl(C_{s,s_1,s_2}^{\sigma
m}|\psi_{s-1}^{m+\sigma}\rangle\nonumber\\
& &\qquad\qquad-C_{s+1,s_1,s_2}^{-1-\sigma m}|\psi_{s+1}^{m+\sigma}\rangle\Bigr),\\
C_{s,s_1,s_2}^m&=&\eta_{s,s_1,s_2}\sqrt{(s-m)(s-m-1)},\\
D_{s,s_1,s_2}^m&=&\eta_{s,s_1,s_2}\sqrt{s^2-m^2},\\
\eta_{s,s_1,s_2}&=&\sqrt{\frac{[(s_1+s_2+1)^2-s^2][s^2-(s_1-s_2)^2]}{(4s^2-1)s^2}},\nonumber\\
& &\\
 \xi_{s,s_1,s_2}&=&\frac{s_1(s_1+1)-s_2(s_2+1)}{s(s+1)}.
\end{eqnarray}
Note that for $s_1=s_2$, $\xi_{s,s_1,s_1}=0$ and
$\eta_{s,s_1,s_1}$ reduces to Eq. (\ref{eta}) for $\eta_{s,s_1}$.
Also, note that these relations preserve the global spin
properties of $S_z=S_{1,z}+S_{2,z}$ and
$S_{\sigma}=S_{1,\sigma}+S_{2,\sigma}$,
$S_z|\psi_s^m\rangle=m|\psi_s^m\rangle$ and
$S_{\sigma}|\psi_s^m\rangle=A_s^{\sigma
m}|\psi_s^{m+\sigma}\rangle$, as for equal spin dimers.  In
addition, the spins on asymmetric dimers with $s_1\ne s_2$ would
be expected to also have different local anisotropy interactions,
so that ${\cal H}_a$ and ${\cal H}_e$ would become
\begin{eqnarray}
{\cal H}_a'&=&-\sum_{i=1}^2J_{a,i}S_{i,z}^2,\\
{\cal H}_e'&=&-\sum_{i=1}^2J_{e,i}(S_{i,x}^2-S_{i,y}^2).
\end{eqnarray}

\subsection{Specific heat details for $s_1=1/2$}
We first present the numerators of the exact expressions for the
specific heat with $s_1=1/2$  and ${\bm B}||\hat{\bm i}$ for
$i=x,y,z$.  We have
\begin{eqnarray}
{\cal N}_{x,y}&=&F_{x,y}^2+\frac{1}{4}(J+2J_{y,x})^2e^{\beta[2(J_{y,x}-J_{x,y})-J]}\nonumber\\
& &+F_{x,y}\sinh(\beta F_{x,y})e^{-\beta
J_{x,y}}\nonumber\\
& &\times[(J+J_{x,y})e^{-\beta J}+(J_{x,y}-2J_{y,x})e^{2\beta
J_{y,x}}]\nonumber\\
& &+\frac{1}{2}e^{-\beta J_{x,y}}\cosh(\beta
F_{x,y})\nonumber\\
& &\times\Bigl([(J+J_{x,y})^2+F_{x,y}^2]e^{-\beta
J}\nonumber\\
& &+[(2J_{y,x}-J_{x,y})^2+F_{x,y}^2]e^{2\beta J_{y,x}}\Bigr),\\
 {\cal N}_z&=&F_z^2\cosh(2\beta
F_z)+\frac{J^2}{4}e^{-\beta(J+2J_b)}\nonumber\\
& &+\frac{1}{2}\cosh(\beta
F_z)\Bigl(\Delta_z(J_b^2+F_z^2)\nonumber\\
& &\qquad+e^{-\beta(J+J_b)}J(J+2J_b)\Bigr)\nonumber\\
& &+F_z\sinh(\beta F_z)\Bigl(J_b\Delta_z+Je^{-\beta(J+J_b)}\Bigr).
\end{eqnarray}

\subsection{Eigenvalues for $s_1=1$ in the crystal representation}

In the remainder of this appendix, we provide some details of our
exact results for $s_1=1$. The cubic equation for the three $s=1$
eigenvalues is given by
\begin{eqnarray}
\epsilon_n&=&-J-J_b-J_a+\lambda_n,\>\>{\rm for}\>\>n=2,3,4,\\\
 0&=&
-\lambda_n^3-(J_a-J_b)\lambda_n^2+\lambda_n[b^2+(J_d-J_e)^2]\nonumber\\
& &+(J_a-J_b)[b^2\cos^2\theta+(J_d-J_e)^2]\nonumber\\
& &-(J_d-J_e)(b^2\sin^2\theta\cos(2\phi).
\end{eqnarray}
 For  ${\bm B}||\hat{\bm
z}$, the cubic equation is easily solved to yield
\begin{eqnarray}
\lambda_n&=&J_b-J_a,\pm\sqrt{b^2+(J_d-J_e)^2}.
\end{eqnarray}
For  ${\bm B}||\hat{\bm x},\hat{\bm y}$, we have
\begin{eqnarray}
\lambda_n&=&\pm(J_e-J_d),-J_{y,x}\pm\sqrt{b^2+\overline{J}_{x,y}^2},\\
\overline{J}_{x,y}&=&\frac{1}{2}[J_a-J_b\mp(J_d-J_e)],
\end{eqnarray}
where $\overline{J}_x$ ($\overline{J}_y$) corresponds to the upper
(lower) sign.

 The six eigenvalues for
the mixed $s=0,2$ states satisfy
\begin{eqnarray}
\epsilon_n&=&-3J-\frac{4}{3}J_a-2J_b+\lambda_n\label{epsilonnspin1}
\end{eqnarray}
 We first define
\begin{eqnarray}
a&=&-\frac{\sqrt{8}}{3}J_a,\label{Jatilde}\\
b_{\perp}&=&-b\sin\theta e^{-i\phi},\\
b_3&=&\sqrt{\frac{3}{2}}b_{\perp},\\
d&=&-\sqrt{6}(J_d+J_e/3),\\
d_3&=&\sqrt{\frac{3}{2}}d,\\
e&=&-\frac{2}{\sqrt{3}}J_e,\label{Jetilde}\\
\tilde{J}&=&J-\frac{2}{9}J_a,\label{Jtilde}\\
\noalign{\rm and} Q_n^p&=&n(J_b+J_a/3)+pb\cos\theta.
\end{eqnarray}
Then the $\lambda_n$ are the eigenvalues of the Hermitian matrix
$\tensor{\bm M}$ given by
\begin{eqnarray}
\tensor{\bm M}\!\!&=&\!\!\left(\begin{array}{cccccc}Q_{-2}^{-2}&b_{\perp}&d&0&0&e\\
b_{\perp}^{*}&Q_1^{-1}&b_3&d_3&0&0\\
d&b_3^{*}&Q_2^{0}&b_3&d&a\\
0&d_3&b_3^{*}&Q_1^1&b_{\perp}&0\\
0&0&d&b_{\perp}^{*}&Q_{-2}^2&e\\
e&0&a&0&e&Q_2^0+3\tilde{J}\end{array}\right).
\end{eqnarray}

 The
resulting sixth order polynomial for the $\lambda_n$ is given
elsewhere.\cite{ekcondmat}  We note that for ${\bm B}||\hat{\bm
i}$, $\tensor{\bm M}$ is block diagonal, breaking up into matrices
of ranks two and four.  These cases are discussed in detail
elsewhere.\cite{ekcondmat}  Here we only present the simplest data
for which the eigenvalues are determined either by linear or by
quadratic equations.

\subsection{Simple special cases}

When only one of the $J_j\ne0$, the eigenvalues for ${\bm
B}||\hat{\bm z}$ simplify considerably.  For $J_b\ne0$, we have
\begin{eqnarray}
\lambda_n^z&=&3J+2J_b, 2J_b, J_b\pm b, -2J_b\pm 2b.
\end{eqnarray}
For $J_d\ne0$, we find
\begin{eqnarray}
\lambda_n^z&=&0,3J, \pm\sqrt{b^2+9J_d^2}, \pm2\sqrt{b^2+3J_d^2}.
\end{eqnarray}
For $J_a\ne0$, the eigenvalues can also be found analytically,
\begin{eqnarray}
\lambda_n^z&=&-\frac{2J_a}{3}\pm 2b, \frac{J_a}{3}\pm
b,\frac{J_a}{3}+\frac{3J}{2}\pm\sqrt{\frac{9}{4}J^2+J_a^2-JJ_a}.\nonumber\\
\end{eqnarray}
We note that the ground state energy in this case is
\begin{eqnarray}
E_1&=&-\frac{3}{2}J-J_a-\sqrt{\frac{9}{4}J^2+J_a^2-JJ_a},
\end{eqnarray}
which explicitly involves mixing of the $s=m=0$ and the $s=2, m=0$
states.

  At the first level crossing with ${\bm
B}||\hat{\bm z}$, we  have for $J_a\ne0$ and the other $J_j=0$,
\begin{eqnarray} \gamma B_{1,1}^{{\rm
lc},z}&=&\frac{1}{2}\Bigl(J+[9J^2+4J_a^2-4JJ_a]^{1/2}\Bigr).\label{Jastep1}
\end{eqnarray}
We note that the first level crossing for $J_b,J_d\ne0$ for ${\bm
B}||\hat{\bm i}$ is equivalent to that of $s_1=1/2$, given by Eq.
(\ref{Bstephalf}). At the second level crossing, simple formulas
are only obtained for ${\bm B}||\hat{\bm z}$ with one $J_j\ne0$.
For ${\bm B}||\hat{\bm z}$ and $J_a\ne0$, $J_b\ne0$ and $J_d\ne0$,
respectively, we have
\begin{eqnarray}
\gamma B_{2,1}^{{\rm
lc},z}&=&\left\{\begin{array}{c}-2J-J_a,\\
-2J -3J_b,\\
\\
\frac{1}{3}\Bigl(20J^2-30J_d^2+8[4J^4-6J^2J_d^2]^{1/2}\Bigr)^{1/2}.\end{array}\right.\label{secondstep}
\end{eqnarray}

\section{Appendix B}
\subsection{Rotation to the induction representation}
The rotation from the crystal representation to the induction
representation is obtained from
\begin{eqnarray}
\left(\begin{array}{c}\hat{\bm x}\\
\hat{\bm y}\\
\hat{\bm z}\end{array}\right)=\left(\begin{array}{ccc}\cos\theta\cos\phi &-\sin\phi &\sin\theta\cos\phi\\
\cos\theta\sin\phi &\cos\phi &\sin\theta\sin\phi\\
-\sin\theta
&0&\cos\theta\end{array}\right)\left(\begin{array}{c}\hat{\bm
x}'\\
\hat{\bm y}'\\
\hat{\bm z}'\end{array}\right),\label{rotation}\nonumber\\
\end{eqnarray}
leading to ${\bm B}=B\hat{\bm z}'$.\cite{klemm}  This operation is
equivalent to a rotation by $-\pi/2$ about the $z$ axis, a
rotation by $\theta$ about the transformed $x$ axis, and then a
rotation by $\pi/2-\phi$ about the transformed $z$
axis.\cite{Goldstein} In effect, in using the above rotation
matrix, we made the arbitrary choice that the rotated $z$ axis
lies in the $x'z'$ plane. After the above rotation, it is still
possible to rotate the crystal by an arbitrary angle $\chi$ about
the $z'$ axis, keeping ${\bm B}||\hat{\bm z}'$.  Hence, there are
in effect an infinite number of equivalent rotations leading to
${\bm B}=B\hat{\bm z}'$.  The resulting Hamiltonian matrix will
then have off-diagonal elements that depend upon $\chi$.
 However, all such rotations necessarily lead to the identical, $\chi$-independent, set
of eigenvalues of the resulting diagonalized Hamiltonian matrix.
We have explicitly checked that the above rotation gives the exact
cubic expression, Eq. (\ref{halfeigen}), for the $s_1=1/2$
eigenvalues, and also leads to the correct eigenstate energies
second order in each of the $J_j$ for $s_1=1$.  We also showed
explicitly that $\chi$ does not enter the eigenstate energies
second order in $J_b$ for arbitrary $s, s_1, m$.

\subsection{Global Hamiltonian}
The global axial and azimuthal anisotropy interactions in the
rotated frame are
\begin{eqnarray}{\cal
H}_b'&=&-J_b\Bigl(S_{z'}^2\cos^2\theta+S_{x'}^2\sin^2\theta\nonumber\\
& &\qquad-\sin(2\theta)\{S_{x'},S_{z'}\}/2\Bigr),\\
{\cal
H}_d'&=&-J_d\Bigl[\cos(2\phi)\Bigl(S_{x'}^2\cos^2\theta+S_{z'}^2\sin^2\theta-S_{y'}^2\nonumber\\
&
&\qquad+\sin(2\theta)\{S_{x'},S_{z'}\}/2\Bigr)\nonumber\\
&
&-\sin(2\phi)\Bigl(\cos\theta\{S_{x'},S_{y'}\}+\sin\theta\{S_{y'},S_{z'}\}\Bigr)\Bigr],\nonumber\\
\end{eqnarray}
where $\{A,B\}=AB+BA$ is the anticommutator.

\subsection{Global Hamiltonian
matrix elements}

 The operations of the rotated global anisotropy interactions upon
these states may be written as
\begin{eqnarray}
{\cal
H}_b'|\varphi_s^m\rangle&=&-\frac{J_b}{4}\Bigl[\Bigl(4m^2+2[s(s+1)-3m^2]\sin^2\theta\Bigr)|
\varphi_s^m\rangle\nonumber\\
& &-\sin(2\theta)\sum_{\sigma=\pm1}(2m+\sigma)A_s^{\sigma
m}|\varphi_s^{m+\sigma}\rangle\nonumber\\
& &+\sin^2\theta\sum_{\sigma=\pm1}F_s^{\sigma
m}|\varphi_s^{m+2\sigma}\rangle\Bigr],
\label{Hbprime}\end{eqnarray} and
\begin{eqnarray} {\cal
H}_d'|\varphi_s^m\rangle&=&-\frac{J_d}{4}\Bigl(2\sin^2\theta\cos(2\phi)[3m^2-s(s+1)]|\varphi_s^m\rangle\nonumber\\
& &+2\sin\theta\sum_{\sigma=\pm1}(2m+\sigma)A_s^{\sigma
m}\nonumber\\
& &\times[\cos\theta\cos(2\phi)+i\sigma\sin(2\phi)]|\varphi_s^{m+\sigma}\rangle\nonumber\\
& &+\sum_{\sigma=\pm1}F_s^{\sigma
m}\bigl[(1+\cos^2\theta)\cos(2\phi)\nonumber\\
&
&+2i\sigma\cos\theta\sin(2\phi)\bigr]|\varphi_s^{m+2\sigma}\rangle\Bigr),\label{Hdprime}
\end{eqnarray}
where $F_s^x$ is defined by  Eq. (\ref{Fsx}).  We note that in
this representation, both ${\cal H}_b$ and ${\cal H}_d$ preserve
the global spin quantum number $s$, but allow $\Delta m=\pm1,\pm2$
transitions.

\subsection{Local Hamiltonian}

With regard to the local spin anisotropy terms in the rotated
coordinate system, we write
\begin{eqnarray}
{\cal H}_a'&=&-J_a\Bigl({\cal O}_{1}\cos^2\theta+{\cal
O}_{2}\sin^2\theta-\frac{\sin(2\theta)}{2}{\cal O}_{3}\Bigr)
\end{eqnarray}
and \begin{eqnarray} {\cal
H}_e'&=&-J_e\Bigl[\cos(2\phi)\Bigl({\cal O}_1\sin^2\theta+{\cal
O}_2\cos^2\theta\nonumber\\
& &\qquad+\frac{1}{2}\sin(2\theta){\cal O}_3-{\cal O}_4\Bigr)\nonumber\\
& &-\sin(2\phi)\Bigl({\cal O}_5\cos\theta+ {\cal
O}_6\sin\theta\Bigr)\Bigr],
\end{eqnarray}
where
\begin{eqnarray}
{\cal O}_{1}&=&\sum_{i=1}^2S_{iz'}^2,\label{O1}\\
{\cal O}_{2}&=&\sum_{i=1}^2S_{ix'}^2,\label{O2}\\
{\cal
O}_{3}&=&\sum_{i=1}^2(S_{iz'}S_{ix'}+S_{ix'}S_{iz'}),\label{O3}\\
{\cal O}_4&=&\sum_{i=1}^2S_{iy'}^2,\\
{\cal O}_5&=&\sum_{i=1}^2(S_{ix'}S_{iy'}+S_{iy'}S_{ix'}),
\end{eqnarray}
and\begin{eqnarray}
 {\cal
O}_6&=&\sum_{i=1}^2(S_{iy'}S_{iz'}+S_{iz'}S_{iy'}).
\end{eqnarray}

\subsection{Local Hamiltonian matrix element components}

 The operations of these interactions are
given by
\begin{eqnarray}
{\cal
O}_{1}|\varphi_s^m\rangle&=&\frac{1}{2}\Bigl(G_{s,s_1}^m|\varphi_s^m\rangle\nonumber\\
&
&+\sum_{\sigma'=\pm1}H_{s,s_1}^{m,\sigma'}|\varphi_{s+2\sigma'}^m\rangle\Bigr),\\
{\cal
O}_{2}|\varphi_s^m\rangle&=&\frac{1}{8}\Bigl(M_{s,s_1}^m|\varphi_s^m\rangle-\sum_{\sigma'=\pm1}N_{s,s_1}^{m,\sigma'}|\varphi_{s+2\sigma'}^m\rangle\nonumber\\
& &+\sum_{\sigma=\pm1}L_{s,s_1}^{\sigma m}|\varphi_s^{m+2\sigma}\rangle\nonumber\\
& &+\sum_{\sigma,\sigma'=\pm1}K_{s,s_1}^{\sigma
m,\sigma'}|\varphi_{s+2\sigma'}^{m+2\sigma}\rangle\Bigr),\\
{\cal
O}_{3}|\varphi_s^m\rangle&=&\frac{1}{4}\sum_{\sigma=\pm1}\Bigl(P_{s,s_1}^{m,\sigma}|\varphi_s^{m+\sigma}\rangle\nonumber\\
& & -\sum_{\sigma'=\pm1}\sigma\sigma' R_{s,s_1}^{m,\sigma,\sigma'}|\varphi_{s+2\sigma'}^{m+\sigma}\rangle\Bigr),\\
{\cal
O}_4|\varphi_s^m\rangle&=&\frac{1}{8}\Bigl(M_{s,s_1}^m|\varphi_s^m\rangle-\sum_{\sigma'=\pm1}N_{s,s_1}^{m,\sigma'}|\varphi_{s+2\sigma'}^m\rangle\nonumber\\
& &-\sum_{\sigma=\pm1}L_{s,s_1}^{\sigma m}|\varphi_s^{m+2\sigma}\rangle\nonumber\\
& &-\sum_{\sigma,\sigma'=\pm1}K_{s,s_1}^{\sigma
m,\sigma'}|\varphi_{s+2\sigma'}^{m+2\sigma}\rangle\Bigr),\\
{\cal
O}_5|\varphi_s^m\rangle&=&\frac{1}{4i}\sum_{\sigma=\pm1}\sigma\Bigl(L_{s,s_1}^{\sigma m}|\varphi_s^{m+2\sigma}\rangle\nonumber\\
& &+\sum_{\sigma'=\pm1}K_{s,s_1}^{\sigma
m,\sigma'}|\varphi_{s+2\sigma'}^{m+2\sigma}\rangle\Bigr),\\
{\cal
O}_6|\varphi_s^m\rangle&=&\frac{1}{4i}\sum_{\sigma=\pm1}\Bigl(\sigma P_{s,s_1}^{m,\sigma}|\varphi_s^{m+\sigma}\rangle\nonumber\\
& &-\sum_{\sigma'=\pm1}\sigma'
R_{s,s_1}^{m,\sigma,\sigma'}|\varphi_{s+2\sigma'}^{m+\sigma}\rangle\Bigr),
\end{eqnarray}
where \begin{eqnarray} M^m_{s,s_1}&=&-4m^2\alpha_{s,s_1}+4[s(s+1)-1](1-\alpha_{s,s_1}),\nonumber\\
& &\\
 N_{s,s_1}^{m,\sigma'}&=&\sum_{\sigma=\pm1}C_{s+(\sigma'+1)/2,s_1}^{-\sigma\sigma'm-(\sigma'+1)/2}
C_{s+(3\sigma'+1)/2,s_1}^{\sigma\sigma'm+(\sigma'-1)/2},\label{Nss1msigma}\\
P_{s,s_1}^{m,\sigma}&=&2A_s^{\sigma m}(2m+\sigma)\alpha_{s,s_1},\\
\noalign{\rm and}
 R_{s,s_1}^{m,\sigma,\sigma'}&=&C_{s+(\sigma'+1)/2,s_1}^{-m\sigma\sigma'-(\sigma'+1)/2}D_{s+(3\sigma'+1)/2,s_1}^{m+\sigma}\nonumber\\
&
&+C_{s+(3\sigma'+1)/2,s_1}^{-m\sigma\sigma'-(\sigma'+1)/2}D_{s+(\sigma'+1)/2,s_1}^m,\label{Rss1xsigma}
\end{eqnarray}  where $\eta_{s,s_1}$, $G_{s,s_1}^m$, $H_{s,s_1}^{m,\sigma'}$,
$K_{s,s_1}^{x,\sigma'}$, and $L_{s,s_1}^x$ are given by Eqs.
(\ref{eta}) and (\ref{Gss1m})-(\ref{Lss1}), respectively.  We note
that for $\sigma'=\pm1$,
$N_{s,s_1}^{m,\sigma'}=2H_{s,s_1}^{m,\sigma'}$.

\section{Appendix C}

\subsection{Inelastic neutron scattering cross-section}
Here we outline the derivation and provide the important results
for the low-$T$ inelastic neutron scattering cross-section,
$S({\bm B},{\bm q},\omega)$. At low $T$, we set
$E_{s,s_1}^m\approx E_s^{m,(0)}+E_{s,s_1}^{m,(1)}$ given by Eqs.
(\ref{E0}) and (\ref{Esm1}). Since this energy was evaluated in
the induction representation, it is easiest to evaluate $S({\bm
q},\omega)$ by choosing the axes $\hat{\bm e}_{\alpha}=\hat{\bm
i}'$, where $i=x,y,z$.  The $\hat{\bm i}'$ can be obtained from
the $\hat{\bm i}$ using the inverse of the rotation matrix in Eq.
(\ref{rotation}). Then, for $\hat{\bm
q}=(\sin\theta_q\cos\phi_q,\sin\theta_q\sin\phi_q,\cos\theta_q)$
in the crystal representation, we have
\begin{eqnarray}
\hat{\bf
q}&=&(\sin\theta_{b,q}\cos\phi_{b,q},\sin\theta_{b,q}\sin\phi_{b,q},\cos\theta_{b,q})
\end{eqnarray} in the induction representation, where
\begin{eqnarray}
\cos\theta_{b,q}&=&\cos\theta\cos\theta_q+\sin\theta\sin\theta_q\cos(\phi-\phi_q),\nonumber\\
\sin\theta_{b,q}\cos\phi_{b,q}&=&\sin\theta_q\cos\theta\cos(\phi-\phi_q)-\sin\theta\cos\theta_q,\nonumber\\
\sin\theta_{b,q}\sin\phi_{b,q}&=&-\sin\theta_q\sin(\phi-\phi_q).
\end{eqnarray}
Then, expanding the wave functions to first order in the
anisotropy interactions,
\begin{eqnarray}
|\tilde{\varphi}_s^m\rangle&=&|\varphi_s^m\rangle+\sum_{{s',m'}\atop{(s',m')\ne(s,m)}}|\varphi_{s'}^{m'}\rangle\Bigl(\langle\varphi_{s'}^{m'}|N_{s,m}^{(1)}\rangle+\ldots\Bigr),\nonumber\\
& &\\
&=&|\varphi_s^{m}\rangle\nonumber\\
& &+\frac{1}{\gamma B}\sum_{\sigma=\pm1}\Bigl(\sigma{\cal
U}_{s,s_1}^{m,\sigma}|\varphi_s^{m+\sigma}\rangle+\frac{\sigma}{2}{\cal
V}_{s,s_1}^{m,\sigma}|\varphi_s^{m+2\sigma}\rangle\Bigr)\nonumber\\
& &+\sum_{\sigma'=\pm1}\frac{{\cal
W}_{s,s_1}^{m,\sigma'}}{J[\sigma'(2s+1)+2]}|\varphi_{s+2\sigma'}^m\rangle\nonumber\\
& &+\sum_{\sigma,\sigma'=\pm1}\Bigl(\frac{{\cal
X}_{s,s_1}^{m,\sigma,\sigma'}}{\sigma\gamma
B+J[\sigma'(2s+1)+2]}|\varphi_{s+2\sigma'}^{m+\sigma}\rangle\nonumber\\
& &+\frac{{\cal Y}_{s,s_1}^{m,\sigma,\sigma'}}{2\sigma\gamma
B+J[\sigma'(2s+1)+2]}|\varphi_{s+2\sigma'}^{m+2\sigma}\rangle\Bigr)+\ldots,\nonumber\\
\end{eqnarray}
where
\begin{eqnarray}
{\cal
U}_{s,s_1}^{m,\sigma}(\theta,\phi)&=&\frac{1}{4}(2m+\sigma)A_s^{\sigma
m}\Bigl[
\sin(2\theta)\Bigl(\tilde{J}_{b,a}^{s,s_1}\nonumber\\
& &\qquad-\tilde{J}_{d,e}^{s,s_1}\cos(2\phi)\Bigr)\nonumber\\
& &-2i\sigma\tilde{J}_{d,e}^{s,s_1}\sin\theta\sin(2\phi)\Bigr],\label{calU}\\
{\cal V}_{s,s_1}^{m,\sigma}(\theta,\phi)&=&-\frac{1}{4}F_s^{\sigma
m}\Bigl[\tilde{J}_{b,a}^{s,s_1}\sin^2\theta\nonumber\\
& &+\tilde{J}_{d,e}^{s,s_1}(1+\cos^2\theta)\cos(2\phi)\nonumber\\
& &+2i\sigma\tilde{J}_{d,e}^{s,s_1}\cos\theta\sin(2\phi)\Bigr],\\
{\cal
W}_{s,s_1}^{m,\sigma'}(\theta,\phi)&=&-\frac{1}{2}H_{s,s_1}^{m,\sigma'}[J_a\cos^2\theta+J_e\sin^2\theta\cos(2\phi)]\nonumber\\
& &+\frac{1}{8}N_{s,s_1}^{m,\sigma'}\sin^2\theta[J_a-J_e\cos(2\phi)],\\
{\cal
X}_{s,s_1}^{m,\sigma,\sigma'}(\theta,\phi)&=&-\frac{\sigma\sigma'}{8}R_{s,s_1}^{m,\sigma,\sigma'}\Bigl(\sin(2\theta)[J_a-J_e\cos(2\phi)]\nonumber\\
& &\qquad-2i\sigma J_e\sin\theta\sin(2\phi)\Bigr),\\
\noalign{\rm and} & &\nonumber\\ {\cal
Y}_{s,s_1}^{m,\sigma,\sigma'}(\theta,\phi)&=&-\frac{1}{8}K_{s,s_1}^{\sigma
m,\sigma'}\Bigl[J_a\sin^2\theta\nonumber\\
& &
\qquad+J_e\Bigl((1+\cos^2\theta)\cos(2\phi)\nonumber\\
& &\qquad+2i\sigma\cos\theta\sin(2\phi)\Bigr)\Bigr],\label{calY}
\end{eqnarray}
where $H_{s,s_1}^{m,\sigma'}$ and $K_{s,s_1}^{x,\sigma'}$ are
given by Eqs. (\ref{Hss1msigma}) and (\ref{Kss1xsigma}),
respectively, $N_{s,s_1}^{m,\sigma'}$ and
$R_{s,s_1}^{m,\sigma,\sigma'}$ are given by Eqs.
(\ref{Nss1msigma}) and (\ref{Rss1xsigma}), respectively, and
$\tilde{J}_{b,a}^{s,s_1}$ and $\tilde{J}_{d,e}^{s,s_1}$ are given
by Eqs. (\ref{tildeJba}) and (\ref{tildeJde}), respectively.  We
note that for $\sigma'=\pm1$,
$N_{s,s_1}^{m,\sigma'}=2H_{s,s_1}^{m,\sigma'}$.

Now, we evaluate the matrix elements
\begin{eqnarray}
M_{\overline{s},s,\alpha}^{\overline{m},m}({\bm
q})&=&\langle\tilde{\varphi}_{\overline{s}}^{\overline{m}}\bigl|S^{\dag}_{\alpha}({\bf
q},0)\bigr|\tilde{\varphi}_s^m\rangle, \end{eqnarray}
 including
the leading corrections to the wave functions due to the
anisotropy interactions. We then find to first order in the
anisotropy interactions,
\begin{eqnarray}
M_{\overline{s},s,z'}^{\overline{m},m}({\bm q})&=&\cos({\bm
q}\cdot{\bm
d})\Bigl(m\delta_{\overline{s},s}\delta_{\overline{m},m}+m\langle
N_{\overline{s},\overline{m}}^{(1)}\bigl|\varphi_s^m\rangle\nonumber\\
&
&+\overline{m}\langle\varphi_{\overline{s}}^{\overline{m}}\bigl|N_{s,m}^{(1)}\rangle\Bigr)-i\sin({\bm
q}\cdot{\bm
d})\sum_{\sigma'=\pm1}\nonumber\\
& &\times\Bigl(D_{s+(\sigma'+1)/2,s_1}^m\delta_{\overline{s},s+\sigma'}\delta_{\overline{m},m}\nonumber\\
& &+D_{s+(\sigma'+1)/2,s_1}^m\langle
N_{\overline{s},\overline{m}}^{(1)}\bigl|\varphi_{s+\sigma'}^m\rangle\nonumber\\
&
&+D_{\overline{s}+(\sigma'+1)/2,s_1}^{\overline{m}}\langle\varphi_{\overline{s}+\sigma'}^{\overline{m}}\bigl|N_{s,m}^{(1)}\rangle
\Bigr),\nonumber\\
& &\\
 M_{\overline{s},s,x'}^{\overline{m},m}({\bm q})&=&
\frac{1}{2}\sum_{\sigma=\pm1}\biggl[\cos({\bm q}\cdot{\bm
d})\Bigl(A_s^{\sigma
m}\delta_{\overline{s},s}\delta_{\overline{m},m+\sigma}\nonumber\\
& &+A_s^{\sigma m}\langle
N_{\overline{s},\overline{m}}^{(1)}\bigl|\varphi_s^{m+\sigma}\rangle+A_{\overline{s}}^{-1+\sigma\overline{m}}\langle
\varphi_{\overline{s}}^{\overline{m}-\sigma}\bigl|N_{s,m}^{(1)}\rangle\Bigr)\nonumber\\
& &-i\sigma\sin({\bm q}\cdot{\bm
d})\sum_{\sigma'=\pm1}(-\sigma')\nonumber\\
& &\times\Bigl(C_{s+(\sigma'+1)/2,s_1}^{-\sigma\sigma'
m-(\sigma'+1)/2}\delta_{\overline{s},s+\sigma'}\delta_{\overline{m},m+\sigma}\nonumber\\
& &+C_{s+(\sigma'+1)/2,s_1}^{-\sigma\sigma'
m-(\sigma'+1)/2}\langle
N_{\overline{s},\overline{m}}^{(1)}\bigl|\varphi_{s+\sigma'}^{m+\sigma}\rangle\nonumber\\
& &
-C_{\overline{s}+(\sigma'+1)/2,s_1}^{\sigma\sigma'\overline{m}-(\sigma'+1)/2}\langle\varphi_{\overline{s}+\sigma'}^{\overline{m}-\sigma}\bigl|N_{s,m}^{(1)}\rangle\Bigr)\biggr],
\end{eqnarray}
and $M_{\overline{s},s,y'}^{\overline{m},m}({\bm q})$ is obtained
from  $M_{\overline{s},s,x'}^{\overline{m},m}({\bm q})$ by
multiplying the entire quantity inside the summation over $\sigma$
by $-i\sigma$.

Now, it is convenient to  break up the contributions from
$\bigl|M_{\overline{s},s,x'}^{\overline{m},m}({\bm q})\bigr|^2$
and $\bigl|M_{\overline{s},s,y'}^{\overline{m},m}({\bm
q})\bigr|^2$ into their symmetric and antisymmetric parts, with
\begin{eqnarray}
\bigl|M_{\overline{s},s,{\rm sym}}^{\overline{m},m}({\bm
q})\bigr|^2&=&\frac{1}{2}\Bigl(\bigl|M_{\overline{s},s,x'}^{\overline{m},m}({\bm
q})\bigr|^2+\bigl|M_{\overline{s},s,y'}^{\overline{m},m}({\bm
q})\bigr|^2\Bigr),\nonumber\\
\end{eqnarray}
etc. Then, the combined contributions from
$\bigl|M_{\overline{s},s,z'}^{\overline{m},m}({\bm q})\bigr|^2$
and $\bigl|M_{\overline{s},s,{\rm sym}}^{\overline{m},m}({\bm
q})\bigr|^2$
 arising from
the first order perturbations of the wave functions in addition to
the Hartree approximation are
\begin{eqnarray}
S_1^{(1{\rm e})}({\bm B},{\bm
q},\omega)&=&\sum_{s=0}^{2s_1}\sum_{m=-s}^se^{-\beta
E_{s,s_1}^m}\nonumber\\
& &\times\Bigl(\cos^2({\bm q}\cdot{\bm
d})\sin^2\theta_{b,q}{\cal F}_{1,s_1}^{m,s}(\omega,\theta,\phi)\nonumber\\
& &+\sin^2({\bm q}\cdot{\bm
d})\sin^2\theta_{b,q}{\cal F}_{2,s_1}^{m,s}(\omega,\theta,\phi)\nonumber\\
& &+\cos^2({\bm q}\cdot{\bm
d})\frac{2-\sin^2\theta_{b,q}}{4}{\cal F}_{3,s_1}^{m,s}(\omega,\theta,\phi)\nonumber\\
& &+\sin^2({\bm q}\cdot{\bm
d})\frac{2-\sin^2\theta_{b,q}}{4}{\cal F}_{4,s_1}^{m,s}(\omega,\theta,\phi)\Bigr),\nonumber\\
\end{eqnarray}
where
\begin{eqnarray}
{\cal F}_{n,s_1}^{m,s}(\omega,\theta,\phi)&=&\sum_{i=0}^2{\cal
F}_{n,s_1}^{m,s (i)}(\omega,\theta,\phi)
\end{eqnarray}
are the contributions to the ${\cal F}_n$ through second order in
the anisotropy interactions, respectively.  The terms to zeroth
 order in the anisotropy interactions are given by Eqs. (\ref{F10})-(\ref{F40}) in the text.  The
 non-vanishing first order terms are
\begin{eqnarray}
{\cal F}_{2,s_1}^{m,s (1)}&=&f_1^{(1)}(\theta,\phi)\sum_{\sigma'=\pm1}\delta\bigl(\omega+E^m_{s,s_1}-E_{s+\sigma',s_1}^m\bigr)\nonumber\\
& &\times D_{s+(\sigma'+1)/2,s_1}^mQ_{2}^{\sigma'},\\
{\cal F}_{4,s_1}^{m,s
(1)}&=&-f_1^{(1)}(\theta,\phi)\sum_{\sigma,\sigma'=\pm1}\delta\bigl(\omega+E_{s,s_1}^m-E_{s+\sigma',s_1}^{m+\sigma}\bigr)\nonumber\\
& &\qquad\times C_{s+(\sigma'+1)/2,s_1}^{-(\sigma'+1)/2-\sigma\sigma'm}Q_4^{\sigma,\sigma'},\\
Q_{2}^{\sigma'}&=&\sum_{\overline{\sigma}=\pm1}\frac{\overline{\sigma}D_{s+(\sigma'+1+2\overline{\sigma}\sigma')/2,s_1}^mH_{s+\sigma'(1-\overline{\sigma})/2,s_1}^{m,\overline{\sigma}\sigma'}}
{2[\sigma'(2s+1)+1+\overline{\sigma}]},\nonumber\\
& &\\
Q_4^{\sigma,\sigma'}&=&\sum_{\overline{\sigma}=\pm1}\frac{\overline{\sigma}C_{s+(\sigma'+1+2\overline{\sigma}\sigma')/2,s_1}^{\sigma\sigma'm+(\sigma'-1)/2}
H_{s+\sigma'(1-\overline{\sigma})/2,s_1}^{m+\sigma(1-\overline{\sigma})/2,\overline{\sigma}\sigma'}}{2[\sigma'(2s+1)+1+\overline{\sigma}]},\nonumber\\
& &\\
f_1^{(1)}(\theta,\phi)&=&\frac{1}{J}\Bigl(J_a-3[J_a\cos^2\theta+J_e\sin^2\theta\cos(2\phi)]\Bigr),\nonumber\\
\end{eqnarray}
where we  suppress the $\omega,\theta,\phi$ arguments of the
${\cal F}_{n,s_1}^{m,s (i)}$ functions for simplicity of
presentation. The   contributions ${\cal F}_{n,s_1}^{m,s
(2)}(\omega,\theta,\phi)$ second order in the anisotropy
interactions may be written as
\begin{eqnarray}
{\cal F}_{1,s_1}^{m,s
(2)}&=&\sum_{\sigma,\sigma'=\pm1}\sum_{p=0}^2\sum_{p'=0}^1\delta\bigl(\omega+E_{s,s_1}^m-E_{s+2p'\sigma',s_1}^{m+p\sigma}\bigr)\nonumber\\
& &\qquad\times {\cal A}_1^{p,p'}(\theta,\phi),\\
{\cal F}_{2,s_1}^{m,s
(2)}&=&\sum_{\sigma,\sigma'=\pm1}\sum_{p=0}^2\sum_{p'=0}^1\delta\bigl(\omega+E_{s,s_1}^m-E_{s+(2p'+1)\sigma',s_1}^{m+p\sigma}\bigr)\nonumber\\
& &\qquad\times {\cal A}_2^{p,p'}(\theta,\phi),\\
{\cal F}_{3,s_1}^{m,s
(2)}&=&\sum_{\sigma,\sigma'=\pm1}\sum_{p=0}^3\sum_{p'=0}^1\delta\bigl(\omega+E_{s,s_1}^m-E_{s+2p'\sigma',s_1}^{m+p\sigma}\bigr)\nonumber\\
& &\qquad\times {\cal A}_3^{p,p'}(\theta,\phi),\\
{\cal F}_{4,s_1}^{m,s
(2)}&=&\sum_{\sigma,\sigma'=\pm1}\sum_{p=0}^3\sum_{p'=0}^1\delta\bigl(\omega+E_{s,s_1}^m-E_{s+(2p'+1)\sigma',s_1}^{m+p\sigma}\bigr)\nonumber\\
& &\qquad\times {\cal A}_4^{p,p'}(\theta,\phi).
\end{eqnarray}
We note that we have suppressed the $\sigma,\sigma',m,s,s_1$
dependencies of the ${\cal A}_n^{p,p'}(\theta,\phi)$ for brevity.
These functions are given by
\begin{eqnarray}
{\cal A}_1^{0,0}&=&{\cal A}_1^{0,1}=0,\\
{\cal
A}_1^{1,0}&=&\bigl[f_1^{(2)}(\theta,\phi)+2\alpha_{s,s_1}f_2^{(2)}(\theta,\phi)\nonumber\\
&
&+\alpha^2_{s,s_1}f_3^{(2)}(\theta,\phi)\bigr]\bigl(G_{1,1}^{\sigma}\bigr)^2,\\
{\cal
A}_1^{2,0}&=&\bigl[f_4^{(2)}(\theta,\phi)+2\alpha_{s,s_1}f_5^{(2)}(\theta,\phi)\nonumber\\
&
&+\alpha^2_{s,s_1}f_6^{(2)}(\theta,\phi)\bigr]\bigl(G_{1,2}^{\sigma}\bigr)^2,\\
{\cal
A}_1^{1,1}&=&f_3^{(2)}(\theta,\phi)\bigl(G_{1,4}^{\sigma,\sigma'}\bigr)^2,\\
{\cal
A}_1^{1,2}&=&f_6^{(2)}(\theta,\phi)\bigl(G_{1,5}^{\sigma,\sigma'}\bigr)^2,
\end{eqnarray}
\begin{eqnarray}
{\cal
A}_2^{0,(1\pm1)/2}&=&f_7^{(2)(\theta,\phi)}\big(G_{2,3}^{\pm\sigma',\sigma'}\bigr)^2,\\
 {\cal
A}_2^{1,0}&=&\bigl(G_{2,1}^{\sigma,\sigma'}\bigr)^2\bigl[f_1^{(2)}(\theta,\phi)+2\alpha_{s,s_1}f_2^{(2)}(\theta,\phi)\nonumber\\
&
&+\alpha^2_{s,s_1}f_3^{(2)}(\theta,\phi)\bigr]+\bigl(G_{2,4}^{\sigma,\sigma',-\sigma'}\bigr)^2f_3^{(2)}(\theta,\phi)\nonumber\\
&
&+2G_{2,1}^{\sigma,\sigma'}G_{2,4}^{\sigma,\sigma',-\sigma'}\bigl[f_2^{(2)}(\theta,\phi)+\alpha_{s,s_1}f_3^{(2)}(\theta,\phi)\bigr],\nonumber\\
& &\\
{\cal
A}_2^{2,0}&=&\bigl(G_{2,2}^{\sigma,\sigma'}\bigr)^2\bigl[f_4^{(2)}(\theta,\phi)+2\alpha_{s,s_1}f_5^{(2)}(\theta,\phi)\nonumber\\
&
&+\alpha^2_{s,s_1}f_6^{(2)}(\theta,\phi)\bigr]+\bigl(G_{2,5}^{\sigma,\sigma',-\sigma'}\bigr)^2f_6^{(2)}(\theta,\phi)\nonumber\\
&
&+2G_{2,2}^{\sigma,\sigma'}G_{2,5}^{\sigma,\sigma',-\sigma'}\bigl[f_5^{(2)}(\theta,\phi)+\alpha_{s,s_1}f_6^{(2)}(\theta,\phi)\bigr],\nonumber\\
& &\\
{\cal
A}_2^{1,1}&=&f_3^{(2)}(\theta,\phi)\bigl(G_{2,4}^{\sigma,\sigma',\sigma'}\bigr)^2,\\
{\cal
A}_2^{2,1}&=&f_6^{(2)}(\theta,\phi)\bigl(G_{2,5}^{\sigma,\sigma',\sigma'}\bigr)^2,
\end{eqnarray}
\begin{eqnarray}
{\cal
A}_3^{1\pm1,0}&=&\bigl[f_1^{(2)}(\theta,\phi)+2\alpha_{s,s_1}f_2^{(2)}(\theta,\phi)\nonumber\\
&
&+\alpha^2_{s,s_1}f_3^{(2)}(\theta,\phi)\bigr]\bigl(G_{3,1}^{\sigma,\pm\sigma}\bigr)^2,\\
{\cal A}_3^{2\pm1,0}&=&\bigl[f_4^{(2)}(\theta,\phi)+2\alpha_{s,s_1}f_5^{(2)}(\theta,\phi)\nonumber\\
&
&+\alpha^2_{s,s_1}f_6^{(2)}(\theta,\phi)\bigr]\bigl(G_{3,2}^{\sigma,\pm\sigma}\bigr)^2,\\
{\cal
A}_3^{1\pm1,1}&=&f_3^{(2)}(\theta,\phi)\bigl(G_{3,4}^{\sigma,\sigma',\pm\sigma}\bigr)^2,\\
 {\cal
A}_3^{1,1}&=&\bigl(G_{3,3}^{\sigma,\sigma'}\bigr)^2f_7^{(2)}(\theta,\phi)+\bigl(G_{3,5}^{\sigma,\sigma',-\sigma}\bigr)^2f_6^{(2)}(\theta,\phi)\nonumber\\
&
&+2G_{3,3}^{\sigma,\sigma'}G_{3,5}^{\sigma,\sigma',-\sigma}f_9^{(2)}(\theta,\phi)\Bigr],\\
{\cal
A}_3^{3,1}&=&f_6^{(2)}(\theta,\phi)\bigl(G_{3,5}^{\sigma,\sigma',\sigma}\bigr)^2,
\end{eqnarray}
\begin{eqnarray}
{\cal
A}_4^{1\pm1,0}&=&\bigl(G_{4,1}^{\sigma,\sigma',\pm\sigma}\bigr)^2
\bigl[f_1^{(2)}(\theta,\phi)\nonumber\\
&
&+2\alpha_{s,s_1}f_2^{(2)}(\theta,\phi)+\alpha^2_{s,s_1}f_3^{(2)}(\theta,\phi)\bigr]\nonumber\\
& &+\bigl(G_{4,4}^{\sigma,\sigma',\pm\sigma,-\sigma'}\bigr)^2f_3^{(2)}(\theta,\phi)\nonumber\\
& &+2G_{4,1}^{\sigma,\sigma',\pm\sigma}G_{4,4}^{\sigma,\sigma',\pm\sigma,-\sigma'}\nonumber\\
& &\times\bigl[f_2^{(2)}(\theta,\phi)+\alpha_{s,s_1}f_3^{(2)}(\theta,\phi)\bigr],\\
{\cal A}_4^{1,0}&=&\bigl(G_{4,2}^{\sigma,\sigma',-\sigma}\bigr)^2\bigl[f_4^{(2)}(\theta,\phi)\nonumber\\
& &+2\alpha_{s,s_1}f_5^{(2)}(\theta,\phi)+\alpha^2_{s,s_1}f_6^{(2)}(\theta,\phi)\bigr]\nonumber\\
&
&+\bigl(G_{4,5}^{\sigma,\sigma',-\sigma,-\sigma'}\bigr)^2f_6^{(2)}(\theta,\phi)\nonumber\\
&
&+2G_{4,2}^{\sigma,\sigma',-\sigma}G_{4,5}^{\sigma,\sigma',-\sigma,-\sigma'}\nonumber\\
& &\times\bigl[f_5^{(2)}(\theta,\phi)+\alpha_{s,s_1}f_6^{(2)}(\theta,\phi)\bigr]\nonumber\\
&
&+\bigl(G_{4,3}^{\sigma,\sigma',-\sigma'}\bigr)^2f_7^{(2)}(\theta,\phi)+2G_{4,3}^{\sigma,\sigma',-\sigma'}\nonumber\\
&
&\times\Bigl(G_{4,2}^{\sigma,\sigma',-\sigma}\bigl[f_8^{(2)}(\theta,\phi)+\alpha_{s,s_1}f_{9}^{(2)}(\theta,\phi)\bigr]\nonumber\\
&
&+G_{4,5}^{\sigma,\sigma',-\sigma,-\sigma'}f_{9}^{(2)}(\theta,\phi)\Bigr),\nonumber\\
& &\\
{\cal A}_4^{3,0}&=&\bigl(G_{4,2}^{\sigma,\sigma',\sigma}\bigr)^2\bigl[f_4^{(2)}(\theta,\phi)\nonumber\\
& &+2\alpha_{s,s_1}f_5^{(2)}(\theta,\phi)+\alpha^2_{s,s_1}f_6^{(2)}(\theta,\phi)\bigr]\nonumber\\
&
&+\bigl(G_{4,5}^{\sigma,\sigma',\sigma,-\sigma'}\bigr)^2f_6^{(2)}(\theta,\phi)+2G_{4,2}^{\sigma,\sigma',\sigma}G_{4,5}^{\sigma,\sigma',\sigma,-\sigma'}\nonumber\\
&
&\times\bigl[f_5^{(2)}(\theta,\phi)+\alpha_{s,s_1}f_6^{(2)}(\theta,\phi)\bigr],\\
{\cal
A}_4^{1\pm1,1}&=&f_3^{(2)}(\theta,\phi)\bigl(G_{4,4}^{\sigma,\sigma',\pm\sigma,\sigma'}\bigr)^2,\\
{\cal A}_4^{1,1}&=&\bigl(G_{4,5}^{\sigma,\sigma',-\sigma,\sigma'}\bigr)^2f_6^{(2)}(\theta,\phi)\nonumber\\
& &+\bigl(G_{4,3}^{\sigma,\sigma',\sigma'}\bigr)^2f_7^{(2)}(\theta,\phi)\nonumber\\
&
&+2G_{4,3}^{\sigma,\sigma',\sigma'}G_{4,5}^{\sigma,\sigma',-\sigma,\sigma'}f_9^{(2)}(\theta,\phi),\\
{\cal
A}_4^{3,1}&=&f_6^{(2)}(\theta,\phi)\bigl(G_{4,5}^{\sigma,\sigma',\sigma,\sigma'}\bigr)^2,
\end{eqnarray}
where
  the second order angular functions are given by
\begin{eqnarray}
f_1^{(2)}(\theta,\phi)&=&\frac{\sin^2\theta}{J^2}\Bigl([J_b-\cos(2\phi)J_d]^2\cos^2\theta\nonumber\\
& &\qquad\qquad+J_d^2\sin^2(2\phi)\Bigr),\label{f1}\\
f_2^{(2)}(\theta,\phi)&=&\frac{\sin^2\theta}{J^2}\Bigl([J_b-J_d\cos(2\phi)][J_a-J_e\cos(2\phi)]\nonumber\\
& &\qquad\times\cos^2\theta  +J_dJ_e\sin^2(2\phi)]\Bigr),\\
f_3^{(2)}(\theta,\phi)&=&\frac{\sin^2\theta}{J^2}\Bigl([J_a-\cos(2\phi)J_e]^2\cos^2\theta\nonumber\\
& &\qquad\qquad+J_e^2\sin^2(2\phi)\Bigr),\\
f_4^{(2)}(\theta,\phi)&=&\frac{1}{J^2}\Bigl(J_b\sin^2\theta+J_d(1+\cos^2\theta)\cos(2\phi)\Bigr)^2\nonumber\\
& &+4\frac{J_d^2}{J^2}\cos^2\theta\sin^2(2\phi),\\
f_5^{(2)}(\theta,\phi)&=&\frac{1}{J^2}[J_b\sin^2\theta+J_d(1+\cos^2\theta)\cos(2\phi)]\nonumber\\
&
&\times[J_a\sin^2\theta+J_e(1+\cos^2\theta)\cos(2\phi)]\nonumber\\
& &+4\frac{J_dJ_e}{J^2}\cos^2\theta\sin^2(2\phi),\\
f_6^{(2)}(\theta,\phi)&=&\frac{1}{J^2}\Bigl(J_a\sin^2\theta+J_e(1+\cos^2\theta)\cos(2\phi)\Bigr)^2\nonumber\\
& &+4\frac{J_e^2}{J^2}\cos^2\theta\sin^2(2\phi),\label{f6}\\
f_7^{(2)}(\theta,\phi)&=&\frac{1}{J^2}\Bigl(J_a-3[J_a\cos^2\theta+J_e\sin^2\theta\cos(2\phi)]\Bigr)^2,\label{f7}\nonumber\\
f_8^{(2)}(\theta,\phi)&=&\frac{1}{J^2}\Bigl(J_a-3[J_a\cos^2\theta+J_e\sin^2\theta\cos(2\phi)]\Bigr)\nonumber\\
& &\times
\Bigl(J_b\sin^2\theta+J_d(1+\cos^2\theta)\cos(2\phi)\Bigr),\\
f_9^{(2)}(\theta,\phi)&=&\frac{1}{J^2}\Bigl(J_a-3[J_a\cos^2\theta+J_e\sin^2\theta\cos(2\phi)]\Bigr)\nonumber\\
& &\times
\Bigl(J_a\sin^2\theta+J_e(1+\cos^2\theta)\cos(2\phi)\Bigr),\label{f9}
\end{eqnarray}
and the constants are given by
\begin{eqnarray}
G_{1,1}^{\sigma}&=&\frac{J(2m+\sigma)}{2\gamma
B}A_s^{\sigma m},\nonumber\\
& &\\ G_{1,2}^{\sigma}&=&\frac{JF_s^{\sigma m}}{4\gamma
B},\\
G_{1,4}^{\sigma,\sigma'}&=&\frac{
R_{s,s_1}^{m,\sigma,\sigma'}}{4\Bigl(\sigma\gamma
B+J[\sigma'(2s+1)+2]\Bigr)},\nonumber\\
& &\\
G_{1,5}^{\sigma,\sigma'}&=&\frac{K_{s,s_1}^{\sigma
m,\sigma'}}{4\Bigl(2\sigma\gamma
B+J[\sigma'(2s+1)+2]\Bigr)},\nonumber\\
& &\\
 G_{2,1}^{\sigma,\sigma'}&=&\frac{J\sigma(2m+\sigma)}{2\gamma
B}\sum_{\overline{\sigma}=\pm1}\Bigl(\overline{\sigma}A_{s-\sigma'(\overline{\sigma}-1)/2}^{\sigma
m}\nonumber\\
& &\qquad\times D_{s+(1+\sigma')/2,s_1}^{m+\sigma(\overline{\sigma}+1)/2}\Bigr),\\
G_{2,2}^{\sigma,\sigma'}&=&-\frac{J\sigma}{8\gamma
B}\sum_{\overline{\sigma}=\pm1}\overline{\sigma}
D_{s+(1+\sigma')/2,s_1}^{m+\sigma(\overline{\sigma}+1)}\nonumber\\
& &\qquad\times F_{s-\sigma'(\overline{\sigma}-1)/2,s_1}^{\sigma
m},\\
G_{2,3}^{\sigma,\sigma'}&=&\sum_{\overline{\sigma}=\pm1}\biggl(\frac{\overline{\sigma}D_{s+\sigma'(\overline{\sigma}+1)+(\sigma+1)/2,s_1}^m}
{4[\sigma'(2s+1)+2+\sigma\sigma'(1-\overline{\sigma})]}\nonumber\\
& &\qquad\times H_{s+\sigma(1-\overline{\sigma})/2,s_1}^{m,\sigma'}\biggr),\\
G_{2,4}^{\sigma,\sigma',\sigma''}&=&-\frac{1}{4}\sum_{\overline{\sigma}=\pm1}\biggl(\sigma\sigma'\overline{\sigma}
D_{s+\sigma'(\overline{\sigma}+1)+(\sigma''+1)/2,s_1}^{m+\sigma(\overline{\sigma}+1)/2}\nonumber\\
 & &\times\frac{R_{s+\sigma''(1-\overline{\sigma})/2,s_1}^{
m,\sigma,\sigma'}}{\sigma\gamma
B/J+\sigma'(2s+1+\sigma''-\sigma''\overline{\sigma})+2}\biggr),\nonumber\\
& &\\
G_{2,5}^{\sigma,\sigma',\sigma''}&=&-\frac{1}{8}\sum_{\overline{\sigma}=\pm1}\biggl(\overline{\sigma}
K_{s+\sigma''(1-\overline{\sigma})/2,s_1}^{\sigma m,\sigma'}
\nonumber\\
& &\times\frac{D_{s+(\overline{\sigma}+1)\sigma'+(\sigma''+1)/2,s_1}^{m+\sigma(\overline{\sigma}+1)}}{2\sigma\gamma B/J+\sigma'(2s+1+\sigma''-\sigma''\overline{\sigma})+2}\biggr),\nonumber\\
& &\\
 G_{3,1}^{\sigma,\sigma'}&=&\frac{\sigma'J}{2\gamma
B}\Bigl((2m+\sigma')A_s^{\sigma(m+\sigma')}A_s^{\sigma'm}\nonumber\\
& &-(2m+2\sigma+\sigma')A_s^{\sigma'(m+\sigma)}A_s^{\sigma m}\Bigr),\\
G_{3,2}^{\sigma,\sigma'}&=&-\frac{\sigma J}{8\gamma
B}\Bigl(A_s^{\sigma'(m+2\sigma)}F_s^{\sigma m}\nonumber\\
& &-A_s^{\sigma'm}F_s^{\sigma(m+\sigma')}\Bigr),\\
G_{3,3}^{\sigma,\sigma'}&=&\sum_{\overline{\sigma}=\pm1}\frac{\overline{\sigma}A_{s+(\overline{\sigma}+1)\sigma'}^{\sigma
m}H_{s,s_1}^{m-\sigma(\overline{\sigma}-1)/2,\sigma'}}{4[\sigma'(2s+1)+2]},\\
G_{3,4}^{\sigma,\sigma',\sigma''}&=&-\sum_{\overline{\sigma}=\pm1}\biggl(\frac{J\sigma\sigma'\overline{\sigma}A_{s+(\overline{\sigma}+1)\sigma'}^{\sigma''m
+\sigma\sigma''(\overline{\sigma}+1)/2}}{4\Bigl(\sigma\gamma
B+J[\sigma'(2s+1)+2]\Bigr)}\nonumber\\
& &\times
R_{s,s_1}^{m-\sigma''(\overline{\sigma}-1)/2,\sigma,\sigma'}\biggr),\\
G_{3,5}^{\sigma,\sigma',\sigma''}&=&-\sum_{\overline{\sigma}=\pm1}\biggl(\frac{J\overline{\sigma}A_{s+(\overline{\sigma}+1)\sigma'}^{\sigma''m+
(1+\overline{\sigma})\sigma\sigma''}}{8\Bigl(2\sigma\gamma
B+J[\sigma'(2s+1)+2]\Bigr)}\nonumber\\
& &\times K_{s,s_1}^{\sigma
m+(1-\overline{\sigma})\sigma\sigma''/2,\sigma'}\biggr),
\end{eqnarray}
\begin{eqnarray}
G_{4,1}^{\sigma,\sigma',\sigma''}&=&\frac{\sigma'\sigma''J}{2\gamma
B}\sum_{\overline{\sigma}=\pm1}\overline{\sigma}[2m+\sigma''+\sigma(1-\overline{\sigma})]\nonumber\\
& &\times
C_{s+(\sigma'+1)/2,s_1}^{-\sigma\sigma'm-\sigma\sigma'\sigma''(\overline{\sigma}+1)/2-(\sigma'+1)/2}\nonumber\\
& &\times
A_{s+\sigma'(1-\overline{\sigma})/2}^{\sigma''m+\sigma\sigma''(1-\overline{\sigma})/2},\\
G_{4,2}^{\sigma,\sigma',\sigma''}&=&-\frac{\sigma'\sigma''J}{8\gamma
B}\sum_{\overline{\sigma}=\pm1}\overline{\sigma}F_{s+\sigma'(1-\overline{\sigma})/2}^{\sigma m+\sigma\sigma''(1-\overline{\sigma})/2}\nonumber\\
& &\times
C_{s+(\sigma'+1)/2,s_1}^{-\sigma'\sigma''m-\sigma\sigma'\sigma''(\overline{\sigma}+1)-(\sigma'+1)/2},\nonumber\\
& &\\
G_{4,3}^{\sigma,\sigma',\sigma''}&=&\sum_{\overline{\sigma}=\pm1}\biggl(\frac{\sigma''\overline{\sigma}
H_{s+\sigma''(1-\overline{\sigma})/2,s_1}^{m+\sigma(1-\overline{\sigma})/2,\sigma'}}{4[\sigma'(2s+1)+2+\sigma'\sigma''(1-\overline{\sigma})]}\nonumber\\
& &\times
C_{s+(\sigma''+1)/2+\sigma'(\overline{\sigma}+1),s_1}^{-\sigma\sigma''m-(\sigma''+1)/2}\biggr),\\
G_{4,4}^{\sigma,\sigma',\sigma'',\sigma'''}&=&-\frac{1}{4}\sum_{\overline{\sigma}=\pm1}\biggl(\sigma\sigma'\sigma'''\overline{\sigma}
R_{s+\sigma'''(1-\overline{\sigma})/2,s_1}^{m+\sigma''(1-\overline{\sigma})/2,\sigma,\sigma'}\nonumber\\
& &\times
\frac{C_{s+(\sigma'''+1)/2+\sigma'(\overline{\sigma}+1),s_1}^{-\sigma''\sigma'''m-\sigma\sigma''\sigma'''(\overline{\sigma}+1)/2-(\sigma'''+1)/2}}{\sigma\gamma
B/J+\sigma'(2s+1+\sigma'''-\sigma'''\overline{\sigma})+2}\biggr),\nonumber\\
& &\\
G_{4,5}^{\sigma,\sigma',\sigma'',\sigma'''}&=&-\frac{1}{8}\sum_{\overline{\sigma}=\pm1}\biggl(\sigma'''\overline{\sigma}J
K_{s+\sigma'''(1-\overline{\sigma})/2,s_1}^{\sigma
m+\sigma\sigma''(1-\overline{\sigma})/2,\sigma'}\nonumber\\
& &\times
\frac{C_{s+(\sigma'''+1)/2+\sigma'(\overline{\sigma}+1),s_1}^{-\sigma''\sigma'''m-(\sigma'''+1)/2-\sigma\sigma''\sigma'''(\overline{\sigma}+1)}}{2\sigma\gamma
B/J+\sigma'(2s+1+\sigma'''-\sigma'''\overline{\sigma})+2}\biggr),\label{G45}\nonumber\\
\end{eqnarray}
where we have used the identities  $A_s^{-m-1}=A_s^m$,
$F_s^{-m-2}=F_s^m$,
$H_{s+2\sigma',s_1}^{m,-\sigma'}=H_{s,s_1}^{m,\sigma'}$,
$R_{s+2\sigma',s_1}^{m+\sigma,-\sigma,-\sigma'}=R_{s,s_1}^{m,\sigma,\sigma'}$,
and $K_{s+2\sigma',s_1}^{-2-\sigma m,-\sigma'}=K_{s,s_1}^{\sigma
m,\sigma'}$.

For completeness, we write the contributions to $S({\bm B},{\bm
q},\omega)$ from the off-diagonal and antisymmetric diagonal
matrix element terms that are first order in the anisotropy
interactions.  These could be measured by collecting data at
specific $(\theta_{b,q},\phi_{b,q})$ values.  We have
\begin{eqnarray}
S_2^{(1{\rm e})}&=&\sum_{s=0}^{2s_1}\sum_{m=-s}^se^{-\beta
E_{s,s_1}^m}\sin^2\theta_{b,q}\cos(2\phi_{b,q})\nonumber\\
& &\times\Biggl\{\cos^2({\bm q}\cdot{\bm
d})\sum_{\sigma=\pm1}\delta\bigl(\omega+E_{s,s_1}^m-E_{s,s_1}^{m+\sigma}\bigr)\nonumber\\
& &\times
G_{5,1}^{\sigma}\Bigl(f_2^{(1)}(\theta,\phi)+\alpha_{s,s_1}f_3^{(1)}(\theta,\phi)\Bigr)\nonumber\\
& &+\sin^2({\bm q}\cdot{\bm d})\sum_{\sigma,\sigma'=\pm1}\delta\bigl(\omega+E_{s,s_1}^m-E_{s+\sigma',s_1}^{m+\sigma}\bigr)\nonumber\\
& &\times\biggl[G_{5,2}^{\sigma,\sigma'}\Bigl(f_2^{(1)}(\theta,\phi)+\alpha_{s,s_1}f_3^{(1)}(\theta,\phi)\Bigr)\nonumber\\
& &+G_{5,3}^{\sigma,\sigma'}f_3^{(1)}(\theta,\phi)\biggr]\Biggr\},
\end{eqnarray}
\begin{eqnarray}
S_3^{(1{\rm e})}&=&\sum_{s=0}^{2s_1}\sum_{m=-s}^se^{-\beta
E_{s,s_1}^m}\sin(2\theta_{b,q})\Biggl\{\cos^2({\bm q}\cdot{\bm
d})\nonumber\\
& &\times
\Bigl(h_1(\theta,\phi,\phi_{b,q})+\alpha_{s,s_1}h_2(\theta,\phi,\phi_{b,q})\Bigr)\nonumber\\
& &\times\Bigl[\delta(\omega)G_{6,1}\nonumber\\
&
&+\sum_{\sigma=\pm1}\delta\bigl(\omega+E_{s,s_1}^m-E_{s,s_1}^{m+\sigma}\bigr)G_{6,2}^{\sigma}\Bigr]\nonumber\\
& &+\sin^2({\bm q}\cdot{\bm
d})\biggl[\sum_{\sigma'=\pm1}\delta\bigl(\omega+E_{s,s_1}^m-E_{s+\sigma',s_1}^{m}\bigr)\nonumber\\
& &\times\Bigl\{\Bigl(h_1^{(1)}(\theta,\phi,\phi_{b,q})+\alpha_{s,s_1}h_2^{(1)}(\theta,\phi,\phi_{b,q})\Bigr)\nonumber\\
& &\times G_{6,3}^{\sigma'}+h_2^{(1)}(\theta,\phi,\phi_{b,q})G_{6,4}^{\sigma'}\Bigr\}\nonumber\\
& &-\sum_{\sigma,\sigma'=\pm1}\delta\bigl(\omega+E_{s,s_1}^m-E_{s+\sigma',s_1}^{m+\sigma}\bigr)\nonumber\\
&
&\times\Bigl\{\Bigl(h_1^{(1)}(\theta,\phi,\phi_{b,q})+\alpha_{s,s_1}h_2^{(1)}(\theta,\phi,\phi_{b,q})\Bigr)\nonumber\\
& &\times G_{6,5}^{\sigma,\sigma'}
+h_2^{(1)}(\theta,\phi,\phi_{b,q})G_{6,6}^{\sigma,\sigma'}\Bigr\}\biggr]\Biggr\},\nonumber\\
\end{eqnarray}
and
\begin{eqnarray}
S_4^{(1{\rm e})}&=&\sum_{s=0}^{2s_1}\sum_{m=-s}^se^{-\beta
E_{s,s_1}^m}\sin^2\theta_{b,q}\sin(2\phi_{b,q})\nonumber\\
& &\times\Biggl\{\cos^2({\bm q}\cdot{\bm
d})\sum_{\sigma=\pm1}\delta\bigl(\omega+E_{s,s_1}^m-E_{s,s_1}^{m+\sigma}\bigr)\nonumber\\
& &\times G_{7,1}^{\sigma}\Bigl(f_4^{(1)}(\theta,\phi)+\alpha_{s,s_1}f_5^{(1)}(\theta,\phi)\Bigr)\nonumber\\
& &+\sin^2({\bm q}\cdot{\bm
d})\sum_{\sigma,\sigma'=\pm1}\delta\bigl(\omega+E_{s,s_1}^m-E_{s+\sigma',s_1}^{m+\sigma}\bigr)\nonumber\\
&
&\times\Bigl[G_{7,2}^{\sigma,\sigma'}\Bigl(f_4^{(1)}(\theta,\phi)+\alpha_{s,s_1}f_5^{(1)}(\theta,\phi)\Bigr)\nonumber\\
& &+G_{7,3}^{\sigma,\sigma'}f_5^{(1)}(\theta,\phi)\Bigr]\Biggr\}.
\end{eqnarray}

  The angular functions and coefficients are
given by
\begin{eqnarray}
f_2^{(1)}(\theta,\phi)&=&\frac{1}{J}\Bigl(J_b\sin^2\theta\nonumber\\
& &\qquad+J_d(1+\cos^2\theta)\cos(2\phi)\Bigr),\\
f_3^{(1)}(\theta,\phi)&=&\frac{1}{J}\Bigl(J_a\sin^2\theta\nonumber\\
& &\qquad+J_e(1+\cos^2\theta)\cos(2\phi)\Bigr),\\
f_4^{(1)}(\theta,\phi)&=&\frac{J_d}{J}\cos\theta\sin(2\phi),\\
f_5^{(1)}(\theta,\phi)&=&\frac{J_e}{J}\cos\theta\sin(2\phi),\\
h_1^{(1)}(\theta,\phi,\phi_{b,q})&=&\frac{1}{J}\Bigl(\cos\phi_{b,q}\sin(2\theta)[J_b-J_d\cos(2\phi)]\nonumber\\
& &+2\sin\phi_{b,q}J_d\sin\theta\sin(2\phi)\Bigr),\\
h_2^{(1)}(\theta,\phi,\phi_{b,q})&=&\frac{1}{J}\Bigl(\cos\phi_{b,q}\sin(2\theta)[J_a-J_e\cos(2\phi)]\nonumber\\
& &+2\sin\phi_{b,q}J_e\sin\theta\sin(2\phi)\Bigr),
\end{eqnarray}
\begin{eqnarray}
 G_{5,1}^{\sigma}&=&-\frac{J(2m+\sigma) \bigl(A_s^{\sigma
m}\bigr)^2}{8\gamma
B},\nonumber\\
& &\\
G_{5,2}^{\sigma,\sigma'}&=&-\frac{\sigma J}{16\gamma B}C_{s+(\sigma'+1)/2,s_1}^{-\sigma\sigma'm-(1+\sigma')/2}\nonumber\\
&
&\times\sum_{\overline{\sigma}=\pm1}\overline{\sigma}C_{s+(\sigma'+1)/2,s_1}^{\sigma\sigma'm+(\sigma'+2\overline{\sigma}\sigma'-1)/2}\nonumber\\
& &\qquad\times
F_{s+\sigma'(1-\overline{\sigma})/2}^{\overline{\sigma}\sigma
m-(1-\overline{\sigma})/2},\\
G_{5,3}^{\sigma,\sigma'}&=&\frac{1}{16}C_{s+(\sigma'+1)/2,s_1}^{-\sigma\sigma'm-(1+\sigma')/2}\nonumber\\
&
&\times\sum_{\overline{\sigma}=\pm1}\biggl(\frac{\overline{\sigma}C_{s+(1+\sigma'+2\overline{\sigma}\sigma')/2,s_1}^{-\sigma\sigma'm-(1+\sigma'+2\overline{\sigma}\sigma')/2}}{2\sigma\gamma
B/J+\sigma'(2s+1)+1+\overline{\sigma}}\nonumber\\
& &\qquad\times
K_{s-\sigma'(\overline{\sigma}-1)/2,s_1}^{\overline{\sigma}\sigma
m+(\overline{\sigma}-1)/2,\overline{\sigma}\sigma'}\biggr),\\
G_{6,1}&=&\frac{Jm}{2\gamma B}[s(s+1)-3m^2],\\
G_{6,2}^{\sigma}&=&\frac{J}{8\gamma B}\bigl(A_s^{\sigma
m}\bigr)^2(2m+\sigma),\\
G_{6,3}^{\sigma'}&=&-\frac{\sigma'J}{8\gamma
B}D_{s+(\sigma'+1)/2,s_1}^m\nonumber\\
&
&\times\sum_{\sigma,\overline{\sigma}=\pm1}\Bigl[(2m\overline{\sigma}+\sigma)
A_{s+\sigma'(\overline{\sigma}+1)/2}^{\sigma\overline{\sigma}m}\nonumber\\
& &\qquad\times C_{s+(\sigma'+1)/2,s_1}^{-\sigma\sigma'm-(1+\overline{\sigma}\sigma')/2}\Bigr],\\
G_{6,4}^{\sigma'}&=&\frac{J}{16}D_{s+(\sigma'+1)/2,s_1}^m\sum_{\sigma,\overline{\sigma}=\pm1}\nonumber\\
&
&\times\biggl(\frac{\overline{\sigma}C_{s+(1+\sigma'-2\overline{\sigma}\sigma')/2,s_1}^{\sigma\sigma'm-(1-\overline{\sigma}\sigma')/2}}{\sigma\gamma
B+J[-\sigma'(2s+1)+\overline{\sigma}-1]}\nonumber\\
& &\times R_{s+\sigma'(\overline{\sigma}+1)/2,s_1}^{m,\overline{\sigma}\sigma,-\overline{\sigma}\sigma'}\biggr),\\
G_{6,5}^{\sigma,\sigma'}&=&-\frac{\sigma'J}{8\gamma
B}C_{s+(\sigma'+1)/2,s_1}^{-\sigma\sigma'm-(\sigma'+1)/2}(2m+\sigma)\nonumber\\
&
&\times\sum_{\overline{\sigma}=\pm1}\overline{\sigma}D_{s+(\sigma'+1)/2,s_1}^{m+\sigma(\overline{\sigma}+1)/2}
A_{s-\sigma'(\overline{\sigma}-1)/2}^{\sigma
m},\nonumber\\
& &\\
G_{6,6}^{\sigma,\sigma'}&=&\frac{J}{16}C_{s+(\sigma'+1)/2,s_1}^{-\sigma\sigma'm-(\sigma'+1)/2}\sum_{\overline{\sigma}=\pm1}\nonumber\\
&
&\times\biggl(\frac{\overline{\sigma}R_{s-\sigma'(\overline{\sigma}-1)/2,s_1}^{m+\sigma(1-\overline{\sigma})/2,\overline{\sigma}\sigma,\overline{\sigma}\sigma'}}{\sigma\gamma
B+J[\sigma'(2s+1)+1+\overline{\sigma}]}\nonumber\\
& &\qquad\times
D_{s+(1+\sigma'+2\overline{\sigma}\sigma')/2,s_1}^{m+\sigma(\overline{\sigma}+1)/2}\biggr),\\
G_{7,1}^{\sigma}&=&\frac{J}{4\gamma B}(2m+\sigma)\bigl(A_s^{\sigma
m}\bigr)^2,\nonumber\\
& &\\
 G_{7,2}^{\sigma,\sigma'}&=&-\frac{\sigma J}{8\gamma
B}C_{s+(\sigma'+1)/2,s_1}^{-\sigma\sigma'
m-(\sigma'+1)/2}\nonumber\\
&
&\times\sum_{\overline{\sigma}=\pm1}\biggl(\overline{\sigma}C_{s+(\sigma'+1)/2,s_1}^{\sigma\sigma'
m-(1-\sigma'-2\overline{\sigma}\sigma')/2}\nonumber\\
& &\times F_{s+\sigma'(1-\overline{\sigma})/2}^{\overline{\sigma}\sigma m-(1-\overline{\sigma})/2}\biggr),\\
G_{7,3}^{\sigma,\sigma'}&=&\frac{1}{8}C_{s+(\sigma'+1)/2,s_1}^{-\sigma\sigma'
m-(\sigma'+1)/2}\nonumber\\
&
&\times\sum_{\overline{\sigma}=\pm1}\biggl(\overline{\sigma}K_{s+\sigma'(1-\overline{\sigma})/2,s_1}^{\overline{\sigma}\sigma
m-(1-\overline{\sigma})/2,\overline{\sigma}\sigma'}\nonumber\\
&
&\times\frac{C_{s+(1+\sigma'+2\overline{\sigma}\sigma')/2,s_1}^{-\sigma\sigma'
m-(1+\sigma'+2\overline{\sigma}\sigma')/2}}{2\sigma\gamma
B/J+\sigma'(2s+1)+1+\overline{\sigma}}\biggr).
\end{eqnarray}

\subsection{Higher order corrections to the EPR response}

Here we evaluate the EPR response $\chi_{-\sigma,\sigma}^{''}({\bm
B},\omega)$ including the corrections to the induction
representation wave functions first order in the anisotropy
interactions.  It can be shown that second order perturbations to
the wave functions only contribute to the EPR response to third
(and higher) order in the anisotropy interactions, and are hence
neglected. From Eq. (\ref{chiexact}), we expand the exact wave
functions $|\tilde{\phi}_n\rangle$ to first order in the
anisotropy interactions. To this order, $s$ and $m$ are still good
quantum numbers, so we rewrite
$|\tilde{\phi}_n\rangle=|\tilde{\varphi}_s^m\rangle$.  We then
find
\begin{eqnarray}
\chi^{''}_{-\sigma,\sigma}({\bm
B},\omega)&\approx&\frac{\gamma^2}{Z^{(1)}}\sum_{s=0}^{2s_1}\sum_{m=-s}^s\sum_{s'=0}^{2s_1}\sum_{m'=-s'}^{s'}\exp[-\beta
E_{s,s_1}^m]\nonumber\\
&
&\times\Bigl(\delta(E_{s,s_1}^m-E_{s',s_1}^{m'}+\omega)\nonumber\\
&
&\qquad-\delta(E_{s',s_1}^{m'}-E_{s,s_1}^m+\omega)\Bigr)\nonumber\\
& &\times\biggl[\sum_{\overline{\sigma}=\pm1}\delta_{s',s}\delta_{m',m+\overline{\sigma}\sigma}\bigl|Z_{0}^{\overline{\sigma}\sigma}\bigr|^2\nonumber\\
& &+\delta_{s',s}\sum_{\sigma'=\pm\sigma}\delta_{m',m+\sigma
+\sigma'}\bigl|Z_{0}^{\sigma+\sigma'}\bigr|^2\nonumber\\
 &
&+\delta_{m',m+\sigma}\sum_{\sigma''=\pm\sigma}\delta_{s',s+2\sigma''}\bigl|Z_{2\sigma''}^{\sigma}\bigr|^2\nonumber\\
&
&+\sum_{\sigma',\sigma''=\pm\sigma}\delta_{s',s+2\sigma''}\Bigl(\delta_{m',m+\sigma+\sigma'}
\bigl|Z_{2\sigma''}^{\sigma+\sigma'}\bigr|^2\nonumber\\
&
&\qquad+\delta_{m',m+\sigma+2\sigma'}\bigl|Z_{2\sigma''}^{\sigma+2\sigma'}\bigr|^2\Bigr)\biggr],
\end{eqnarray}
where the amplitudes $Z_{s'-s}^{m'-m}$ are given to first order in
the anisotropy energies by
\begin{eqnarray}
Z_0^{+\sigma}&=&A_s^{\sigma m}+{\cal O}(J_i/J)^2,\\
Z_0^{\sigma+\sigma'}&=&\frac{\sigma'}{\gamma
B}\Bigl(A_s^{\sigma(m+\sigma')}{\cal
U}_{s,s_1}^{m,\sigma'}-A_s^{\sigma m}\bigl({\cal
U}_{s,s_1}^{m+\sigma+\sigma',-\sigma'}\bigr)^{*}\Bigr),\nonumber\\
& &\\
Z_0^{-\sigma}&=&\frac{-\sigma}{2\gamma B}\Bigl(A_s^{-2+\sigma
m}{\cal V}_{s,s_1}^{m,-\sigma}-A_s^{\sigma m}\bigl({\cal
V}_{s,s_1}^{m-\sigma,\sigma}\bigr)^{*}\Bigr),\nonumber\\
& &\\
 Z_{2\sigma''}^{\sigma}&=&\frac{A_{s+2\sigma''}^{\sigma
m}{\cal
 W}_{s,s_1}^{m,\sigma''}-A_s^{\sigma m}\bigl({\cal
 W}_{s+2\sigma'',s_1}^{m+\sigma,-\sigma''}\bigr)^{*}}{J[\sigma''(2s+1)+2]},\nonumber\\
 & &\\
 Z_{2\sigma''}^{\sigma+\sigma'}&=&\frac{1}{\sigma'\gamma
 B+J[\sigma''(2s+1)+2]}\nonumber\\
 & &\times\Bigl(A_{s+2\sigma''}^{\sigma(m+\sigma')}{\cal
 X}_{s,s_1}^{m,\sigma',\sigma''}\nonumber\\
 & &\qquad-A_s^{\sigma m}\bigl({\cal
 X}_{s+2\sigma'',s_1}^{m+\sigma+\sigma',-\sigma',-\sigma''}\bigr)^{*}\Bigr),\nonumber\\
 & &\\
 Z_{2\sigma''}^{\sigma+2\sigma'}&=&\frac{1}{2\sigma'\gamma
 B+J[\sigma''(2s+1)+2]}\nonumber\\
 & &\times\Bigl(A_{s+2\sigma''}^{\sigma(m+2\sigma')}{\cal
 Y}_{s,s_1}^{m,\sigma',\sigma''}\nonumber\\
 & &\qquad-A_s^{\sigma m}\bigl({\cal
 Y}_{s+2\sigma'',s_1}^{m+\sigma+2\sigma',-\sigma',-\sigma''}\bigr)^{*}\Bigr),\nonumber\\
\end{eqnarray}
leading to
\begin{eqnarray}
\bigl|Z_{0}^{\sigma+\sigma'}\bigr|^2&=&\frac{J^2}{(2\gamma
B)^2}\Bigl(A_s^{\sigma(m+\sigma')}(2m+\sigma')A_s^{\sigma'm}\nonumber\\
& &-A_s^{\sigma
m}(2m+2\sigma+\sigma')A_s^{\sigma'(m+\sigma)}\Bigr)^2\nonumber\\
& &\times\Bigl(f_1^{(2)}(\theta,\phi)+2\alpha_{s,s_1}f_2^{(2)}(\theta,\phi)\nonumber\\
& &\qquad\qquad+\alpha_{s,s_1}^2f_3^{(2)}(\theta,\phi)\Bigr),\nonumber\\
& &\\
\bigl|Z_{0}^{-\sigma}\bigr|^2&=&\frac{J^2(2m-\sigma)^2\bigl(A_s^{\sigma
m-1}\bigr)^2}{(4\gamma B)^2}\Bigl(f_4^{(2)}(\theta,\phi)\nonumber\\
& &\qquad+2\alpha_{s,s_1}f_5^{(2)}(\theta,\phi)+\alpha^2_{s,s_1}f_6^{(2)}(\theta,\phi)\Bigr),\nonumber\\
& &\\
\bigl|Z_{2\sigma''}^{\sigma}\bigr|^2&=&\frac{f_7^{(2)}(\theta,\phi)}{16[\sigma''(2s+1)+2]^2}\nonumber\\
&
&\times\Bigl(\sum_{\overline{\sigma}=\pm1}\overline{\sigma}A_{s+\sigma''(1+\overline{\sigma})}^{\sigma
m}H_{s,s_1}^{m+\sigma(1-\overline{\sigma})/2,\sigma''}\Bigr)^2,\nonumber\\
& &\\
\bigl|Z_{2\sigma''}^{\sigma+\sigma'}\bigr|^2&=&\frac{J^2f_3^{(2)}(\theta,\phi)}{16\Bigl(\sigma'\gamma
B+J[\sigma''(2s+1)+2]\Bigr)^2}\nonumber\\
&
&\times\Bigl(\sum_{\overline{\sigma}=\pm1}\overline{\sigma}A_{s+\sigma''(1+\overline{\sigma})}^{\sigma
m+\sigma\sigma'(1+\overline{\sigma})/2}
R_{s,s_1}^{m+\sigma(1-\overline{\sigma})/2,\sigma',\sigma''}\Bigr)^2,\nonumber\\
& &\\
\bigl|Z_{2\sigma''}^{\sigma+2\sigma'}\bigr|^2&=&\frac{J^2f_6^{(2)}(\theta,\phi)}{64\Bigl(2\sigma'\gamma
B+J[\sigma''(2s+1)+2]\Bigr)^2}\nonumber\\
&
&\times\Bigl(\sum_{\overline{\sigma}=\pm1}\overline{\sigma}A_{s+\sigma''(1+\overline{\sigma})}^{\sigma
m+\sigma\sigma'(1+\overline{\sigma})}
K_{s,s_1}^{\sigma'm+\sigma\sigma'(1-\overline{\sigma})/2,\sigma''}\Bigr)^2,\nonumber\\
\end{eqnarray}
 where
$E_{s,s_1}^m=E_s^{m,(0)}+E_{s,s_1}^{m,(1)}$ is given by Eqs.
(\ref{E0}) and (\ref{Esm1}), $A_s^m$, $F_s^x$,
$H_{s,s_1}^{m,\sigma'}$, $K_{s,s_1}^{x,\sigma'}$,
$\alpha_{s,s_1}$, and $R_{s,s_1}^{m,\sigma,\sigma'}$ are given by
Eqs. (\ref{Asm}), (\ref{Fsx}), (\ref{Hss1msigma}),
 (\ref{Kss1xsigma}), (\ref{alphass1}), and (\ref{Rss1xsigma}),
respectively, the $f_n^{(2)}(\theta,\phi)$ are given in Eqs.
(\ref{f1})-(\ref{f7}), and we have used the relations following
Eq. (\ref{G45}). We note that since $\sigma=\pm1$ for clockwise
and counterclockwise oscillatory induction senses, setting
$\sigma',\sigma''=\pm1$ is equivalent to setting them equal to
$\pm\sigma$.

For the three cases with $m'=m$, there are no associated resonant
magnetic inductions, and since the EPR frequency $\omega$ cannot
be easily varied in an experiment, these terms are irrelevant. For
the 10 remaining cases with $m'\ne m$, the resonant magnetic
induction is then found to have the form given by Eq.
(\ref{Bres}).  The term with amplitude $Z_{0}^{2\sigma}$ has
$a_1=1/2$, $b_1=2(m+\sigma)$, and $c_1=0$.  The term with
amplitude $Z_{0}^{-\sigma}$ has $a_2=1$, $b_2=(2m-\sigma)/2$, and
$c_2=0$.  The two terms with amplitudes $Z_{2\sigma''}^{\sigma}$
where $\sigma''=\pm\sigma$ have $a_3=1$,
$b_3=\sigma[2+\sigma''(2s+1)]$, and $c_3=-(2m+\sigma)/2$. The two
terms with amplitudes $Z_{2\sigma''}^{2\sigma}$, where
$\sigma''=\pm\sigma$, have $a_4=2$,
$b_4=2\sigma[2+\sigma''(2s+1)]$, and $c_4=4(m+\sigma)$. The four
terms with amplitudes $Z_{2\sigma''}^{\sigma+2\sigma'}$, where
$\sigma',\sigma''=\pm\sigma$, have $a_5=1/(1+2\sigma\sigma')$,
$b_5=[2+\sigma''(2s+1)]/(\sigma+2\sigma')]$, and
$c_5=m+\sigma'+\sigma/2$.

\section{Appendix D}

\subsection{Second order induction representation Hamiltonian}

In this appendix, we evaluate the corrections to the eigenstate
energies second order in the four anisotropy interaction energies
$J_j$ for $j=a,b,d,e$. The operations of the rotated Hamilatonian
${\cal H}'$ upon the eigenstates $|\varphi_s^m\rangle$ may be
written as
\begin{eqnarray}
{\cal
H}'|\varphi_s^m\rangle&=&(E_{s}^{m,(0)}+E_{s,s_1}^{m,(1)})|\varphi_s^m\rangle
+\sum_{\sigma'=\pm1}{\cal W}_{s,s_1}^{m,\sigma'}|\varphi_{s+2\sigma'}^m\rangle\nonumber\\
& &+\sum_{\sigma=\pm1}\Bigl({\cal
U}_{s,s_1}^{m,\sigma}|\varphi_s^{m+\sigma}\rangle+{\cal
V}_{s,s_1}^{m,\sigma}|\varphi_s^{m+2\sigma}\rangle\Bigr)\nonumber\\
& &+\sum_{\sigma,\sigma'=\pm1}\Bigl({\cal
X}_{s,s_1}^{m,\sigma,\sigma'}|\varphi_{s+2\sigma'}^{m+\sigma}\rangle\nonumber\\
& &\qquad+{\cal
Y}_{s,s_1}^{m,\sigma,\sigma'}|\varphi_{s+2\sigma'}^{m+2\sigma}\rangle\Bigr),\label{Hprime}
\end{eqnarray}
where $E_{s,s_1}^{m,(0)}$ and $E_{s,s_1}^{m,(1)}$ are given by
Eqs. (\ref{E0}) and (\ref{Esm1}), respectively, and the ${\cal U}$
to ${\cal Y}$ functions are given by Eqs. (\ref{calU}) to
(\ref{calY}), respectively.

\subsection{Second order eigenstate energies}

 From Eq. (\ref{Hprime}), the second order
eigenstate energies may be written as
\begin{eqnarray}
E_{s,s_1}^{m,(2)}&=&\frac{1}{\gamma
B}\sum_{\sigma=\pm1}\sigma\Bigl(\Bigl|{\cal
U}_{s,s_1}^{m,\sigma}\Bigr|^2+\frac{1}{2}\Bigl|{\cal
V}_{s,s_1}^{m,\sigma}\Bigr|^2\Bigr)\nonumber\\
& &+\sum_{\sigma'=\pm1}\frac{|{\cal
W}_{s,s_1}^{m,\sigma'}|^2}{J[2+(2s+1)\sigma']}\nonumber\\
& &+\sum_{\sigma,\sigma'\pm1}\Bigl(\frac{|{\cal
X}_{s,s_1}^{m,\sigma,\sigma'}|^2}{J[2+(2s+1)\sigma']+\sigma\gamma
B}\nonumber\\
& &+\frac{|{\cal
Y}_{s,s_1}^{m,\sigma,\sigma'}|^2}{J[2+(2s+1)\sigma']+2\sigma\gamma
B}\Bigr).
\end{eqnarray}

For simplicity, we rewrite this as

\begin{eqnarray}
E_{s,s_1}^{m,(2)}&=&E_{s,s_1}^{m,(2){\cal
U}}+E_{s,s_1}^{m,(2){\cal
V}}+E_{s,s_1}^{m,(2){\cal W}}\nonumber\\
& &+E_{s,s_1}^{m,(2){\cal X}}+E_{s,s_1}^{m,(2){\cal Y}},\\
E_{s,s_1}^{m,(2){\cal U}}&=&\frac{m\sin^2\theta}{2\gamma
B}[4s(s+1)-8m^2-1]\nonumber\\
&
&\times\Bigl(\cos^2\theta[\tilde{J}_{b,a}^{s,s_1}-\cos(2\phi)\tilde{J}_{d,e}^{s,s_1}]^2\nonumber\\
& &\qquad+\sin^2(2\phi)(\tilde{J}_{d,e}^{s,s_1})^2\Bigr),\label{Esm2}\\
E_{s,s_1}^{m,(2){\cal V}}&=&-\frac{m}{8\gamma B}[2s(s+1)-2m^2-1]\nonumber\\
&
&\times\Bigl([\sin^2\theta\tilde{J}_{b,a}^{s,s_1}+(1+\cos^2\theta)\cos(2\phi)\tilde{J}_{d,e}^{s,s_1}]^2\nonumber\\
&
&\qquad+4\cos^2\theta\sin^2(2\phi)(\tilde{J}_{d,e}^{s,s_1})^2\Bigr),\\
E_{s,s_1}^{m,(2){\cal
W}}&=&\frac{d_{s,s_1}^m}{16J}\Bigl(J_a-3[J_a\cos^2\theta+J_e\sin^2\theta\cos(2\phi)]\Bigr)^2,\nonumber\\
& &\label{Esm2W}\\
E_{s,s_1}^{m,(2){\cal
X}}&=&\frac{f_{s,s_1}^m(\gamma B/J)\sin^2\theta}{2J}\nonumber\\
& &\times\Bigl([J_a-J_e\cos(2\phi)]^2\cos^2\theta\nonumber\\
& &\qquad+\sin^2(2\phi)J_e^2\Bigr),\label{Esm2X}\\
E_{s,s_1}^{m,(2){\cal
Y}}&=&\frac{g_{s,s_1}^m(\gamma B/J)}{64J}\Bigl[\Bigl(J_a\sin^2\theta\nonumber\\
& &\qquad+J_e(1+\cos^2\theta)\cos(2\phi)\Bigr)^2\nonumber\\
& &\qquad+4\cos^2\theta\sin^2(2\phi)J_e^2\Bigr],\label{Esm2Y}
\end{eqnarray}
where
\begin{eqnarray}
d_{s,s_1}^m&=&-\frac{(s^2-m^2)[(s-1)^2-m^2]\eta_{s,s_1}^2\eta_{s-1,s_1}^2}{(2s-1)}\nonumber\\
&&+\eta_{s+2,s_1}^2\eta_{s+1,s_1}^2\nonumber\\
& &\times
\frac{[(s+1)^2-m^2][(s+2)^2-m^2]}{(2s+3)},\nonumber\\
f_{s,s_1}^m(x)&=&-\frac{\eta_{s,s_1}^2\eta_{s-1,s_1}^2(s^2-m^2)}{(2s-1)^2-x^2}\nonumber\\
& &\times\Bigl((2s-1)[(s-1)(s-2)+m^2]\nonumber\\
& &\qquad-m(2s-3)x\Bigr)\nonumber\\
&
&+\frac{\eta_{s+2,s_1}^2\eta_{s+1,s_1}^2[(s+1)^2-m^2]}{(2s+3)^2-x^2}\nonumber\\
& &\times\Bigl((2s+3)[(s+2)(s+3)+m^2]\nonumber\\
& &\qquad-m(2s+5)x\Bigr),\\
g_{s,s_1}^m(x)&=&-\frac{2\eta_{s,s_1}^2\eta_{s-1,s_1}^2}{(2s-1)^2-4x^2}\Bigl((2s-1)[m^4\nonumber\\
& &+m^2(6s^2-18s+11)\nonumber\\& &+s(s-1)(s-2)(s-3)]\nonumber\\
& &-4mx(2s-3)(s^2-3s+1+m^2)\Bigr)\\
& &
+\frac{2\eta_{s+2,s_1}^2\eta_{s+1,s_1}^2}{(2s+3)^2-4x^2}\Bigl((2s+3)[m^4\nonumber\\
& &+m^2(6s^2+30s+35)\nonumber\\
& &+(s+1)(s+2)(s+3)(s+4)]\nonumber\\
& &-4mx(2s+5)(s^2+5s+5+m^2)\Bigr).
\end{eqnarray}

There is a remarkable amount of symmetry in the angular dependence
of the eigenstate energies.  We note that $E_{s,s_1}^{m,(2){\cal
X}}(\theta,\phi)$ and $E_{s,s_1}^{m,(2){\cal U}}(\theta,\phi)$
have the same forms, differing in the replacements of the
interactions $\tilde{J}_{b,a}^{s,s_1}$ and
$\tilde{J}_{d,e}^{s,s_1}$ with $J_a$ and $J_e$, respectively, and
with  different overall constant functions. The same comparison
can also be made with $E_{s,s_1}^{m,(2){\cal Y}}(\theta,\phi)$ and
$E_{s,s_1}^{m,(2){\cal V}}(\theta,\phi)$.  In addition, we note
that there is a remarkable similarity in the $\theta,\phi$
dependence of $E_{s,s_1}^{m,(2){\cal W}}$ with that of the local
spin anisotropy part of $B_{s,s_1}^{{\rm lc}(1)}(\theta,\phi)$
given by Eq. (\ref{Bstep}).

The contributions to the $s$th level crossing second order in the
anisotropy interactions are calculated as indicated in Eq.
(\ref{Bstep2}), and are found to be
\begin{eqnarray}\Bigl(E_{s,s_1}^{s,(2)}-E_{s-1,s_1}^{s-1,(2)}\Bigr)\Bigr|_{B=-Js/\gamma}&=&J\sum_{n=1}^7a_n(s,s_1)f_n^{(2)}(\theta,\phi),\label{level2}\nonumber\\
\end{eqnarray}
where the $f_n^{(2)}$ are given by Eqs. (\ref{f1})-(\ref{f7}),
respectively.

\subsection{Second order level crossing coefficients}

The coefficients are given by
\begin{eqnarray}
a_1(s,s_1)&=&\frac{(9-20s+12s^2)}{2s},\\
a_2(s,s_1)&=&\frac{[a_{2,0}(s)+4s_1(s_1+1)a_{2,1}(s)]}{s(2s+1)(2s+3)},\\
a_{2,0}(s)&=&3(9+s-23s^2+4s^3+12s^4),\nonumber\\
a_{2,1}(s)&=&9-8s-4s^2,\\
a_3(s,s_1)&=&a_3^{\cal U}(s,s_1)+a_3^{\cal X}(s,s_1),\\
a_3^{\cal
U}(s,s_1)&=&\frac{[3-3s-3s^2+4s_1(s_1+1)]^2}{2(2s+3)^2}\nonumber\\
& &-\frac{(s-1)[3+3s-3s^2+4s_1(s_1+1)]^2}{2s(2s+1)^2},\nonumber\\
& &\\
a_3^{\cal
X}(s,s_1)&=&\frac{[s(s+2)-4s_1(s_1+1)]}{2(s+1)(s+3)(2s+1)^2(2s+3)^2}\nonumber\\
& &\times\frac{a_{3,0}^{\cal X}(s)+4s_1(s_1+1)a_{3,1}^{\cal X}(s)}{(2s+5)(3s+1)},\nonumber\\
a_{3,0}^{\cal X}(s)&=&(s+1)(s+3)(51+114s\nonumber\\
& &+209s^2+302s^3+164s^4+24s^5),\\
a_{3,1}^{\cal X}(s)&=&129+318s+395s^2\nonumber\\
& &+442s^3+300s^4+72s^5,\nonumber\\
a_{4}(s,s_1)&=&\frac{(4s-3)}{8s},\\
a_5(s,s_1)&=&-\frac{3[a_{5,0}(s)+4s_1(s_1+1)]}{4s(2s+1)(2s+3)},\\
a_{5,0}(s)&=&3+3s-5s^2-4s^3,\\
a_6(s,s_1)&=&a_6^{\cal V}(s,s_1)+a_6^{\cal Y}(s,s_1),\\
 a_6^{\cal
 V}(s,s_1)&=&\frac{[3-3s-3s^2+4s_1(s_1+1)]^2}{8(2s-1)(2s+3)^2}\nonumber\\
 &
 &-\frac{(s-1)[3+3s-3s^2+4s_1(s_1+1)]^2}{8s(2s-3)(2s+1)^2},\nonumber\\
 & &\\
 a_6^{\cal
Y}(s,s_1)&=&\frac{1}{4(2s-3)(2s-1)(2s+1)^2(2s+3)^2}\nonumber\\
&
&\times\frac{\sum_{n=0}^2a_{6,n}^{\cal Y}(s)[4s_1(s_1+1)]^n}{(2s+5)(4s+1)(4s+3)},\nonumber\\
a_{6,0}^{\cal Y}(s)&=&s\Bigl(-216+837s+5052s^2 +3521s^3\nonumber\\
& &-12414s^4-21876s^5-7464 s^6\nonumber\\
& &+6720s^7+5376s^8+1024s^9\Bigr),\\
a_{6,1}^{\cal Y}(s)&=&3\Bigl(81+102s+10s^2\nonumber\\
& &+196s^3+296s^4+80s^5\Bigr),\\
a_{6,2}^{\cal Y}(s)&=&189+258s-100s^2+152s^3\nonumber\\
& &+640s^4+256s^5,\\
a_7(s,s_1)&=&\frac{[s(s+2)-4s_1(s_1+1)]}{4(2s+1)^3(2s+3)^3(2s+5)}\nonumber\\
&
&\times\Bigl(4s_1(s_1+1)(-1+38s+60s^2+24s^3)\nonumber\\
& &+(s+1)(3+67s+94s^2+44s^3+8s^4)\Bigr).\nonumber\\
\end{eqnarray}

By expanding the solutions in the crystal representation to second
order in the $J_j$, we have explicitly checked these expressions
for $s_1=1/2, s=1$, and for $s_1=1$, $s=1,2$.  We note that for
$s_1=1/2$, only $a_1$ and $a_4$ are non-vanishing.

\subsection{Two-level thermodynamic coefficients $a_{1,s}$}
Here we present the expression for $a_{1,s}$ to ${\cal
O}(J_j/J)^2$ appearing in Eqs. (\ref{epsilonsb}) and
(\ref{peak})-(\ref{slope}) in the text. We have
\begin{eqnarray}
a_{1,s}&=&s+\sum_{n=1}^6b_n(s,s_1)f_n^{(2)}(\theta,\phi)+{\cal O}(J_j/J)^3,\nonumber\\
\end{eqnarray} where the $f_n(\theta,\phi)$ are given by Eqs.
(\ref{f1})-(\ref{f6}), and
\begin{eqnarray}
b_1(s,s_1)&=&\frac{(2s-1)^2}{2s}\Theta(s-1),\\
b_2(s,s_1)&=&\frac{(2s-1)^2}{s}\alpha_{s,s_1}\Theta(s-1),\\
b_3(s,s_1)&=&\frac{(2s-1)^2}{2s}\alpha^2_{s,s_1}\Theta(s-1)+b_3^{\cal X}(s,s_1),\\
b_3^{\cal
X}(s,s_1)&=&-\frac{s(27+44s+27s^2+6s^3)}{6(s+1)^2(s+3)^2(2s+3)^2(2s+5)}\nonumber\\
&
&\times\Bigl([4s_1(s_1+1)-s(s+2)]\nonumber\\
& &\times[4s_1(s_1+1)-(s+1)(s+3)]\Bigr),\\
b_4(s,s_1)&=&\frac{(2s-1)}{8s}\Theta(s-1),\\
b_5(s,s_1)&=&\frac{(2s-1)}{4s}\alpha_{s,s_1}\Theta(s-1),\\
b_6(s,s_1)&=&\frac{(2s-1)}{8s}\alpha^2_{s,s_1}\Theta(s-1)+b_6^{\cal Y}(s,s_1),\\
b_6^{\cal
Y}(s,s_1)&=&\frac{s(s-1)}{8(4s^2-1)}[(s^2-1-4s_1(s_1+1)]\nonumber\\
& &\times[s(s-2)-4s_1(s_1+1)]\nonumber\\
&
&-\frac{sb_{6,1}(s)[(s(s+2)-4s_1(s_1+1)]}{72(2s+1)(2s+3)^2(2s+5)(4s+3)^2}\nonumber\\
& &\times[(s+1)(s+3)-4s_1(s_1+1)],\\
b_{6,1}(s)&=&369+1011s+1420s^2+1076s^3\nonumber\\
& &+416s^4+64s^5, \end{eqnarray}
 where $\Theta(x)=1$ for $x\ge0$,
$\Theta(x)=0$ for $x<0$ is the Heaviside step function.


\begin{thebibliography}{99}
\bibitem{background}
R. Sessoli, D. Gatteschi, A. Caneschi, and M. A. Novak, Nature
(London) {\bf 365}, 141 (1993).

\bibitem{sarachik}
J. R. Friedman, M. P. Sarachik, J. Tejada, and R. Ziolo, Phys.
Rev. Lett. {\bf 76}, 3830 (1996).

\bibitem{loss}
M. N. Leuenberger and D. Loss, Nature (London) {\bf 410}, 789
(2001).

\bibitem{Fe8}
W. Wernsdorfer, T. Ohm, C. Sangregorio, R. Sessoli, D. Mailly, and
C. Paulsen, Phys. Rev. Lett. {\bf 82}, 3903 (1999).

\bibitem{WS}
W. Wernsdorfer and R. Sessoli, Science {\bf 284}, 133 (1999).

\bibitem{Fe8spin9}
D. Zipse, J. M. North, N. S. Dalal, S. Hill, and R. S. Edwards,
Phys. Rev. B {\bf 68}, 184408 (2003).

\bibitem{Fering1}
M. Affronte, A. Cornia, A. Lascialfari, F. Borsa, D. Gatteschi, J.
Hinderer, M. Horvati{\'c}, A. G. M. Jansen, and M.-H. Julien,
Phys. Rev. Lett. {\bf 88}, 167201 (2002).

\bibitem{Fe6ring}
O. Waldmann, J. Sch{\"u}lein, R. Koch, P. M{\"u}ller, I. Bernt, R.
W. Saalfrank, H. P. Andres, H. U. G{\"u}del, and P. Allenspach,
Inorg. Chem. {\bf 38},  5879 (1999).

\bibitem{Fe8ring}
O. Waldmann, R. Koch, S. Schromm, J. Sch{\"u}lein, P. M{\"u}ller,
I. Bernt, R. W. Saalfrank, F. Hempel, and E. Balthes, Inorg. Chem.
{\bf 40},  2986 (2001).

\bibitem{Fering2}
H. Nakano and S. Miyashita, J. Phys. Chem. Solids {\bf 63}, 1519
(2002).

\bibitem{V2neutron}
D. A. Tennant, S. E. Nagler, A. W. Garrett, T. Barnes, and C. C.
Torardi, Phys. Rev. Lett. {\bf 78}, 4998 (1997).
\bibitem{Gudel}
 H. U. G{\"u}del, Neutron News {\bf 7}, 24 (1996)

\bibitem{V2P2O9}
A. W. Garrett, S. E. Nagler, D. A. Tennant, B. C. Sales, and T.
Barnes, Phys. Rev. Lett. {\bf 79}, 745 (1997).

\bibitem{ek}
D. V. Efremov and R. A. Klemm, Phys. Rev. B {\bf 66}, 174427
(2002).

\bibitem{Fe2}
F. Le Gall, F. Fabrizi de Biani, A. Caneschi, P. Cinelli, A.
Cornia, A. C. Fabretti, and D. Gatteschi, Inorg. Chim. Acta {\bf
262}, 123 (1997).

\bibitem{Fe2mag}
Y. Shapira, M. T. Liu, S. Foner, R. J. Howard, and W. H.
Armstrong, Phys. Rev. B {\bf 63}, 094422 (2001).

\bibitem{Fe2Cl}
Y. Shapira, M. T. Liu, S. Foner, C. E. Dub{\'e}, and P. J.
Bonitatebus, Jr., Phys. Rev. B {\bf 59}, 1046 (1999).


\bibitem{taft}
K. L. Taft, C. D. Delfs, G. C. Papaefthymiou, S. Foner, D.
Gatteschi, and S. J. Lippard, J. Am. Chem. Soc. {\bf 116}, 823
(1994).

\bibitem{Fe2Cl3}
J. D. Walker and R. Poli, Inorg. Chem. {\bf 29}, 756 (1990).

\bibitem{Fe2Clnew}
J. A. Bertrand, J. L. Breece, and P. G. Eller, Inorg. Chem. {\bf
13}, 125 (1974).


\bibitem{Mn4dimer}
R. Tiron, W. Wernsdorfer, D. Foguet-Albiol, N. Aliaga-Alcalde, and
G. Christou, Phys. Rev. Lett. {\bf 91}, 227203 (2003).


\bibitem{Mn4dimerDalal}
J. M. North, N. S. Dalal, D. Foguet-Albiol, A. Vinslava, and G.
Christou, Phys. Rev. B {\bf 69}, 174419 (2004).


\bibitem{WaldmannNi}
O. Waldmann, J. Hassmann, P. M{\"u}ller, D. Volkmer, U. S.
Schubert, and J.-M. Lehn, Phys. Rev. B {\bf 58}, 3277 (1998).

\bibitem{Almenar}
J. J. Borr{\'a}s-Almenar, J. M. Clemente-Juan, E. coronado, and B.
S. Tsukerblat, Inorg. Chem. {\bf 38},  6081 (1999).

\bibitem{ekcondmat}
D. V. Efremov and R. A. Klemm, cond-mat/0409168.

\bibitem{white}
R. M. White, {\it Quantum Theory of Magnetism} (McGraw-Hill, New
York, 1970), Chapters 1, 5, 8.

\bibitem{park}
K. Park, M. A. Novotny, N. S. Dalal, S. Hill, and P. A. Rikvold,
Phys. Rev. B {\bf 66}, 144409 (2002).

\bibitem{klemm}
R. A. Klemm and J. R. Clem, Phys. Rev. B {\bf 21}, 1868 (1980); R.
A. Klemm, SIAM J. Appl. Math. {\bf 55}, 986 (1995).
\bibitem{Goldstein}
 H. Goldstein, {\it Classical Mechanics},
(Addison-Wesley, Reading, MA, 1965), p. 109.


\end{thebibliography}
\end{document}